\documentclass[11pt]{article}

    \usepackage{amsthm}
    \usepackage{amssymb}
    \usepackage{amsmath}
    \usepackage{eucal}
    \usepackage{feynmf}
    \usepackage{hyperref}
    \usepackage{mathrsfs}
    \usepackage{yfonts}
    \usepackage{bbm}
    \usepackage[T1]{fontenc}
    \usepackage{bbm}
    \usepackage{geometry}
    \usepackage[all]{xy}

  \newcommand{\nn}{\nonumber}
  
  \newcommand{\al}{\alpha}
  
  \newcommand{\ga}{\gamma}
  \newcommand{\Ga}{\Gamma}

  \newcommand{\eps}{\varepsilon}
  \newcommand{\ka}{\kappa}

  \newcommand{\la}{\lambda}
  \newcommand{\La}{\Lambda}
  \newcommand{\si}{\sigma}
  \newcommand{\vf}{\varphi}
  \newcommand{\om}{\omega}

  \renewcommand{\.}{\cdot} 
  \newcommand{\ut}{\tilde u}
  \renewcommand{\d}{\partial}
  \newcommand{\be}{\begin{equation}}
  \newcommand{\ee}{\end{equation}}

  \newcommand{\NN}{\mathbbmss{N}} 
  \newcommand{\RR}{\mathbbmss{R}} 
  \newcommand{\CC}{\mathbbmss{C}} 
  \newcommand{\supp}{\mathrm{supp}}

  \newcommand{\Dcal}{\mathcal{D}}

  \newcommand{\Hcal}{\mathcal{H}}

  \newcommand{\Pcal}{\mathcal{P}}

  \DeclareMathAlphabet{\mathpzc}{OT1}{pzc}{m}{it}

   \author{\null\\Michael D\"utsch$^{(1)}$\\
 \null\\
        \small{$^{(1)}$ Institut f\"ur Theoretische Physik, Universit\"at G\"ottingen,}\\
\small{Friedrich-Hund-Platz 1, D-37077 G\"ottingen, Germany}\\
\small{michael.duetsch@theorie.physik.uni-goe.de}}

    \title{Massive vector bosons: is the geometrical interpretation as a spontaneously 
  broken gauge theory possible at all scales?}

\begin{document}

   \maketitle

  \begin{abstract}
The usual derivation of the Lagrangian of a
model for massive vector bosons, by spontaneous 
symmetry breaking of a gauge theory, implies that the prefactors 
of the various interaction terms are uniquely determined functions of the coupling 
constant(s) and the masses.
Since, under the renormalization group (RG) flow,
different interaction terms get different 
loop-corrections, it is uncertain, whether these functions remain fixed under 
this flow. We investigate this question for the $U(1)$-Higgs-model to 1-loop order 
in the framework of Epstein-Glaser renormalization. Our main result reads: choosing the 
renormalization mass scale(s) in a way corresponding 
to the minimal subtraction scheme, the geometrical interpretation as a spontaneously 
broken gauge theory gets lost under the RG-flow. This holds also for the
clearly stronger property of BRST-invariance of the Lagrangian. On the other
hand we prove that physical consistency, which is a weak form of BRST-invariance
of the time-ordered products, is maintained under the RG-flow.
  \end{abstract}

    \theoremstyle{plain}
    \newtheorem{df}{Definition}[section]
    \newtheorem{thm}[df]{Theorem}
    \newtheorem{prop}[df]{Proposition}
    \newtheorem{cor}[df]{Corollary}
    \newtheorem{lemma}[df]{Lemma}

    \theoremstyle{plain}
    \newtheorem*{Main}{MainTheorem}
    \newtheorem*{MainT}{MainTechnical Theorem}

    \theoremstyle{definition}
    \newtheorem{rem}[df]{Remark}
    \newtheorem{example}[df]{Example}

   \theoremstyle{definition}
    \newtheorem{ass}{\underline{\textit{Assumption}}}[section]


  \tableofcontents
  \markboth{Contents}{Contents}
  \section{Introduction}

The classical geometrical concepts of fibre bundles and group theory have been crucial for the 
development of quantum gauge theories, and the Higgs mechanism was the key to incorporate the 
electroweak interaction into the framework of renormalizable field theory. 

However, to the best 
of our knowledge, the Higgs mechanism is not well understood on a purely quantum level. 
And it is not needed: starting with massive BRST-invariant free fields, making a general 
renormalizable ansatz for the interaction and requiring physical consistency (PC) 
\cite{Kugo-Ojima,DuetschSchroer2000,Grigore2000} or perturbative gauge invariance (PGI) 
\cite{DS99,AsteDuetschScharf99,Scharf2001,GraciaBondia2010, Scharf2010}\footnote{PGI was 
first introduced in \cite{DHKS94}; in \cite{GraciaBondia2010} it is called
  'causal gauge invariance'.} 
one obtains a consistent perturbative quantum theory of massive vector bosons.
(Some obvious properties as Poincar{\'e} invariance and relevant discrete symmetries are also taken into 
account.) PC is the condition that the free BRST-charge%
\footnote{That is the charge implementing the BRST-transformation of the asymptotic free fields.}
commutes with the ``$S$-matrix'' in the adiabatic limit, in order that the latter induces a 
well-defined operator on the physical subspace; PGI is a refinement of this condition
which is formulated independently of the adiabatic limit -- a sufficient (but in general not necessary)
condition for PC.
If the ansatz for the interaction contains only trilinear and quadrilinear fields\footnote{Throughout
  this paper we use the words bilinear, trilinear and quadrilinear in the sense of bi-, tri- and
  quadrilinear in the basic fields.} the resulting Lagrangian is essentially unique and agrees 
  with what one obtains from spontaneous symmetry breaking of a gauge theory; in particular the 
presence of Higgs particles and chirality of fermionic interactions
can be understood in this way without recourse to any geometrical or
group theoretical concepts \cite{StoraVienna1997,DS99}. These derivations of the interaction
from basic QFT-principles use PGI (or PC) only on the level of tree diagrams (PGI-tree).

In the literature the geometrical interpretation of the Standard Model of electroweak interactions
as a spontaneously broken gauge theory is frequently used at several (or even all) scales.
This is evident for the cosmological models relying on the ``electroweak phase transition''.
Or, looking carefully at the geometrical derivations of a value for the Higgs mass of Connes et 
al. (\cite{Connes2007} and references cited therein) and 
Tolksdorf \cite{Tolk2007}, we realized that in these papers
the geometrical interpretation is used at two very 
distinct scales: at the $Z$-mass and at the unification scale.

This paper was initiated by serious doubts about the geometrical interpretability at all scales, 
which rely on the following: this interpretation is 
possible iff the prefactors of the various interaction terms
(i.e.~of the vertices) are prescribed functions of the coupling constant(s) and masses. Since
different vertices get different loop-corrections it is uncertain, whether these functions remain
fixed under the RG-flow.

Similarly to the conventional literature \cite{Sibold}, our RG-flow depends strongly
on the renormalization scheme. Naively one might think that this is not so,
because we define the RG-flow by a scaling transformation \cite{HW03,DF04,BDF09}.%
\footnote{This seems to be the obvious way to introduce the
  RG-flow in the Epstein-Glaser framework \cite{EG73}. Namely, in this
  framework renormalization is the extension of distributions (see
footnote \ref{fn:extension}) and, as long as
  the adiabatic limit \eqref{adlim} is not performed, 
renormalization in this sense cannot be interpreted as a redefinition of fields, 
masses and coupling constants depending on a mass scale.}
But the scheme dependence comes in by the choice of the renormalization mass scale(s) $M$:
the scaling transformation may act on $M$ or it may not, and different choices for 
different Feynman diagrams are possible. 

An important result of this paper is that 
\textit{PC is maintained under the renormalization group (RG) flow} (Sect.~\ref{sec:stab-PC}). 
It is well known that also renormalizability (by power counting) is preserved.
But, our original hope that these two properties yield enough information about the running 
interaction to answer the geometrical interpretability, turned out to be too optimistic.
Due to the presence of bilinear fields, PC and renormalizability 
are much less restrictive than in the above mentioned calculations 
involving only tri- and quadrilinear fields.

For this reason we proceed in a less elegant way: we answer the geometrical interpretability
by means of a lot of explicit 1-loop computations of the RG-flow (Sect.~\ref{sec:1-loop}). Since,
up to a few scalar field examples in \cite{DF04,BDF09}, such calculations have not yet been 
done in the framework of Epstein-Glaser renormalization, we explain them in detail
(see Sects.~\ref{ssec:explicit-comput}-\ref{ssec:e=e} and Appendices 
\ref{app:violation-scaling}-\ref{app:z(L)-coefficients}).

To get information about the important question whether PGI is maintained under the 
RG-flow, we analyze PGI-tree for the running interaction (Sec.~\ref{sec:PGI-tree}).

BRST-invariance of the Lagrangian is a property which is truly stronger than the geometrical 
interpretabilty and also stronger than PGI-tree. We investigate whether it can be preserved 
under the RG-flow by a suitable renormalization prescription (Sects.~\ref{sec:BRST} and 
\ref{sec:1-loop}).

  We assume that the reader is familiar with the formalism for 
Epstein-Glaser renormalization (also called ``causal perturbation theory'')
  given in \cite{DF04}, in particular we will use the Main Theorem, 
which is the basis for our definition of the RG-flow, and the scaling and mass expansion
\cite{HW02,Duetsch2014}.

  \section{Precise formulation of the question}

\textbf{The Lagrangian of the model:}
  to simplify the calculations we study only one massive vector field $A^\mu$, the 
  corresponding St\"uckelberg field $B$, a further real scalar field $\vf$ 
(usually called ``Higgs field'') and the 
  Fadeev-Popov ghost fields $(u\,,\,\tilde u)$. We work with the free Lagrangian 
  \begin{align} 
  L_0^{\mathbf{m},\La} &= -\frac{1}4\,F^2 + \frac{m^2}{2} (A\. A) 
  + \frac{1}{2} (\d B\. \d B) -\frac{m_B^2}{2}\, B^2 
  - \frac{\La}{2} (\d A)^2 \nn
  \\[\jot] 
  &\qquad + \frac{1}{2} (\d\vf\. \d\vf) - \frac{m_H^2}{2}\, \vf^2 
  + \d\ut\. \d u - m_u^2\, \ut u \ , \label{L-free}
  \end{align} 
  where $F^2:=(\d^\mu A^\nu-\d^\nu A^\mu)(\d_\mu A_\nu-\d_\nu A_\mu)$,
$\mathbf{m}:=(m,m_B,m_u,m_H)$ denotes the masses of the various basic fields
and $\La$ is the gauge-fixing parameter.

For the moment we do not care about any notion of gauge symmetry and
admit interactions of the form
  \begin{align}\label{L-int}
  L^{\mathbf{m},\La}_{\ka,\mathbf{\la}}
  &= \ka\Bigl( m(A\.A)\vf - \frac{\la_{10}\,m_u^2}{m}\ut u\vf + \la_1 B(A\.\d\vf)\nn\\ 
  &\quad - \la_2 \vf(A\.\d B) - \frac{\la_3 m_H^2}{2m} \vf^3
  - \frac{\la_4 m_H^2}{2m} B^2\vf\Bigr)\nn \\
  &\quad + \ka^2\Bigl(\frac{\la_5}{2} (A\.A) \vf^2 + \frac{\la_6}{2} (A\.A) B^2
  - \frac{\la_7m_H^2}{8m^2} \vf^4 \nn\\
  &\quad - \frac{\la_8m_H^2}{4m^2} \vf^2 B^2
  - \frac{\la_9m_H^2}{8m^2} B^4+\la_{11}\,(A\.A)^2\Bigr)\nn\\
  &\quad+((\la_{12}-1)m+\sqrt{\La}\,m_B)\,(A\cdot\d B)\ ,
  \end{align}
  where $\ka$ is the coupling constant and
$\mathbf{\la}:=(\la_1,...,\la_{12})$ are arbitrary real parameters. 
Apart from the last, bilinear term, each field monomial in $L^{\mathbf{m},\La}_{\ka,\mathbf{\la}}$ has its
own, independent coupling constant $\ka\la_j$ or $\ka^2\la_j$.
The reason for the complicated definition of 
  $\la_{12}$ will become clear below in \eqref{parameter-geom}-\eqref{geometric}.
 The free Lagrangian is parametrized by $\mathbf{m}$ and $\La$;  
the interaction $L$ has 13 additional parameters: $\ka$ and the dimensionless coupling parameters
  $\mathbf{\la}$. We point out that at the present stage
we do {\bf not} assume the usual mass relations 
  $m_B=m_u=\frac{m}{\sqrt{\La}}$, we consider $m,\, 
  m_B$ and $m_u$ as independent parameters.

The set of monomials appearing in $L^{\mathbf{m},\La}_{\ka,\mathbf{\la}}$ \eqref{L-int} is the minimal set 
with the following properties:
\begin{itemize}
\item $(L_0^{\mathbf{m},\La}+L^{\mathbf{m},\La}_{\ka,\mathbf{\la}})$ contains all monomials which appear in the
Lagrangian of the $U(1)$-Higgs model;
\item computing the RG-flow for the model given by $(L_0^{\mathbf{m},\La}+L^{\mathbf{m},\La}_{\ka,\mathbf{\la}})$,
there do not appear any new field monomials in the running interaction, except for a constant field
$k\in\CC$ (see \eqref{z(L)}), i.e.~the
set of field monomials appearing in $(L_0^{\mathbf{m},\La}+L^{\mathbf{m},\La}_{\ka,\mathbf{\la}})$ is stable
under the RG-flow.
\end{itemize}

We point out that each term in $L_0$ and $L$ is even under the field parity transformation
  \be\label{fieldparity}
  (A,B,\vf,u,\ut)\mapsto (-A,-B,\vf,u,\ut)\ .
  \ee
Setting $\ut:=0$ and $u:=0$ and ignoring the $A\d B$-term, the set of monomials appearing in 
\eqref{L-int} can be characterized as follows: apart from $B\vf\d A=\d(B\vf A)-BA\d\vf-\vf A\d B$,
these are all trilinear and quadrilinear field monomials which are Lorentz invariant, have mass dimension
$\leq 4$ and respect the symmetry \eqref{fieldparity}.

\textbf{Geometrical interpretation:} by the
  ``classical version'' of the model $L_0+L$ we mean $L_0+L-L_\mathrm{gf}-L_\mathrm{ghost}$, where 
 \be\label{L-gf}
  L_\mathrm{gf}^{m_B,\La}:=-\frac{\La}{2} \biggl( \d\. A + \frac{m_B}{\sqrt{\La}} B \biggr)^2
  \ee  
is the gauge-fixing term and
\be\label{L-ghost}
  L_{\mathrm{ghost}\,\ka\la_{10}}^{m_u}:= \d\ut\. \d u - m_u^2\, \ut u
   - \frac{\ka\la_{10}\,m_u^2}{m}\ut u\vf
  \ee 
is the ghost term, which is the sum of all terms in $L_0+L$ containing the ghost fields $\ut,\,u$.

There is a distinguished choice of the parameters $\mathbf{\la}$:
by straightforward calculation we find that the classical 
version of $L_0+L$  can be geometrically interpreted as
  a spontaneously broken $U(1)$-gauge model 
iff the parameters $\mathbf{\la}$ have the values
  \be\label{parameter-geom}
  \la_1=...=\la_9=1\ ,\quad \la_{11}=\la_{12}=0\ .
  \ee
  Explicitly, these values of the parameters are equivalent to
  \be\label{geometric}
  L_0+L-L_\mathrm{gf}-L_\mathrm{ghost}-\sqrt{\La}\,m_B\,\d_\mu(A^\mu B)
  =-\frac{1}4\,F^2+\frac{1}2\,(D^\mu\Phi)^*D_\mu\Phi-V(\Phi)\ ,
  \ee
  where 
  \be
  \Phi:=iB+\frac{m}{\ka}+\vf\ ,\quad D^\mu:=\d^\mu-i\ka\,A^\mu
  \ee
  and 
  \be
  V(\Phi):=\frac{\ka^2m_H^2}{8m^2}\,(\Phi^*\Phi)^2-\frac{m_H^2}{4}\,(\Phi^*\Phi)+\frac{m_H^2 m^2}{8\ka^2}\ .
  \ee
  The minima of the potential $V(\Phi)$
  are on the circle $\Phi=\frac{m}{\ka}\,e^{i\al}\ ,\ \al\in [0,2\pi)$.
  The choice of a minima, usually one takes $\Phi_\mathrm{min}=\frac{m}{\ka}$, breaks 
  the $U(1)$ symmetry 'spontaneously' and the fields $\vf$ and $B$ are the deviations from $\Phi_\mathrm{min}$
  in radial and tangential direction.

  Besides $m,\,m_H,\,\ka$ and $\La$, also the parameters
  $m_B,\,m_u$ and $\la_{10}$ are not restricted by the geometrical interpretation \eqref{geometric}.
  The latter are usually fixed as follows:
  \begin{itemize}
  \item the bilinear mixed term $\sim A\d B$ in $L_0+L$ hampers the particle interpretation. 
  For $\la_{12}=0$ (as required by the geometrical interpretation \eqref{parameter-geom}),
  the condition that the $A\d B$-term vanishes is equivalent to the mass relation
  \be\label{m-mB}
  m_B=\frac{m}{\sqrt{\La}}\ .
  \ee
  \item In the next section we will see that {\it BRST-invariance of the total Lagrangian}
  $L_0+L$ implies the geometrical interpretation \eqref{geometric}, however it restricts also the ghost parameters.
  Explicitly, BRST-invariance of the total Lagrangian is equivalent to the parameter values
  \eqref{parameter-geom} and
  \be\label{parameter-BRS}
  m_u^2=\frac{m_B\, m}{\sqrt{\La}}\qquad\text{and}\qquad \la_{10}=1\ ;
  \ee
  note that this holds also in the presence of an $A\d B$-term, i.e.~the 
  validity of \eqref{m-mB} is not assumed here.
  \end{itemize}

  The main aim of this paper is the following: we will start with the $U(1)$-Higgs model,
i.e.~with the parameter values \eqref{parameter-geom}, \eqref{m-mB} and \eqref{parameter-BRS},
and with that we will investigate {\it whether the parameter values \eqref{parameter-geom}
  are stable under the RG-flow generated by scaling transformations}, i.e.~we 
  study the question {\it whether the geometrical interpretation \eqref{geometric} is possible 
  'at all scales'}. 

\textbf{Definition of the RG-flow:}
from now on we will use the just mentioned initial values \eqref{parameter-geom}, \eqref{m-mB} 
and \eqref{parameter-BRS}. With that we have only two independent masses
$\mathbf{m}:=(m,m_H)$, and the interaction $L\equiv L^{\mathbf{m}}\equiv L^{\mathbf{m},\La}_{\ka}$ is
of the form
\be\label{L1+L2}
L=\ka\,L_1+\ka^2\,L_2\ .
\ee
In view of Epstein-Glaser renormalization \cite{EG73}, we introduce an adiabatic switching of 
the coupling constant by a test function $g\in\mathcal{D}(\RR ^4)$:
\be\label{L(g)}
L(g)\equiv L^{\mathbf{m}}(g):=\int dx\,\Bigl(\ka\,g(x)\,L_1(x)+\bigl(\ka\,g(x)\bigr)^2\,L_2(x)\Bigr)\ .
\ee

Following \cite{HW03,DF04,BDF09} we define the RG-flow by means of a scaling transformation of the fields
  \be\label{scaltrafo}
  \si_\rho^{-1}(\phi(x))=\rho\,\phi(\rho x)\ ,\quad \phi=A^\mu,B,\vf,u,\ut\ ,\quad\rho > 0\ ,
  \ee
  and a simultaneous scaling of the masses 
  $\mathbf{m}\mapsto\rho^{-1}\mathbf{m}=(\rho^{-1} m,\rho^{-1} m_H)$;
see \cite{DF04} for the precise definition of $\si_\rho$.
Under this transformation the classical action is invariant
(up to a scaling of the switching function $g$); namely, due to 
$\si_\rho^{-1}\,L^{\rho\mathbf{m}}(x)=\rho^4\,L^{\mathbf{m}}(\rho x)$ and the same for $L_0$,
we have
  \be\label{scaling-classical}
  \int dx\,L_0^\mathbf{m}(x)+L^\mathbf{m}(g)=
  \si_\rho^{-1}\Bigl(\int dx\,L_0^{\rho\mathbf{m}}(x)+
  L^{\rho\mathbf{m}}(g_\rho)\Bigr)\ ,\quad\quad g_\rho(x):=  g(\rho x)\ , 
  \ee
where the parameters $\La,\ka$ are suppressed since they are not affected by the scaling transformation.

  In QFT scaling invariance is broken by quantum effects. To explain this more in detail, 
we introduce the generating functional $S(iL(g))$ of the 
  time ordered products of $L(g)$, i.e.
\be\label{T-product}
T_n(L(g)^{\otimes n})=\frac{d^n}{i^n\,d\eta^n}\vert_{\eta =0}\,S(i\eta \,L(g))
\quad\text{or more generally}\quad T_n=S^{(n)}(0)\ ,
\ee
which we construct inductively by Epstein-Glaser renormalization \cite{EG73}.
  We use that, for a purely massive model and with a suitable (re)normalization of $S(iL(g))$, 
  the adiabatic limit 
  \be\label{adlim}
  \mathbf{S}[L]:=\lim_{\eps\downarrow 0}S(iL(g_\eps))\ ,
\quad g_\eps(x):=g(\eps x)\ ,
  \ee
  exists, where $g(0)=1$ is assumed \cite{EG73,EG76}.%
\footnote{In this paper we treat the adiabatic limit 
  on a heuristic level, for a rigorous treatment we refer to the mentioned papers of Epstein and Glaser,
  in which it is shown that for purely massive models the adiabatic limit \eqref{adlim} 
  exists (in the strong operator sense) and is unique (i.e. independent of the choice of $g$).}
  Now, computing $\mathbf{S}[L]$ for the scaled fields
  $\si_\rho^{-1}(\phi(x))$ \eqref{scaltrafo} and transforming the result back by $\si_\rho$, we
  obtain a result which differs in general from $\mathbf{S}[L]$ by a change of the renormalization 
prescription. The Main Theorem of perturbative  renormalization \cite{SP82,DF04,HW03}
   implies that the transformation $\mathbf{S}_{\mathbf{m}}[L^{\mathbf{m}}]\mapsto 
\sigma_\rho(\mathbf{S}_{\rho^{-1}\mathbf{m}}[\si_\rho^{-1}(L^{\mathbf{m}})])$
  can equivalently be expressed by a renormalization of the interaction $L^{\mathbf{m}}
\mapsto z_\rho(L^{\mathbf{m}})$,
  explicitly
  \be\label{main-theorem-adlim}
  \sigma_\rho(\mathbf{S}_{\rho^{-1}\mathbf{m}}[\si_\rho^{-1}(L^{\mathbf{m}})])=
\mathbf{S}_{\mathbf{m}}[ z_\rho(L^{\mathbf{m}})]\ ,
  \ee
where the lower index $\mathbf{m}$ of $\mathbf{S}_{\mathbf{m}}$ denotes the masses of the Feynman propagators.
  This is explained more in detail in sect.~\ref{sec:stab-PC}. 

\textbf{The form of the running interaction:}
in Sect.~\ref{sec:stab-PC} we will see that,
  with a slight restriction on the (re)normalization of $S(iL(g))$, the
  new interaction $z_\rho(L)$ has the form
  \begin{align}\label{z(L)}
   z_\rho(L^{\mathbf{m},\La}_{\ka}) & \simeq \hbar^{-1}\Bigl[ k_\rho   
   -\frac{1}4 a_{0\rho}\,F^2 + \frac{m^2}{2}a_{1\rho} (A\. A) - \frac{a_{2\rho}}{2} (\d A\.\d A)\nn
  \\[\jot] &\qquad
  + \frac{1}{2} b_{0\rho} (\d B\. \d B) - \frac{m^2}{2\,\La}\, b_{1\rho}\, B^2 
  + \frac{1}{2} c_{0\rho}\,(\d\vf\. \d\vf) - \frac{m_H^2}{2}c_{1\rho}\, \vf^2 \nn
  \\[\jot] 
  &\qquad
   -\frac{m^2}{\La} \,c_{2\rho}\, \ut u+b_{2\rho}\,m\,(A\cdot\d B)\nn
  \\[\jot] 
  &\quad +\ka\Bigl((1+ l_{0\rho})\,m(A\.A)\vf -\frac{m}{\La}\,\ut u\vf +(1+ l_{1\rho})\, B(A\.\d\vf)\nn
  \\[\jot] 
  &\qquad - (1+ l_{2\rho})\, \vf(A\.\d B) - \frac{(1+ l_{3\rho}) m_H^2}{2m}\, \vf^3
  - \frac{(1+l_{4\rho}) m_H^2}{2m}\, B^2\vf\Bigr)
  \nn \\[\jot]
  &\quad +\ka^2\Bigl( \frac{(1+ l_{5\rho})}{2} (A\.A)\, \vf^2 + \frac{(1+ l_{6\rho})}{2} (A\.A)\, B^2
  - \frac{(1+ l_{7\rho})m_H^2}{8m^2}\, \vf^4 \nn
  \\[\jot] 
  &\qquad - \frac{(1+ l_{8\rho})m_H^2}{4m^2}\, \vf^2 B^2
  - \frac{(1+ l_{9\rho})m_H^2}{8m^2}\, B^4 +l_{11\rho}\,(A\. A)^2\Bigr)\Bigr]\ ,
  \end{align}
  where $k_\rho\in\hbar\,\CC[[\hbar]]$ is a constant field (it is the contribution of the
vacuum diagrams) and
  $\simeq$ means 'equal up to the addition of terms of type $\d^a A$', 
where $|a|\geq 1$ and $A$ is a local field polynomial; such a
$\d^a A$-term vanishes in the adiabatic limit. 
It is a peculiarity of this model that a term $\sim\d\ut \d u$
  does not appear in $z_\rho(L)$ (if not added ``by hand'' -- see Remark \ref{rem:fin-ren})
and that there are also no trilinear and quadrilinear 
  terms in $(z_\rho(L)-L)$ containing $\ut u$. 

The dimensionless, $\rho$-dependent coefficients
  $k_\rho,\,a_{j\rho},\,b_{j\rho},\,c_{j\rho}$ and $l_{j\rho}$ 
will collectively be denoted by $e_\rho$. In principle these coefficients are computable -- 
at least to lowest orders (see the 1-loop computations in 
Sects.~\ref{ssec:explicit-comput}-\ref{ssec:e=e} and Appendices 
\ref{app:violation-scaling}-\ref{app:z(L)-coefficients}); 
however, at the present stage they are
unknown. The $e_\rho$'s are of order $\mathcal{O}(\hbar)$
  (i.e.~they are loop corrections), more precisely they are formal power series in $\ka^2\hbar$ with 
  vanishing term of zeroth order,
  \be\label{powerseries}
  e_\rho=\sum_{n=1}^\infty e_\rho^{(n)}\,(\ka^2\hbar)^n\ ,\quad e=k,a_j,\,b_j,\,c_j,\,l_j\ .
  \ee
  Due to $z_{\rho=1}(L)=L/\hbar$, all
functions $\rho\mapsto e_\rho$ have the initial value $0$ at $\rho=1$. 

 \textit{Proof of \eqref{powerseries}:} 
That $e_\rho$ is a formal power series of the form \eqref{powerseries} can be seen as follows.
  To every $e_\rho$ there corresponds a class of Feynman diagrams with external legs according to
  \eqref{z(L)}. For example, the diagrams contributing to $b_{0\rho}$ have $2$ external legs, both are 
  $B$-fields with $0$ or $1$ partial derivative. The vertices are given by $L$ \eqref{L-int}, 
  i.e.~we have trilinear vertices $\sim\ka$ and quadrilinear vertices $\sim\ka^2$.
  For each vertex there is a factor $\hbar^{-1}$ and for each inner line a factor $\hbar$.
  A diagram with $r$ trilinear vertices, $s$ quadrilinear vertices, $p$ inner and $q$ external
  lines satisfies
  \be
  3r+4s-2p=q
  \ee
  and, hence, its contribution to $z_\rho(L)$ \eqref{z(L)} is   
  \be
  \sim \ka^{r+2s}\,\hbar^{r/2+s-q/2}\ .
  \ee
  If $q$ is odd, $q=2q_0+1$, also $r$ is odd, $r=2r_0+1$ ($q_0,r_0\in\NN_0$), and with that we obtain 
  the factor 
  \be\label{odd}
  \ka\,\hbar^{-q_0}\,(\ka^2\hbar)^{r_0+s}\ .
  \ee
  If $q$ is even, $q=2q_0$, also $r$ is even, $r=2r_0$ ($q_0,r_0\in\NN_0$), and with that we get
  \be\label{even}
  \hbar^{-q_0}\,(\ka^2\hbar)^{r_0+s}\ .
  \ee
  The contributing diagrams satisfy $n:=r_0+s\geq 1$ for $q=2,3$ and $n:=r_0+s-1\geq 1$ for
  $q=4$. With that we obtain \eqref{powerseries} -- the additional factors $\hbar^{-1}$ (for $q=2$),
  $\hbar^{-1}\ka$ (for $q=3$) and  $\hbar^{-1}\ka^2$ (for $q=4$) in \eqref{odd}-\eqref{even}
  agree precisely with the prefactors in $z_\rho(L)$ \eqref{z(L)}. $\quad\qed$

\textbf{Renormalization of the wave functions, masses, gauge-fixing parameter and the coupling constants:}
  except for the $A\d B$-term, all bilinear terms of $z_\rho(L)$ do not appear in $L$. However,
  introducing new fields, which are of the form
\be\label{new-fields}
\phi_\rho(x)=f_\phi(\rho)\,\phi(x)\ ,\quad \phi=A,\,B,\,\vf,
\ee
where $f_\phi:\,(0,\infty)\to\CC$ is a $\phi$-dependent function,
and introducing a running gauge-fixing parameter $\La_\rho$,
  running masses $\mathbf{m}_\rho\equiv (m_\rho,\,m_{B\rho},\,m_{u\rho},\,m_{H\rho})$ and 
  running coupling constants $\ka_\rho,\,\la_{j\rho}$, we can 
achieve that $L_0+z_\rho(L)-k_\rho$ has the same form as $L_0+L$,
  in particular we absorb the novel bilinear interaction terms in the free Lagrangian:
  \be\label{parameter-ren}
  \bigl(L_0^{\mathbf{m},\La}+z_\rho(L^{\mathbf{m},\La}_{\ka,\mathbf{\la}})-k_\rho\bigr)(A,B,\vf,u,\ut)=
  \bigl(L_0^{\mathbf{m}_\rho,\La_\rho}+
  L^{\mathbf{m}_\rho,\La_\rho}_{\ka_\rho,\mathbf{\la}_\rho}\bigr)
  (A_\rho,B_\rho,\vf_\rho,u,\ut)\ .
  \ee
We will use the shorthand notation 
$$
L_0+z_\rho(L)-k_\rho=L_0^\rho+L^\rho
$$ 
for this equation.
Since every new field is of the form \eqref{new-fields},
the condition \eqref{parameter-ren} is an equation for polynomials in the old fields; 
equating the coefficients the implicit definition \eqref{parameter-ren}
of the running quantities turns into the following explicit equations:\\
  - for the wave functions
  \be\label{wf-ren}
  A^\mu_\rho=\sqrt{1+a_{0\rho}}\,A^\mu\ ,\quad
  B_\rho=\sqrt{1+b_{0\rho}}\,B\ ,\quad
  \vf_\rho=\sqrt{1+c_{0\rho}}\,\vf\ ;
  \ee
  - for the gauge-fixing parameter
  \be\label{gauge-ren}
  \La_\rho=\frac{\La+a_{2\rho}}{1+a_{0\rho}}\ ;
  \ee
  - for the masses
  \begin{align}\label{mass-ren}
  m_\rho=\sqrt{\frac{1+a_{1\rho}}{1+a_{0\rho}}}\,m\ ,&\qquad
  m_{H\rho}=\sqrt{\frac{1+c_{1\rho}}{{1+c_{0\rho}}}}\,m_H\ ,\nn\\
  m_{B\rho}=\sqrt{\frac{1+b_{1\rho}}{{1+b_{0\rho}}}}\,\frac{m}{\sqrt{\La}}\ ,&\qquad
  m_{u\rho}=\sqrt{1+c_{2\rho}}\,\frac{m}{\sqrt{\La}}\ ;
  \end{align}
  - for the coupling constant
  \be\label{cc-ren}
  \kappa_\rho=\frac{1+l_{0\rho}}{\sqrt{(1+a_{0\rho})(1+a_{1\rho})(1+c_{0\rho})}}\,\kappa\ ; 
  \ee
  and the running coupling parameters $\mathbf{\la}_\rho$ are implicitly determined by
  \begin{align}\label{cc-parameter-ren}
  \frac{\ka\,m}{\La}\,\ut u\vf & =\frac{\ka_\rho\la_{10\rho}\,m_{u\rho}^2}{m_\rho}\,\ut u\vf_\rho\ ,\nn\\
  \ka(1+l_{1\rho})\, B(A\.\d\vf) & =\ka_\rho\la_{1\rho}\, B_\rho(A_\rho\.\d\vf_\rho)\ ,\nn\\
  ........ & = .......................\nn\\
  \frac{\ka (1+l_{4\rho}) m_H^2}{m} B^2\vf & =
  \frac{\ka_\rho\la_{4\rho} m_{H\rho}^2}{m_\rho} B_\rho^2\vf_\rho\ ,\nn\\
  \ka^2(1+l_{5\rho})\, (A\.A) \vf^2 & =\ka_\rho^2\la_{5\rho}\, (A_\rho\.A_\rho) \vf_\rho^2\ ,\nn\\
  ........ & = .......................\nn\\
  \kappa^2\,l_{11\rho}\,(A\.A)^2 & =\kappa_\rho^2\la_{11\rho}\,(A_\rho\.A_\rho)^2\ ,\nn\\
  b_{2\rho}\,m\,(A\.\d B) & =\bigl((\la_{12\rho}-1)m_\rho+
  \sqrt{\La_\rho}\,\, m_{B\rho}\bigr)\,(A_\rho\.\d B_\rho)\ .
  \end{align}

  The renormalizations \eqref{wf-ren}-\eqref{cc-parameter-ren} are not diagonal (as one naively 
  might think): the new fields/parameters depend not only on the pertinent old field/parameter, because 
  the coefficients $a_{j\rho},\,b_{j\rho},\,c_{j\rho},\,l_{j\rho}$ are functions of the whole set 
  $\{\mathbf{m},\La,\ka\}$ of old parameters. The renormalization of the wave functions can be 
  interpreted as follows: the field monomials appearing in $L_0+z_\rho(L)$ can be viewed as a basis of 
  a vector space. The redefinitions \eqref{wf-ren} are then a change of the ``unit of lenght on the 
  various coordinate axis''.


\begin{rem}[``Perturbative agreement''] By the renormalization of the wave functions, masses and gauge 
fixing-parameter, we change the splitting of the total Lagrangian $L_0+z_\rho(L)$ into a free and 
interacting part, i.e.~we change the starting point for the perturbative expansion. To justify this,
one has to show that the two pertubative QFTs given by the splittings $L_0+z_\rho(L)$ and
$L_0^\rho+L^\rho$, respectively, have the same physical content. This statement can be viewed as an 
application of the ``Principle of Perturbative Agreement'' of Hollands and Wald, which is used in
\cite{HW05} as an additional renormalization condition. 

The proof that the ``old'' perturbative QFT (given by $L_0+z_\rho(L)$) and the ``new'' one (given by
$L_0^\rho+L^\rho$) are physically equivalent is beyond the scope of this paper. 
For the wave function and mass renormalization in a scalar field theory, the
following conjecture has been formulated (by using the framework of algebraic QFT) and verified for a 
few examples \cite{BDF08}: given a renormalization prescription (i.e.~an $S$-functional
\eqref{T-product}) for the old perturbative QFT, there exists a renormalization prescription for the 
new perturbative QFT, such that the pertinent nets of local observables in the algebraic adiabatic limit
(see \cite{BF00} or \cite{DF04,BDF09}) are equivalent. The corresponding isomorphisms can be chosen such 
that local fields are identified with local fields modulo the free field equation.

For models with spin $1$ fields, the gauge-fixing parameter has also to be renormalized; and there is the 
difficulty that in general the new free theory (given by $L^\rho_0$) is \textit{not} BRST-invariant,
see Remark \ref{bilin-interaction}.
\end{rem}

\textbf{Geometrical interpretation at an arbitrary scale:}
since we have written the running Lagrangian $L_0+z_\rho(L)-k_\rho$
in the form $L_0^\rho+L^\rho$,
the equivalence of \eqref{geometric} and \eqref{parameter-geom} can be 
applied to it:
$L_0^\rho+L^\rho$ can be geometrically interpreted
  iff the  $\la_{j\rho}$ have the values
  \be\label{runningparameter-geom}
  \la_{1\rho}=\la_{2\rho}=...=\la_{9\rho}=1\ ,\quad\la_{11\rho}=\la_{12\rho}=0\ .
  \ee
  To be precise: by 'geometrical interpretation' we mean here that
  \begin{align}\label{L-new}
  \bigl(L_0^{\mathbf{m}_\rho,\La_\rho}+
  L^{\mathbf{m}_\rho,\La_\rho}_{\ka_\rho,\mathbf{\la}_\rho}\bigr)(A_\rho,B_\rho,\vf_\rho,u,\ut)
  & =-\frac{1}4\,F_\rho^2+\frac{1}2\,(D_\rho^\mu\Phi_\rho)^*D_{\rho\mu}\Phi_\rho-V_\rho(\Phi_\rho)\nn\\
  & + L_{\mathrm{gf}}^\rho+L_{\mathrm{ghost}}^\rho+\sqrt{\La_\rho}\,\, m_{B\rho}\,\d_\mu(A_\rho^\mu B_\rho)\ ,
  \end{align}
  where $F^{\mu\nu}_\rho:=\d^\mu A^\nu_\rho-\d^\nu A^\mu_\rho\ $,
  \begin{align}
  \Phi_\rho &  :=iB_\rho+\frac{m_\rho}{\ka_\rho}+\vf_\rho\ ,\quad D_\rho^\mu:=\d^\mu-i\ka_\rho\,A_\rho^\mu\nn\\
  V_\rho(\Phi_\rho)& :=\frac{\ka_\rho^2 m_{H_\rho}^2}{8m_\rho^2}\,(\Phi_\rho^*\Phi_\rho)^2-
  \frac{m_{H_\rho}^2}{4}\,(\Phi_\rho^*\Phi_\rho)+\frac{m_{H_\rho}^2 m_\rho^2}{8\ka_\rho^2}
  \end{align}
  and
  \begin{align}\label{L-gf-ghost-rho}
  L_{\mathrm{gf}}^\rho & :=-\frac{\La_\rho}{2} \biggl( \d\. A_\rho +
  \frac{m_{B\rho}}{\sqrt{\La_\rho}} B_\rho \biggr)^2\ ,\nn\\
  L_{\mathrm{ghost}}^\rho& :=\d\ut\cdot \d u-m_{u\rho}^2\,\ut u
  -\frac{\ka_\rho\,\la_{10\rho}\,m_{u\rho}^2}{m_\rho}\,\ut u\vf_\rho\ .
  \end{align}

  Our main question is whether \eqref{runningparameter-geom} holds true
when starting with the $U(1)$-Higgs-model; 
for simplicity we also assume that initially we are in Feynman gauge: $\La_{\rho=1}=1$.

With these initial values, the geometrical interpretability \eqref{runningparameter-geom} 
is equivalent to the following relations among the coefficients $e_\rho$:
\begin{align}
\la_{1\rho}=1\,\,\,\text{gives}&\quad\quad\quad\frac{1+l_{1\rho}}{1+l_{0\rho}}=
\sqrt{\frac{1+b_{0\rho}}{1+a_{1\rho}}}\ ,\label{e-geom-interpret1}\\
\la_{2\rho}=1\,\,\,\text{gives}&\quad\quad\quad l_{2\rho}=l_{1\rho}\ ,\label{e-geom-interpret2}\\
\la_{3\rho}=1\,\,\,\text{gives}&\quad\quad\quad\frac{1+l_{3\rho}}{1+l_{0\rho}}=
\frac{1+c_{1\rho}}{1+a_{1\rho}}\ ,\label{e-geom-interpret3}\\
\la_{4\rho}=1\,\,\,\text{gives}&\quad\quad\quad\frac{1+l_{4\rho}}{1+l_{3\rho}}=
\frac{1+b_{0\rho}}{1+c_{0\rho}}\ ,\label{e-geom-interpret4}\\
\la_{5\rho}=1\,\,\,\text{gives}&\quad\quad\quad\frac{1+l_{5\rho}}{(1+l_{0\rho})^2}=
\frac{1}{1+a_{1\rho}}\ ,\label{e-geom-interpret5}\\
\la_{6\rho}=1\,\,\,\text{gives}&\quad\quad\quad\frac{1+l_{6\rho}}{1+l_{5\rho}}=
\frac{1+b_{0\rho}}{1+c_{0\rho}}\ ,\label{e-geom-interpret6}\\
\la_{7\rho}=1\,\,\,\text{gives}&\quad\quad\quad\frac{1+l_{7\rho}}{(1+l_{0\rho})^2}=
\frac{1+c_{1\rho}}{(1+a_{1\rho})^2}\ ,\label{e-geom-interpret7}\\
\la_{8\rho}=1\,\,\,\text{gives}&\quad\quad\quad\frac{1+l_{8\rho}}{1+l_{7\rho}}=
\frac{1+b_{0\rho}}{1+c_{0\rho}}\ ,\label{e-geom-interpret8}\\
\la_{9\rho}=1\,\,\,\text{gives}&\quad\quad\quad\frac{1+l_{9\rho}}{1+l_{7\rho}}=
\Bigl(\frac{1+b_{0\rho}}{1+c_{0\rho}}\Bigr)^2\ ,\label{e-geom-interpret9}\\
\la_{11\rho}=0\,\,\,\text{gives}&\quad\quad\quad l_{11\rho}=0\ ,\label{e-geom-interpret11}\\
\la_{12\rho}=0\,\,\,\text{gives}&\quad\quad\quad b_{2\rho}=
\sqrt{(1+a_{2\rho})(1+b_{1\rho})}-\sqrt{(1+a_{1\rho})(1+b_{0\rho})}\ .\label{e-geom-interpret12}
\end{align}
Searching all values for the coefficients $e_\rho$ which solve
this system of equations, we find that this is quite a large set: neglecting $k_\rho$, 
$9$ coefficients can freely be chosen (e.g.~$a_{0\rho},\,a_{1\rho},\,a_{2\rho},\,b_{0\rho},\,
b_{1\rho},\,c_{0\rho},\,c_{1\rho},\,c_{2\rho}$ and $l_{0\rho}$), the other $11$ 
coefficients are then uniquely determined by the
$11$ equations \eqref{e-geom-interpret1}-\eqref{e-geom-interpret12}.

Combining the equations \eqref{e-geom-interpret3}, \eqref{e-geom-interpret5} and \eqref{e-geom-interpret7}
we obtain
\be\label{geom-interpret-crucial}
\frac{1+l_{7\rho}}{1+l_{3\rho}}=\frac{1+l_{5\rho}}{1+l_{0\rho}}\ .
\ee
It will turn out that the conditions \eqref{e-geom-interpret12} and \eqref{geom-interpret-crucial}
are crucial for the geometrical interpretability.

For later purpose we mention that, with the considered initial values, the explicit formula for $\la_{10\rho}$ reads
\be\label{la-10-running}
\la_{10\rho}=\frac{1+a_{1\rho}}{(1+c_{2\rho})(1+l_{0\rho})}\ .
\ee


  \section{BRST-invariance of the Lagrangian}\label{sec:BRST}

  The main result of this Section is that BRST-invariance of the Lagrangian is a sufficient
  (but not necessary) condition for the geometrical interpretation 
-- for both, the initial Lagrangian $L_0+L$ and the running Lagrangian $L_0+z_\rho(L)$.

 The BRST-transformation 
  $s\equiv s_\beta=s_0+\ka\beta\,s_1$ is a graded derivation which commutes with partial derivatives and 
  is given on the basic fields by
  \begin{align}\label{BRS-trafo}
  s\,A^\mu=\d^\mu u\ ,& \quad s\,B=mu+\ka\beta\,u\vf\ ,\quad s\,\vf=-\ka\beta\,Bu\ ,\nn\\
  s\,u=0\ ,& \quad s\,\ut=-\La\,(\d A+\frac{m_B}{\sqrt{\La}}\,B)\ .
  \end{align}
  Since it is a priory not clear which of the coupling constant $\ka,\,\ka\la_j$ in $L$ 
  \eqref{L-int} is equal 
  to the $\ka$ in the BRST-transformation, we have introduced the parameter $\beta$ in $s$.

\textbf{BRST-invariance of the initial Lagrangian:} explicitly this property reads
 \be\label{sL=0}
  s_\beta (L_0+L)\simeq 0
  \ee
(where again $\simeq$ means 'equal up to the addition of derivatives 
of local field polynomials'); it is equivalent to
\be\label{BRS(L)=0}
\beta=1\quad\text{and the parameter values \eqref{parameter-geom}
  and \eqref{parameter-BRS}.}
\ee

  That the parameter values \eqref{BRS(L)=0}
  imply \eqref{sL=0} can be seen by formally interpreting the BRST-transformation 
  of $A^\mu$ and $(\vf,B)$ as an infinitesimal gauge transformation,
  \be\label{s-geom}
  s(A^\mu,\Phi)=\frac{d}{d\al}\vert_{\al =0}\,(A^\mu+\al\,\d^\mu u,\,e^{i\al\ka u}\,\Phi)\ ,
  \ee 
  and by taking into account that $D^\mu$ is a pertinent covariant derivative. With that we immediately 
see that  $s(F^2)=0\ ,\ s((D^\mu\Phi)^*D_\mu\Phi)=0\ ,\ s(V(\Phi))=0$, and by 
  using \eqref{geometric} and a simple calculation we obtain
  \be\label{geom->BRS}
  s(L_0+L)=\sqrt{\La}\,m_B\,s\d(AB)+s(L_\mathrm{gf}+L_\mathrm{ghost})=
  \d_\mu\bigl(\sqrt{\La}\,m_B\,s(A^\mu B)+(s\ut)(sA^\mu)\bigr)\ .
  \ee
  The proof that \eqref{sL=0} is also sufficient for the parameter values 
  \eqref{BRS(L)=0}, is a straightforward calculation:
  inserting \eqref{L-free} and \eqref{L-int} into \eqref{sL=0} one obtains (after some work)
  these parameter values. The relations \eqref{parameter-BRS} are precisely the condition that
  $s(L_\mathrm{gf}+L_\mathrm{ghost})\simeq 0$, where we assume that $\beta=1$ is already obtained from 
  other parts of the calculation.

\textbf{BRST-invariance of the running Lagrangian $(L_0+z_\rho(L))$:} the property
  \be\label{s(z(L))=0}
  s_\beta(L_0+z_\rho(L))\simeq 0\ ,
  \ee
  where $s_\beta \equiv s$ is given by \eqref{BRS-trafo}, determines $z_\rho(L)-k_\rho$ uniquely
in terms of the three coefficients
  \be\label{BRS-free-param}
  a_\rho:=a_{0\rho}\ ,\quad b_\rho:=b_{0\rho}\ ,\quad l_\rho:=c_{1\rho}\ ,
  \ee
which can freely be chosen.
  More explicitly, the condition \eqref{s(z(L))=0} is equivalent to $\beta=1$ and the 
  following form of $z_\rho(L)$:
  \begin{align}\label{z(L)-BRS}
   z_\rho(L^{\mathbf{m},\La}_{\ka}) & \simeq k_\rho
   -\frac{1}4 a_{\rho}\,F^2 + b_{\rho}\bigl(\frac{m^2}{2} \,A^2-m\,A\d B
  + \frac{1}{2}\, (\d B)^2+ \frac{1}{2}\,(\d\vf)^2\bigr) -
  l_{\rho}\,\frac{m_H^2}{2}\, \vf^2 \nn
  \\[\jot]
  &+\ka\Bigl(-\frac{m}{\La}\ut u\vf +(1+ b_{\rho})\, \bigl(m\,A^2\vf+B\,A\d\vf
  - \vf\,A\d B\bigr)\nn \\[\jot]
  &\qquad - \frac{(1+ l_{\rho}) m_H^2}{2m}\, (\vf^3+ B^2\vf)\Bigr)
  \nn \\[\jot]
  &+\ka^2\Bigl( \frac{(1+ b_{\rho})}{2} (A^2\, \vf^2 + A^2\, B^2)
  - \frac{(1+ l_{\rho})m_H^2}{8m^2}\, (\vf^4 + 2\, \vf^2 B^2+ B^4 )\Bigr)\ .
  \end{align}
One verifies easily that with these relations among the coefficients $e_\rho$,
the equations \eqref{e-geom-interpret1}-\eqref{e-geom-interpret12}
are satisfied, that is, \eqref{s(z(L))=0} implies indeed the geometrical 
  interpretation \eqref{L-new}. However, due to the presence of bilinear terms in $z_\rho(L)$,
  the difference between BRST-invariance of the Lagrangian \eqref{s(z(L))=0} and 
the geometrical interpretation \eqref{L-new} does not only concern the ghost sector,
  as for $L_0+L$ (see \eqref{parameter-BRS}), it is clearly bigger --  the number of 
free coefficients $e_\rho$ is $3$ versus $9$. 

  The proof that the set of solutions of the condition \eqref{s(z(L))=0} is given by $\beta=1$
  and \eqref{z(L)-BRS}, is a somewhat lengthy and straightforward calculation, which is quite boring.
  More instructive is the following understanding of the parameter values \eqref{z(L)-BRS}: the above
  derivation \eqref{s-geom}-\eqref{geom->BRS} of BRST-invariance of $L_0+L$, 
  by using the geometrical interpretation, can only be applied to
  $\ L^\rho_0+L^\rho(=L_0+z_\rho(L)-k_\rho)\ $, if 
the BRST transformation $s$ \eqref{BRS-trafo} expressed in terms 
  of the $\rho$-fields, has the same form as for
  the original fields, up to a global prefactor $\ga$.  Explicitly this requirement reads
  \begin{align}\label{BRS-trafo-rho}
  s\,A_\rho^\mu=\sqrt{1+a_{0\rho}}\,\,s\,A^\mu & =\gamma\,\d^\mu u\ ,\nn\\
  s\,B_\rho=\sqrt{1+b_{0\rho}}\,\,s\,B & =\gamma\,(m_\rho u+\ka_\rho\,u\vf_\rho)\ ,\nn\\
  s\,\vf_\rho=\sqrt{1+c_{0\rho}}\,\,s\,\vf & =-\gamma\,(\ka_\rho\,B_\rho u)\nn\\
  s\,\ut & =-\gamma\,\La_\rho\,(\d A_\rho+\frac{m_{B_\rho}}{\sqrt{\La_\rho}}\,B_\rho)
  \end{align}
  and $s\,u=\ga\,0=0$ is trivially satisfied. From the first equation we obtain 
  \be\label{gamma}
  \ga=\sqrt{1+a_{0\rho}}
  \ee
  and with that the further equations are equivalent to
  \be\label{parameter-BRS1}
  b_{\rho}:=b_{0\rho}=a_{1\rho}=c_{0\rho}=l_{0\rho}\quad\text{and}\quad b_{1\rho}=0=a_{2\rho} \ .
  \ee 
  To take the demand for validity of the geometrical interpretation into account, we insert  
  \eqref{parameter-BRS1} into \eqref{e-geom-interpret1}-\eqref{e-geom-interpret12}, this yields
  \begin{align}\label{parameter-BRS2}
  b_\rho & = l_{1\rho}=l_{2\rho}=l_{5\rho}=l_{6\rho}=-b_{2\rho}\nn\\
  l_\rho:= c_{1\rho} & =l_{3\rho}=l_{4\rho}=l_{7\rho}=l_{8\rho}=l_{9\rho}\nn\\
  l_{11\rho} & =0\ .
  \end{align}
  In addition, the derivation \eqref{s-geom}-\eqref{geom->BRS} needs
  BRST-invariance of $(L_{\mathrm{gf}}^\rho+L_{\mathrm{ghost}}^\rho)$ \eqref{L-gf-ghost-rho}, which is 
  equivalent to
  $m_{u\rho}^2=\frac{m_{B\rho}\, m_\rho}{\sqrt{\La_\rho}}$ and $\la_{10\rho}=1$ (similarly to~\eqref{parameter-BRS}); 
  both equations give
  \be\label{parameter-BRS3}   
  c_{2\rho}=0 
  \ee
  by using \eqref{parameter-BRS1} (and the formulas
 \eqref{gauge-ren}, \eqref{mass-ren}, \eqref{la-10-running} for the running quantities).
  The parameter relations \eqref{parameter-BRS1}, \eqref{parameter-BRS2} and \eqref{parameter-BRS3} 
  agree precisely with \eqref{z(L)-BRS}. 

  As usual, the BRST-transformation $s$ \eqref{BRS-trafo} is nilpotent modulo the field equations
  of $L_0+L$; we point out that this holds also for $s$ expressed in terms of the $\rho$-fields
  (i.e.~\eqref{BRS-trafo-rho}-\eqref{gamma}) w.r.t.~the field equations of the new Lagrangian $L^\rho_0+L^\rho$:
  \be\label{s2=0}
  s^2\ut=-\La_\rho(1+a_\rho)\,\Bigl(\square u+\frac{m_{B_\rho}}{\sqrt{\La_\rho}}
  (m_\rho u+\ka_\rho\,u\vf_\rho)\Bigr)=\La_\rho(1+a_\rho)\,
  \frac{\delta\,\int dx\,(L^\rho_0+L^\rho)}{\delta \ut}\ ,
  \ee
  where we use the preceding relations, i.e.~we assume that \eqref{s(z(L))=0} holds true.

\section{Perturbative gauge invariance (PGI)}\label{sec:PGI}

For the initial model $S(iL(g))$ we admit all renormalization prescriptions which fulfill the 
Epstein-Glaser axioms \cite{EG73,DF04} and a suitable version of BRST-invariance. 
The latter should be well adapted to the inductive Epstein-Glaser 
construction of the time-ordered products and to our definition of the RG-flow. We will see that 
PGI \cite{DHKS94,DS99} fulfills these criteria.

\textbf{Physical consistency (PC).}
To motivate PGI we start with PC, which is a somewhat weaker condition 
\cite{Kugo-Ojima,DuetschSchroer2000,Grigore2000}. Let $Q$ be the charge implementing 
the free BRST-transformation $s_0:=s|_{\ka =0}$, explicitly
  \be
  [Q,\phi]^\mp_\star\approx\,i\hbar\, s_0\phi\ ,\quad\phi=A^\mu\,,\,B\,,\,\vf\,,\,u\,,\,\ut\ ,
  \ee
where $[\cdot\,,\,\cdot]^\mp_\star$ denotes the graded commutator w.r.t.~the $\star$-product
and $\approx$ means 'equal modulo the free field equations'.
  The nilpotency $Q^2\approx 0$ reflects $s_0^2\approx 0$.
  For our model with Feynman gauge $\La=1$, the charge $Q$ is given by the somewhat heuristic 
formula\footnote{A rigorous definition of $Q$ is given in \cite{Dutsch1999}.}
  \be
  Q=\int_{x^0=\mathrm{constant}}d^3x\,\bigl((\d A+mB)\,\d^0u-\d^0(\d A+mB)\,u\bigr)\ .
  \ee

For the asymptotic free fields, the ``subspace'' of physical states can be described as
 $$
\Hcal_{\mathrm{phys}}:=\frac{\mathrm{ker}\,Q}{\mathrm{ran}\,Q}\ .
$$
The operator $\mathbf{S}[L]$ \eqref{adlim} induces a well defined operator from
$\Hcal_{\mathrm{phys}}$ into itself iff
 $$
[Q,\mathbf{S}[L]]_\star\vert_{\mathrm{ker}\,Q}\approx 0\ ,
$$
see e.g.~\cite{DuetschSchroer2000}. This is the reason to require
\be\label{PC}
\text{physical consistency (PC):}\quad 0\approx [Q,\mathbf{S}[L]]_\star\equiv
\lim_{\eps\downarrow 0}[Q,S(iL(g_\eps)/\hbar)]_\star\ .
\ee

\textbf{Perturbative gauge invariance (PGI):}
to fulfill PC in the inductive Epstein-Glaser construction of time-ordered products,
we need a version of PC \textit{before} the adiabatic limit $g\to 1$ is taken: precisely 
for this purpose PGI was introduced in \cite{DHKS94}.


  PGI is the condition that to a given local interaction
  \be
  \mathcal{ L}(g):=\hbar^{-1}\>
  \sum_{k=1}^\infty \kappa^k\,\int dx\,(g(x))^k\, \mathcal{ L}_{(k)}(x)\ ,\label{Lcal(g)}
  \ee
  there exists a ``$Q$-vertex''
  \be
  \mathcal{ P}^{\nu}(g;f):=\sum_{k=1}^\infty \kappa^k\,\int dx\,(g(x))^{(k-1)}\,
  \mathcal{ P}_{(k)}^\nu (x)\,f(x)\label{P(g)}
  \ee
  (where $g,f\in\Dcal(\RR^4)$ and $\mathcal{ L}_{(k)},\,\mathcal{ P}_{(k)}$ are local field polynomials) 
  and a renormalization of the time-ordered products such that
  \be\label{PGI}
  [Q,S\bigl(i\, \mathcal{L}(g)\bigr)]_\star\approx \frac{d}{d\eta}\vert_{\eta =0}\,
  S\bigl(i\, \mathcal{ L}(g)+\eta\, \mathcal{P}^\nu(g;\d_\nu g)\bigr)\ .
  \ee

That PGI implies PC, is easy to see (on the heuristic level on which we treat 
the adiabatic limit in this paper): the r.h.s.~of \eqref{PGI} vanishes in the adiabatic limit,
since it is linear
in the $Q$-vertex, the latter is linear in $\d_\nu g$ and $\d_\nu g_\eps\sim\eps$.    

For time-ordered products $T_n$ of order $n\geq 2$, PGI is a renormalization condition --
it is a particular case of the 'Master BRST Identity', which is the application of the 'Master Ward 
Identity' to the conservation of the free BRST-current, see \cite{Dutsch2002,DF03}.

It is well-known that the $U(1)$-Higgs model is anomaly-free. 
Hence, our initial model can be renormalized such that PGI 
\eqref{PGI} holds to all orders in $\kappa$, where $\mathcal{ L}(g):=L(g)$ is given by \eqref{L(g)}
and
\be\label{eq:PGI-P}
  \mathcal{ P}^{\nu}(g;f):=\int dx\,\Bigl(\kappa \,P^\nu_1(x)+\kappa^2g(x)\,P^\nu_2(x)\Bigr) f(x) ,
  \ee
with
\be\label{eq:LPoriginal}
 P_1^{\nu}  =  m\,A^\nu u\vf-\d^\nu B\,u\vf +Bu\,\d^\nu\vf \ ,\quad\quad
  P_2^{\nu}  = A^\nu u \vf^2+A^\nu u B^2        \ ,
\ee
where $L_k$ \eqref{L(g)} and
$P_k^\nu$ are $(k+2)$-linear in the basic fields and Feynman gauge $\La =1$ is chosen.

To apply PGI to the running interaction $z_\rho(L)$, we insert the power series \eqref{powerseries}
for the coefficients $e_\rho$ into $z_\rho(L)$ \eqref{z(L)}, to write the latter as a power series in $\ka$,
\be\label{z(L)-ka}
z_\rho(L)=\hbar^{-1}\sum_{k=1}^\infty z_{\rho\,k}(L)\,\,\ka^k\ .
\ee
So, for this interaction, $\mathcal{L}(g)$  \eqref{Lcal(g)} is given by
\be\label{z(L)(g)}
\mathcal{L}(g)=z_\rho(L)(g):=\hbar^{-1}\sum_{k=1}^\infty \int dx\,
z_{\rho\,k}(L)(x)\,(\ka\,g(x))^k\ .
\ee
Explicitly, with $\La_{\rho=1}=1$ we have
  \be\label{L-j}
  \mathcal{ L}_{(1)}=L_1\ ,\quad \mathcal{ L}_{(2)}=L_2+\hbar\, L_0^{(1)}\ ,\quad
  \mathcal{ L}_{(3)}=\hbar\,L_1^{(1)}\ ,\quad \mathcal{ L}_{(4)}=\hbar\,L_2^{(1)}+\hbar^2 L_0^{(2)} 
  \ee 
  etc., where $L_k^{(j)}$ is $(k+2)$-linear in the basic fields
  and the upper index $j$ denotes the order in $\hbar$; explicitly  
  \begin{align}\label{L-k-j}
  L_0^{(j)}  = & -\frac{1}4 a^{(j)}_{0\rho}\,F^2 + \frac{m^2}{2}a^{(j)}_{1\rho}\,A^2 - \frac{a^{(j)}_{2\rho}}{2} (\d A)^2
  + \frac{1}{2} b^{(j)}_{0\rho} (\d B)^2 - \frac{m^2}{2}\, b^{(j)}_{1\rho}\, B^2\nn\\
  &+ \frac{1}{2} c^{(j)}_{0\rho}\,(\d\vf)^2 - \frac{m_H^2}{2}c^{(j)}_{1\rho}\, \vf^2 
   -m^2 \,c^{(j)}_{2\rho}\, \ut u+m\,b^{(j)}_{2\rho}\,A\d B\ ,\nn \\
  L_1^{(j)}  = & m\,l_{0\rho}^{(j)}\,A^2\vf +l_{1\rho}^{(j)}\,B(A\d\vf)-l_{2\rho}^{(j)}\, \vf(A\d B) 
  - \frac{m_H^2}{2m} (l_{3\rho}^{(j)}\,\vf^3+l_{4\rho}^{(j)}\,B^2\vf) \ ,\nn \\
  L_2^{(j)}  = &  \frac{1}{2}\Bigl(l_{5\rho}^{(j)}\,A^2\vf^2 + l_{6\rho}^{(j)}\,A^2 B^2\Bigr)-
  \frac{m_H^2}{8m^2} (l_{7\rho}^{(j)}\,\vf^4+2\,l_{8\rho}^{(j)}\, \vf^2 B^2+l_{9\rho}^{(j)}\,B^4)
  +l_{11\rho}^{(j)}\,(A^2)^2         
  \end{align}
  for $j\geq 1$. The pertinent $\mathcal{ P}_{(k)}$ in \eqref{P(g)} must have a similar structure
  \be\label{P-j}
  \mathcal{ P}_{(1)}=P_1\ ,\quad \mathcal{ P}_{(2)}=P_2+\hbar\, P_0^{(1)}\ ,\quad
  \mathcal{ P}_{(2)}=\hbar\,P_1^{(1)}\ ,\quad \mathcal{ P}_{(2)}=\hbar\,P_2^{(1)}+\hbar^2 P_0^{(2)}
  \ee
  etc., where the indices of $P_k^{(j)}$ have the same meaning as for $L_k^{(j)}$.


  \section{Stability of physical consistency under the renormalization group flow}
\label{sec:stab-PC}

\textbf{Stability of PC:} 
it is hard to find out whether PGI is maintained under the RG-flow, i.e.~whether PGI
for $\mathcal{L}(g)=L(g)$ \eqref{L(g)} implies PGI for $\mathcal{L}(g)=z_\rho(L)(g)$
\eqref{z(L)(g)}. In Sect.~\ref{sec:PGI-tree} we 
show that PGI for $S\bigl(iz_\rho(L)(g)\bigr)$
can be fulfilled on the level of tree diagrams (with vertices $z_\rho(L)(g)$), if
one takes only the 1-loop contributions $e_\rho^{(1)}$ \eqref{powerseries} to $z_\rho(L)$
into account. But this depends on the renormalization prescription 
for $S\bigl(iL(g)\bigr)\,$: using a prescription  corresponding 
to the minimal subtraction scheme, PGI
gets lost under the RG-flow, already at the level of tree diagrams.

However, the somewhat weaker property of PC is maintained under the RG-flow; more precisely
we will prove that
  \be\label{PC-RG}
  \bigl[Q,\mathbf{S}[L]\bigr]_\star\approx 0\quad\Rightarrow\quad 
\bigl[Q,\mathbf{S}[z_\rho(L)]\bigr]_\star\approx 0\ .
  \ee
Hence, at least in this weak form, BRST-invariance of the time-ordered products is stable under the RG-flow.
We point out that \eqref{PC-RG} is a model-independent result; only rather weak assumptions are needed, 
which will be given in the course of the proof.

\textbf{Construction of $z_\rho(L)$:} to prove \eqref{PC-RG}, we need to understand precisely how
$z_\rho(L)$ is constructed. We use the formalism of \cite{DF04} (see also \cite{BDF09}), 
in particular we apply the Main Theorem \cite{DF04,HW03}: assuming that
$S$ fulfills the axioms of Epstein-Glaser renormalization, this holds 
also for the scaled time-ordered products $\si_\rho\circ S\circ\si_\rho^{-1}$; therefore, there exists 
a unique map $Z_\rho\equiv Z_{\rho,\mathbf{m}}$ from the space of local interactions into itself such that
  \footnote{We use the convention for $Z_\rho$ given in \cite{BDF09}, which differs by factors $i$
  from the definition $\tilde Z_\rho(F):=D_\rho(e_\otimes^F)$ in \cite{DF04}, namely:
  $Z_\rho (iF)=i\,\tilde Z_\rho (F)\ $.}
  \be\label{main-theorem}
  \si_\rho\circ S_{\rho^{-1}\mathbf{m}}\circ\si_\rho^{-1}=S_{\mathbf{m}}\circ Z_{\rho,\mathbf{m}}
  \ee
  (the lower index $\mathbf{m}$ on $S$ and $Z_\rho$  denotes the masses of the underlying $\star$-product,
  i.e.~the masses of the Feynman propagators). 

In addition $Z_\rho$ is of the following form \cite[Prop.~4.3]{DF04}: let $\mathcal{P}$
be the space of local field polynomials, $h\in\mathcal{D}(\RR^4)$ and 
$A(h)=\int dx\,\,A(x)\,h(x)$ for $A\in\mathcal{P}$. Given $Z_\rho$, 
there exist linear and symmetric maps
$d^\rho_{n,a}\,:\,\mathcal{P}^{\otimes n}\to\mathcal{P}$ for $n\geq 2,\,a\equiv
(a_1,\ldots,a_n)\in(\NN_0^4)^n$, such that
\be\label{Z-form}
Z_\rho(A(h))=A(h)+\sum_{n=2}^\infty\frac{1}{n!}\sum_a\int dx\,\,d^\rho_{n,a}(A^{\otimes n})(x)\,
\prod_{l=1}^n(\d^{a_l}h(x))\ .
\ee
The  expressions $d^\rho_{n,a}(A^{\otimes n})$ are uniquely determined if one requires
$d^\rho_{n,a}(A^{\otimes n})\in\Pcal_\mathrm{bal}$, where $\Pcal_\mathrm{bal}\subset\Pcal$
is the subspace of ``balanced fields'', defined in \cite{DF04}.

Applying \eqref{Z-form} to $A(h)=iL(g)/\hbar=(i/\hbar)\sum_{j=1,2}\int dx\,\,(\ka g(x))^j\,L_j(x)$
\eqref{L(g)}, we get
  \begin{align}\label{Z(L)-form}
  Z_\rho(iL(g)/\hbar)=iL(g)/\hbar+&\sum_{n=2}^\infty\frac{i^n}{n!\,\hbar^n}
\sum_a\sum_{j_1,\ldots,j_n=1,2}\ka^{j_1+\cdots+j_n}\notag\\
&\cdot\int dx\,\,d^\rho_{n,a}(L_{j_1}\otimes\cdots \otimes L_{j_n})(x)\,
\prod_{l=1}^n\d^{a_l}(g(x))^{j_l}\ .
  \end{align}
In view of the adiabatic limit and $\d g_\eps(x)=\mathcal{O}(\eps)$, we cut off the 
terms with derivatives of $g$:
\be\label{Z(L)-adlim}
Z_\rho(iL(g_\eps)/\hbar)=i\,z_\rho(L)(g_\eps)+\mathcal{O}(\eps)\ ,
\ee
where
\be
z_\rho(L)(g)=\frac{1}{\hbar}\Bigl( L(g)+\sum_{n=2}^\infty\frac{i^{n-1}}{n!\,\hbar^{n-1}}
\sum_{j_l=1,2}\int dx\,\,d^\rho_{n,0}(L_{j_1}\otimes\cdots \otimes L_{j_n})(x)\,
(\ka g(x))^{j_1+\cdots+j_n}\Bigr)\ .
\ee
Hence, $z_\rho(L)(g)$ is indeed of the form \eqref{z(L)(g)} with
\be\label{z-k}
z_{\rho\,k}(L)=L_k+\sum_{n=2}^k\frac{i^{n-1}}{n!\,\hbar^{n-1}}
\sum_{j_1+\cdots +j_n=k} d^\rho_{n,0}(L_{j_1}\otimes\cdots \otimes L_{j_n})\ ,
\ee
where $L_k:=0$ for $k\geq 3$. Finally, $z_\rho(L)$ is
obtained from \eqref{z-k} by means of \eqref{z(L)-ka}. From \eqref{z(L)(g)}
and \eqref{Z(L)-adlim} we see that $z_{\rho\,k}(L)$ is uniquely determined up to
the addition of terms $\d^a A,\,\,|a|\geq 1,\,\,A\in\Pcal$ -- as claimed 
in \eqref{z(L)}.

From \eqref{Z(L)-adlim} and (multi-)linearity of the time-ordered products,
we conclude that the adiabatic limit of \eqref{main-theorem} applied
to $iL(g)$ gives indeed \eqref{main-theorem-adlim}:
  \begin{align}\label{MThmadlim}
  \sigma_\rho(\mathbf{S}_{\rho^{-1}\mathbf{m}}[\si_\rho^{-1}(L^{\mathbf{m}})])&:=
\lim_{\eps\downarrow 0}\si_\rho\circ S_{\rho^{-1}\mathbf{m}}\circ\si_\rho^{-1}(iL^{\mathbf{m}}(g_\eps))
  =\lim_{\eps\downarrow 0}S_{\mathbf{m}}\bigl(Z_\rho(iL^{\mathbf{m}}(g_\eps))\bigl)\nn\\
  & =\lim_{\eps\downarrow 0}S_{\mathbf{m}}\bigl(i\,z_\rho(L^{\mathbf{m}})(g_\eps)\bigl)=:
\mathbf{S}_{\mathbf{m}}[ z_\rho(L^{\mathbf{m}})]\ .
  \end{align}
Here we assume that $S_\mathbf{m}(iL(g))$ is renormalized such that the adiabatic limit
$\eps\downarrow 0$ exists and is unique for 
  $\si_\rho\circ S_{\rho^{-1}\mathbf{m}}\circ\si_\rho^{-1}(iL(g_\eps))\ $ for all $\rho > 0$;
hence, this limit exists also for $S\bigl(i\,z_\rho(L)(g_\eps)\bigr)$.

\textbf{Proof of stability of PC \eqref{PC-RG}:} by using the Main Theorem, the relations
  \be
  \si_\rho^{-1}(L^\mathbf{m}(g))=L^{\rho^{-1}\mathbf{m}}(g_{1/\rho})\,\,\, 
\quad (\,\text{again}\,\,g_\la(x):=g(\la x)\,)
  \ee
  and
  \be 
  \si_\rho(F\star_{\rho^{-1}\mathbf{m}}G)= \si_\rho(F)\star_{\mathbf{m}}\si_\rho(G)\ ,\quad
  \rho\,\si_\rho\circ Q_{\rho^{-1}\mathbf{m}}=Q_{\mathbf{m}}\ ,
  \ee
  we obtain
  \begin{align}\label{Q-zrhoL}
  [Q_{\mathbf{m}},S_{\mathbf{m}}(Z_\rho(iL^\mathbf{m}(g_\eps)))]_{\star_{\mathbf{m}}} & =
  [Q_{\mathbf{m}},\si_\rho\circ S_{\rho^{-1}\mathbf{m}}(iL^{\rho^{-1}\mathbf{m}}(g_{\eps/\rho}))]_{\star_{\mathbf{m}}}\nn\\
  & = \rho\,\si_\rho\Bigl([Q_{\rho^{-1}\mathbf{m}}, S_{\rho^{-1}\mathbf{m}}(iL^{\rho^{-1}\mathbf{m}}(g_{\eps/\rho}))
  ]_{\star_{\rho^{-1}\mathbf{m}}}\Bigr)\ .
  \end{align}
  Now, assuming that $S_\mathbf{m}(iL^\mathbf{m}(g))$ fulfills PC \eqref{PC}
for all values $m,m_H>0$ of the masses, we conclude that the adiabatic limit $\eps\downarrow 0$ of the 
  last expression in \eqref{Q-zrhoL} vanishes. (Here we use that
  it does not matter whether we perform the adiabatic limit with $g$ or $g_{1/\rho}$, since it is unique.) 
  With that and with \eqref{Z(L)-adlim} we obtain the assertion \eqref{PC-RG}:
  \be
  0\approx\lim_{\eps\downarrow 0}\,[Q,S(Z_\rho(iL(g_\eps)))]_\star=
  \lim_{\eps\downarrow 0}\,[Q,S(i\,z_\rho(L)(g_\eps))]_\star =\bigl[Q,\mathbf{S}[z_\rho(L)]\bigr]_\star\ .
\quad\qed
  \ee

\textbf{Completion of the derivation of the form of $z_\rho(L)$:} having given the construction of 
$z_\rho(L)$ \eqref{z-k} (see also Sect.~5 of \cite{DF04}), we are able to explain why on 
the r.h.s.~of \eqref{z(L)} precisely these field monomials appear and no others:
\begin{itemize}
  \item each term appearing in $z_\rho(L)$ is Lorentz invariant, has ghost number $=0$ and
  its mass dimension is $\leq 4$ (see formula (5.5) in \cite{DF04}).

  \item Since the only interaction term containing $\ut u$ is $m\ut u\vf$, each term in $(z_\rho(L)-L)$
  which is bilinear in the ghost fields has a factor $m^2$ and, hence, its  mass dimension is $\leq 2$.
  This excludes a $\d\ut\,\d u$-term and non-trivial trilinear and 
  quadrilinear terms containing $\ut u$.

  \item The property that $L$ is even under the field parity transformation \eqref{fieldparity} 
  goes over to each diagram contributing to $S(iL(g))$ and, hence, each term appearing in $z_\rho(L)$
  has also this property. This reduces the number of possible terms in $z_\rho(L)$ quite strongly.

  \item To discuss the appearance of one-leg terms $\sum_ac_a\,\d^a\phi$ in $z_\rho(L)\ $ 
(where $\phi=A^\mu,\,B,\,\vf$ and $c_a\in\CC$),
we write \eqref{main-theorem} to $n$-th order, by using the chain rule:
\begin{align}\label{eq:Zn}
Z_{\rho,\mathbf{m}}^{(n)}\bigl(L(g_\eps)^{\otimes n}\bigr)=&\sigma_\rho \circ T_{n\,\mathbf{m}/\rho}\bigl(
(\sigma_\rho^{-1}\,L(g_\eps))^{\otimes n}\bigr)-T_{n\,\mathbf{m}}\bigl(L(g_\eps)^{\otimes n}\bigr)\notag\\
&-\sum_{P\in\mathrm{Part}(\{1,...,n\},\,n>|P|>1} T_{|P|\,\mathbf{m}}\bigl(\otimes_{I\in P}
Z_{\rho,\mathbf{m}}^{|I|}(L(g_\eps)^{\otimes|I|})\bigr)\ ,
\end{align}
where $Z_\rho^{(n)}:=Z_\rho^{(n)}(0)$ is the $n$-th derivative of $Z_\rho(F)$ at $F=0$ and
the two terms with $|P|=n$ and $|P|=1$, resp., are explicitly written out. 
Taking \eqref{Z(L)-form}
into account, we see that each one-leg term appearing on the r.h.s.~of \eqref{eq:Zn} is 
a sum of terms of the form
\be\label{1-leg}
\int dx_1\ldots dx_k\,G_1(\eps x_1)\ldots G_k(\eps x_k)\sum_b\d^b\phi(x_k)
t_b(x_1-x_k,\ldots,x_{k-1}-x_k)\ ,
\ee
where $k=n$ or $k=|P|$, the testfunctions $G_j$ are of the form $G_j(x)=\prod_{l=1}^{n_j}\d^{a_{jl}}g(x)$
and $t_a=\om_0\bigl(T_k(\ldots)\bigr)$ is the vacuum expectation value of a time-ordered product.
The expression \eqref{1-leg} can be written as an integral in momentum space: up to a power of 
$(2\pi)$ as prefactor it is equal to
\begin{align}\label{1-leg-p}
\int dp_1\ldots dp_k&\,\hat G_1(p_1)\ldots \hat G_k(p_k)\,\,\hat\phi(-\eps(p_1+\ldots +p_k))\notag\\
\cdot\sum_b &(-i\eps(p_1+\dots+p_k))^b\,\hat t_b(-\eps p_1,\ldots,-\eps p_{k-1})\ .
\end{align}
From \cite{EG73} we know that $\hat t_b(p)$ is analytic in a neighbourhood of $p=0$, since all 
fields are massive. Hence, in the adiabatic limit $\eps\downarrow 0$ of \eqref{1-leg-p}, 
the $(|b|>0)$-terms vanish and, hence, do not contribute to $z_\rho(L)$.

To avoid the appearance of a $(b=0)$-term in $z_\rho(L)$, 
we first mention that we only have to consider 
the case in which the singular order of $t:=t_{b=0}$ is $\om(t)\geq 0$;%
\footnote{\label{fn:extension} For $t\in\Dcal'(\RR^l)$ or $t\in\Dcal'(\RR^l\setminus\{0\})$, 
the singular order is defined as $\om(t):=\mathrm{sd}(t)-l$, 
where $\mathrm{sd}(t)$ is Steinmann's scaling degree of $t$, 
which measures the UV-behaviour of $t$ \cite{Ste71}. In the Epstein-Glaser framework, 
renormalization is the extension of a distribution
$t^\circ\in\mathcal{D}'(\RR^l\setminus\{0\})$ to a distribution $t\in
\mathcal{D}'(\RR^l)$, with the condition that $\mathrm{sd}(t)=\mathrm{sd}(t^{\circ})$.
In the case $\mathrm{sd}(t^{\circ})<l$,
the extension is unique, due to the scaling degree requirement, and obtained by
``direct extension'', see \cite[Theorem 5.2]{BF00}, \cite[Appendix B]{DF04} and 
\cite[Theorem 4.1]{DFKR14}.}
for the following reason: a term with $\om(t)<0$ is non-local, i.e.
$\supp(t)\not\subset\{0\}$. But the l.h.s.~of \eqref{eq:Zn} is local; hence, the
$(\om(t)<0)$-terms appearing on the r.h.s.~of \eqref{eq:Zn} must cancel, when 
restricted to $\Dcal(\RR^{4k}\setminus\Delta_k$) 
(where $\Delta_k:=\{(x_1,\ldots,x_k)\in\RR^{4k}\,|\,x_1=\ldots =x_k\}$).
Since for these terms, the extension to $\Dcal(\RR^{4k})$ is unique, 
they cancel also on $\Dcal(\RR^{4k})$. 

Obviously, the finite renormalization
\be\label{ren-1-leg}
\hat t(p)\mapsto \hat t(p)-\hat t(0)\ ,\quad t\equiv t_{b=0}\ ,
\ee
which is admitted due to $\om(t)\geq 0$, 
removes the possible one-leg terms in $z_\rho(L)$.
This renormalization preserves PGI, because of
\be
[Q,\phi]\approx i\,\d_\mu\phi_1^\mu\ ,\,\,\,\text{where}\,\,\, 
\phi_1^\mu= 0,\,-\tfrac{\d^\mu u}{m},\,
g^{\mu\nu}u\,\,\,\text{for}\,\,\, \phi= \vf,\,B,\,A^\nu\ ,\,\text{resp.};
\ee
in detail:
\be
[Q\,,\,\hat t(0)\,\phi(x_k)\,\delta(x_1-x_k,\ldots,x_{k-1}-x_k)]\approx
i\sum_{l=1}^k\d_\mu^{x_l}\Bigl(\hat t(0)\,\phi_1^\mu(x_k)\,
\delta(x_1-x_k,\ldots,x_{k-1}-x_k)\Bigr)\ .
\ee
We point out that when we perform the finite renormalization \eqref{ren-1-leg}
for a $t$ belonging to $T_{n\,\mathbf{m}}$, then the corresponding $t$ belonging to 
$\sigma_\rho \circ T_{n\,\mathbf{m}/\rho}\circ (\sigma_\rho^{-1})^{\otimes n}$ is 
automatically modified by precisely the same finite renormalization,
because the renormalization condition $\hat t(0)=0$ is scaling invariant.

If one does not perform the finite renormalization \eqref{ren-1-leg},
one-leg terms may appear in $z_\rho(L)$; however, only in second and higher loop 
orders. Namely, they fulfill \eqref{odd} with $q_0=0$ and $n:=r_0+s\geq 1$, hence 
they appear in \eqref{z(L)} as
\be
z_\rho(L)=\hbar^{-1}\Bigl(\ka^{-1}\sum_{n=2}^\infty e_\rho^{(n)}\,
(\ka^2\hbar)^n\,\phi+\ldots\Bigr)\ .
\ee
\end{itemize}

\section{Geometrical interpretation at all scales to 1-loop order}\label{sec:1-loop}

In this section we explain, how one can fulfill the geometrical interpretation at all scales,
i.e.~the equations \eqref{e-geom-interpret1}-\eqref{e-geom-interpret12}, on 1-loop level.
For this purpose we derive a lot of results about the 1-loop coefficients $e_\rho^{(1)}$ \eqref{powerseries}
of the running interaction $z_\rho(L)$ \eqref{z(L)}. Throughout we choose Feynman gauge 
$\La =1$ for the initial $U(1)$-Higgs model. The conventions for the signs and factors $i,\,2\pi$ 
are fixed in \eqref{conventions}.

\subsection{The two ways to renormalize}
\label{ssec:explicit-comput}

Renormalizing a 1-loop Feynman diagram, there are two 
crucially different methods to choose the renormalization mass scale. 
We explain this in terms of the computation of the 1-loop coefficient
$c_{2\rho}^{(1)}$, which is the one that is most easily to compute.

\textbf{Computation of $c_{2\rho}^{(1)}$:} we recall that 
$Z_\rho\bigl(i\,L(g)/\hbar\bigr)$ is a formal Taylor series,
\be\label{Z-series}
Z_\rho\bigl(i\,L(g)/\hbar\bigr)=i\,L(g)/\hbar+\sum_{n=2}^\infty\frac{i^n}{\hbar^n\,n!}\,
Z_\rho^{(n)}(L(g)^{\otimes n})\ ;
\ee
according to \eqref{eq:Zn} the $(n=2)$-term is obtained by
\be\label{Z-2}
Z_\rho^{(2)}(L^\mathbf{m}(g)^{\otimes 2})=\si_\rho\circ T_{2\,\rho^{-1}\mathbf{m}}
\bigl(\si^{-1}_\rho(L^\mathbf{m}(g))^{\otimes 2}\bigr)-
T_{2\,\mathbf{m}}\bigl(L^\mathbf{m}(g)^{\otimes 2}\bigr)\ .
\ee
To compute $c_{2\rho}^{(1)}$ we select the term with external legs $\ut u$, which is $\sim\ka^2$:
\be
T_{2\,\mathbf{m}}\bigl(L_1(x_1)\otimes L_1(x_2)\bigr)=m^2\,
\bigl(t^{\ut u}_\mathbf{m}(x_1-x_2)\,\ut(x_1)u(x_2)+(x_1\leftrightarrow x_2)+...\bigr)\ ,
\ee
where
\be
t^{\ut u}_\mathbf{m}(x_1-x_2):=\omega_0\bigl(T_{2\,\mathbf{m}}
(u(x_1)\vf(x_1)\otimes\ut(x_2)\vf(x_2))\bigr)
\ee
and $\omega_0$ denotes the vacuum state. The corresponding
contribution to $Z_\rho(\hbar^{-1}i\,L(g))$ reads
\begin{align}\label{ut-u}
Z_\rho(i\,L(g)/\hbar)=& \hbar^{-1}i\,L(g)-\frac{\ka^2\,m^2}{\hbar^2}\int dx_1\,dx_2\,\,g(x_1)g(x_2)\,\nn\\
&\cdot \bigl(\rho^4\,t^{\ut u}_{\rho^{-1}\mathbf{m}}(\rho(x_1-x_2))-t^{\ut u}_\mathbf{m}(x_1-x_2)\bigr)
\,\ut(x_1)u(x_2)+\ldots\ .
\end{align}
We will see that
\be\label{ut-u-scaling}
\rho^4\,t^{\ut u}_{\rho^{-1}\mathbf{m}}(\rho y)-t^{\ut u}_\mathbf{m}(y)=
\hbar^2\,C_\mathrm{fish}\,\log\rho\,\,\,\delta(y)
\ee
with a constant $C_\mathrm{fish}\in i\,\RR$. 
Inserting \eqref{ut-u-scaling} into \eqref{ut-u} 
and using \eqref{Z(L)-form}, \eqref{z-k}, we end up with
\be\label{c2}
c^{(1)}_{2\rho}=-iC_\mathrm{fish}\,\log\rho\ .
\ee

To derive \eqref{ut-u-scaling} and to compute the number $C_\mathrm{fish}$, 
we start with the unrenormalized
version of $t^{\ut u}_\mathbf{m}$: the restriction of $t^{\ut u}_\mathbf{m}(y)$ to 
$\mathcal{D}(\RR^4\setminus\{0\})$ agrees with
\be\label{t-uu}
t^{\ut u\,\circ}_\mathbf{m}(y):=\hbar^2\,t_{m,m_H}(y)\,\quad t_{m,m_H}(y):=
\Delta^F_m(y)\,\Delta^F_{m_H}(y)\in\mathcal{D}'(\RR^4\setminus\{0\})\ ,
\ee
where $\Delta^F_m$ is the Feynman propagator to the mass $m$. Due to
$\rho^2\,\Delta^F_{\rho^{-1}m}(\rho y)=\Delta^F_m(y)$,
the unrenormalized distribution $t^{\ut u\,\circ}_\mathbf{m}$ scales homogeneously,
\be\label{scal-hom}
\rho^4\,t^{\ut u\,\circ}_{\rho^{-1}\mathbf{m}}(\rho y)=t^{\ut u\,\circ}_\mathbf{m}(y)\ .
\ee
The question is, whether this property can be maintained in the process of renormalization
(i.e.~extension, see footnote \ref{fn:extension}). 

To construct the extension $t^{\ut u}_\mathbf{m}\in
\mathcal{D}'(\RR^4)$ we use the scaling and mass expansion 
(shortly 'sm-expansion') \cite{Duetsch2014}; in the present
case this means that we split $t^{\ut u\,\circ}_\mathbf{m}(y)$ into 
the corresponding massless distribution $-\hbar^2 t_\mathrm{fish}^\circ(y)$
and a remainder $r^\circ_\mathbf{m}(y)$, which is of order $r^\circ_\mathbf{m}=\mathcal{O}(m^2,m_H^2)$:
\be\label{ut-u-sm}
t^{\ut u\,\circ}_\mathbf{m}(y)=\hbar^2 t_\mathrm{fish}^\circ(y)+r^\circ_\mathbf{m}(y)\ ,\quad
t_\mathrm{fish}^\circ(y):=(D^F(y))^2\,,\quad 
\mathrm{sd}(t_\mathrm{fish}^\circ)=\mathrm{sd}(t^\circ_\mathbf{m})=4\ ,
\ee
where $D^F:=\Delta^F_{m=0}$ is the massless Feynman propagator. The remainder
$r^\circ_\mathbf{m}$ has a unique extension $r_\mathbf{m}\in \mathcal{D}'(\RR^4)$ with
$\mathrm{sd}(r_\mathbf{m})=\mathrm{sd}(r^\circ_\mathbf{m})=2$, which is obtained 
by direct extension; it preserves the homogeneous scaling \eqref{scal-hom}. 

The unrenormalized massless part $t_\mathrm{fish}^\circ$ scales 
homogeneously in $y$, but this property cannot be preserved: the extension
needs a mass scale $M >0$ and with that homogeneous scaling in $y$ is broken at least 
by a logarithmic term. All extensions with such a minimal breaking can be obtained by 
differential renormalization:
\be\label{diffren}
t_\mathrm{fish}^M(y)=\frac{-1}{64\,\pi^4}\,\square_y\Bigl(
\frac{\log(-M^2(y^2-i0))}{y^2-i0}\Bigr)\in\mathcal{D}'(\RR^4)  \ ,
\quad M>0\,\,\text{arbitrary},
\ee
see e.g.~appendix B in \cite{DF04}. 

\textbf{The two methods to choose the renormalization mass scale.}
Whether homogeneous scaling in $y$ and $(m^{-1},m_H^{-1})$ \eqref{scal-hom} is maintained
depends on the following choice:
\begin{itemize}
\item[(A)] If we choose for $M$ a fixed mass scale, which is independent of $m,m_H$, 
homogeneous scaling is broken:
\begin{align}\label{diffren-scal}
\rho^4\,t^{\ut u}_{\rho^{-1}\mathbf{m}}(\rho y)-t^{\ut u}_\mathbf{m}(y)=&
\hbar^2\,\bigl(\rho^4\,t_\mathrm{fish}^M(\rho y)-t_\mathrm{fish}^M(y)\bigr)\notag\\
=&\hbar^2\,C_\mathrm{fish}\,\log\rho\,\,\,\delta(y)\ ,\quad 
C_\mathrm{fish}:=\frac{-i}{8\,\pi^2}\ ,
\end{align}
by using $\square(\frac{1}{y^2-i0})=i\,4\,\pi^2\,\delta(y)\ $.
The breaking term is unique, i.e.~independent of $M$; 
therefore, we may admit different values
of $M$ for different $t$-distributions, however, all $M$'s must be independent of $m,m_H$.

\item[(B)] Homogeneous scaling \eqref{scal-hom} can be maintained by choosing 
$M:=\al_1 m+\al_2 m_H$, where $(\al_1,\al_2)\in(\RR^2\setminus\{(0,0)\})$
may be functions of $\tfrac{m}{m_H}$:
\be\label{diffren-scal-1}
\rho^4\,t^{\ut u}_{\rho^{-1}\mathbf{m}}(\rho y)-t^{\ut u}_\mathbf{m}(y)=
-\hbar^2\,\bigl(\rho^4\,t_\mathrm{fish}^{\rho^{-1}M}(\rho y)-t_\mathrm{fish}^M(y)\bigr)=0\ .
\ee
With that, $t^{\ut u}_\mathbf{m}$ does not contribute to the RG-flow: $c_{2\rho}^{(1)} =0$.
\end{itemize}

\begin{rem} 
When using the renormalization method (B), we have to weaken a bit the
sm-expansion axiom given in \cite{Duetsch2014}. In detail: among other conditions,
this axiom requires that the term $l=0$ in the sm-expansion%
\footnote{In $4$ dimensions only \textit{even} powers of $m$ and $m_H$ appear.} 
$$
t_\mathbf{m}(y)=\sum_{l=0}^L\sum_{l_1,l_2\geq 0,\,l_1+l_2=l}m^{2l_1}m_H^{2l_2}\,u_{l_1,l_2}^{(\mathbf{m})}(y)
+\mathfrak{r}_{2L+2}^{(\mathbf{m})}(y)\ ,
$$
i.e. $u_{0,0}^{(\mathbf{m})}$, is independent of $\mathbf{m}$. 
Only the distributions
$u_{l_1,l_2}^{(\mathbf{m})}$ with $l_1+l_2\geq 1$ may be polynomials in
$(\log\tfrac{m}{M_1},\log\tfrac{m_H}{M_1})$, where $M_1>0$ is a fixed 
mass scale. From \eqref{ut-u-sm}-\eqref{diffren} we explicitly see that this condition
is violated by the method (B); e.g.~for $M:=m$ we have
$$
\square_y\Bigl(\frac{\log(-m^2(y^2-i0))}{y^2-i0}\Bigr)=
\square_y\Bigl(\frac{\log(-M_1^2(y^2-i0))}{y^2-i0}\Bigr)+
8i\pi\,\delta(y)\,\,\log\tfrac{m}{M_1}\ .
$$
So, when using method (A), we keep the original version of the sm-expansion axiom;
but,  when using method (B), we admit that also $u_{0,0}^{(\mathbf{m})}$ is a 
polynomial in $(\log\tfrac{m}{M_1},\log\tfrac{m_H}{M_1})$. Proceeding analogously to
\cite{Duetsch2014}, one verifies that using method (B) and the inductive Epstein-Glaser
construction of time-ordered products, this weakened version of the sm-expansion axiom 
can be fulfilled to all orders of perturbation theory.
\end{rem}

{\bf Conjecture:} \textit{If we renormalize all $t$-distributions in all inductive 
steps of the Epstein-Glaser construction by the choice (B), i.e.~we use as renormalization mass
scale throughout $M:=\al_1 m+\al_2 m_H$ (where $(\al_1,\al_2)$ are as above, different values 
of $(\al_1,\al_2)$ for different diagrams are admitted), then the RG-flow is trivial:}
\be
z_\rho(L)=L/\hbar\quad\quad\forall\rho>0.
\ee

\begin{proof} We prove this Conjecture for massless, primitive divergent diagrams.%
\footnote{That is, massless diagrams $\Ga$ with singular order $\omega(\Ga)\geq 0$ 
(see footnote \ref{fn:extension}) which 
do not contain any subdiagram $\Ga_1\subset\Ga$ with less vertices and with $\omega(\Ga_1)\geq 0$.
For example, the setting sun diagram is a primitive divergent 2-loop diagram.}
This covers all massive 1-loop diagrams with singular order $\om= 0$ or $1$, because
for these diagrams, only the leading term of the sm-expansion, which is the corresponding
massless distribution, contributes to the RG-flow. However, note that also the 
$(\om =2)$-diagrams \eqref{fish-1}-\eqref{fish-2} are covered, because 
their scaling behaviour can be traced back to the scaling behaviour of the 
massless fish-diagram, see Appendix \ref{app:violation-scaling}.

Let $y:=(y_1,\ldots,y_l)$, $Y_j:=y_j^2-i0$; for the considered diagrams
the unrenormalized distribution $t^\circ\in\Dcal'(\RR^{4l}\setminus\{0\})$
scales homogeneously:
\be
\rho^{\om+4l}\,t^\circ(\rho y)=t^\circ(y)\ .
\ee
We work with an analytic regularization \cite{Hollands2007}:
\be
t^{\zeta\circ}(y):=t^\circ(y)\,(M^{2l}Y_1\ldots Y_l)^\zeta\ ,
\ee
where $\zeta\in\CC\setminus\{0\}$ with $|\zeta|$ sufficiently small,
and $M>0$ is a renormalization mass scale. $t^{\zeta\circ}$ scales also 
homogeneously -- by the regularization we gain that the degree (of the scaling)  
is $(\om+4l-2l\zeta)$, which is not an integer. Therefore,
the homogeneous extension $t^\zeta \in\Dcal'(\RR^{4l})$ is unique and can be obtained by differential 
renormalization \cite[Sect.~IV.D]{DFKR14}:
\be
t^\zeta(y)=\frac{1}{\prod_{j=0}^\om (j-\om+2l\zeta)}\sum_{r_1\ldots r_{\om+1}}
\d_{y_{r_{\om+1}}}\ldots\d_{y_{r_1}}\Bigl(\overline{y_{r_1}\ldots y_{r_{\om+1}}\,t^{\zeta\circ}(y)}\Bigr)\ ,
\ee
where $\sum_r \d_{y_r}(y_r\ldots):=\sum_r \d^{y_r}_\mu(y_r^\mu\ldots)$ and the overline denotes the direct 
extension. In order that the limit $\zeta\to 0$ exists, we subtract from the Laurent series $t^\zeta$ 
its principle part. According to \cite[Corollary 4.4]{DFKR14} the term
$\sim\zeta^0$ (``minimal subtraction'') is an admissible extension $t^M$ of $t^\circ$:
\begin{align}
t^M(y)=\frac{(-1)^\om}{\om!}\sum_{r_1\ldots r_{\om+1}}
\d_{y_{r_{\om+1}}}\ldots\d_{y_{r_1}}\Bigl[&\frac{1}{2l}
\Bigl(\overline{y_{r_1}\ldots y_{r_{\om+1}}\,t^\circ(y)\,\log(M^{2l}Y_1\ldots Y_l)}\Bigr)\notag\\
&+(\sum_{j=1}^\om\frac{1}{j})\Bigl(\overline{y_{r_1}\ldots y_{r_{\om+1}}\,t^\circ(y)}\Bigr)\Bigr]\ ,
\end{align}
see \cite[formula (104)]{DFKR14}.
The second term is of the form $\sum_{|a|=\om}C_a\,\d^a\delta(y)$. The first term breaks 
homogeneous scaling in $y$ logarithmically, but we explicitly see that
\be
\rho^{\om+4l}\,t^{\rho^{-1}M}(\rho y)=t^M(y)\ ;
\ee
this proves the Conjecture.
\end{proof}

\begin{rem}\label{AB-PGI} 
We only admit renormalizations of the initial $U(1)$-Higgs model which fulfill PGI.
This requirement is neither in conflict with method (A) nor with method (B), for the following 
reason: we require PGI only for the initial model, i.e.~only at one fixed scale.  
Now, working at one fixed scale, the renormalization constant $M$ 
appearing in \eqref{diffren} may have any value $M>0$ for both methods (A) 
and (B) and, hence, one may choose it such that PGI is satisfied. These methods only prescribe
how $M$ behaves under a scaling transformation: using (A) it remains unchanged, using (B) it 
is also scaled: $M\mapsto\rho^{-1}M$.
\end{rem}


\goodbreak

\textbf{Computation of $b_{0\rho}^{(1)}$:} The purpose of this computation is to illustrate the 
methods (A) and (B) for a 1-loop coefficient having contributions from more than one Feynman diagram; 
in addition this computation is also a preparation for the following Subsection.

To compute $b_{0\rho}^{(1)}$ we have to take the 
following terms of $T_{2\,\mathbf{m}}\bigl(L_1(x_1)\otimes L_1(x_2)\bigr)$ into account:
\be\label{BdB}
t_{\mathbf{m}\,\la\nu}^{\d B\d B}(x_1-x_2)\,\d^\la B(x_1)\d^\nu B(x_2)
+\Bigl(t_{\mathbf{m}\,\nu}^{B\d B}(x_1-x_2)\, B(x_1)\d^\nu B(x_2)+(x_1\leftrightarrow x_2)\Bigr)\ ,
\ee
where $(x_1\leftrightarrow x_2)$ refers only to the $t^{B\d B}$-term and
\begin{align}\label{BdB-1}
&t_{\mathbf{m}\,\la\nu}^{\d B\d B}(x_1-x_2):=\omega_0\Bigl(T_{2\,\mathbf{m}}(\vf A_\la(x_1)
\otimes \vf A_\nu(x_2))\Bigr)\ ,\notag\\
&t_{\mathbf{m}\,\nu}^{B\d B}(x_1-x_2):=-\omega_0\Bigl(T_{2\,\mathbf{m}}(\d^\la \vf A_\la(x_1)
\otimes \vf A_\nu(x_2))\Bigr)\ .
\end {align}
The unrenormalized $t$-distributions read
\begin{align}\label{BdB-unren}
&t_{\mathbf{m}\,\la\nu}^{\d B\d B\,\circ}(y)=
-\hbar^2\,g_{\la\nu}\,\Delta^F_m(y)\,\Delta^F_{m_H}(y)\in\Dcal^\prime(\RR^{4}\setminus\{0\})\ ,\notag\\
&t_{\mathbf{m}\,\nu}^{B\d B\,\circ}(y)=
\hbar^2\,\Delta^F_m(y)\,\d_\nu \Delta^F_{m_H}(y)\in\Dcal^\prime(\RR^{4}\setminus\{0\})\ ;
\end{align}
both scale homogeneously, e.g. $\rho^5 \,t_{\rho^{-1}\mathbf{m}\,\nu}^{B\d B\,\circ}(\rho y)=
t_{\mathbf{m}\,\nu}^{B\d B\,\circ}(y)$. 

We renormalize both diagrams by using method (A). Since $t^{\d B\d B\,\circ}$
essentially agrees with $t^{\ut u\,\circ}$, we know from \eqref{diffren-scal} that
\be\label{dBdB-scal}
\rho^4\,t^{\d B\d B}_{\rho^{-1}\mathbf{m}\,\la\nu}(\rho y)-
t^{\d B\d B}_{\mathbf{m}\,\la\nu}(y)=
-\hbar^2\,g_{\la\nu}\,C_\mathrm{fish}\,\log\rho\,\,\,\delta(y)\ .
\ee

To extend $t^{B\d B\,\circ}_{\mathbf{m}\,\nu}$, we use again the sm-expansion
\be\label{BdB-sm} 
t^{B\d B\,\circ}_{\mathbf{m}\,\nu}(y)=v^\circ_\nu(y)+r^\circ_{\mathbf{m}\,\nu}(y)\ ,\quad
v^\circ_\nu(y):=\hbar^2\,D^F(y)\,\d_\nu D^F(y)=\frac{\hbar^2}{2}\,\d_\nu t_\mathrm{fish}^\circ(y)\,,
\ee
the statements in the preceding example about the remainder $r^\circ_{\mathbf{m}\,\nu}$
and its extension $r_{\mathbf{m}\,\nu}\in \Dcal^\prime(\RR^{4})$
hold true also in the present case, with the exception that now 
$\mathrm{sd}(r^\circ_{\mathbf{m}\,\nu})=3$.
All extensions of $v^\circ_\nu$ with a minimal (i.e.~logarithmic) breaking of homogeneous scaling 
in $y$ can be obtained by differential renormalization:
\be
v^M_\nu(y)=\frac{\hbar^2}{2}\,\d_\nu t_\mathrm{fish}^M(y)\in\mathcal{D}'(\RR^4)\ ,
\quad M>0\,\,\text{arbitrary}.
\ee
Choosing $M$ according to method (A) we get
\be\label{BdB-scal}
\rho^5 \,t_{\rho^{-1}\mathbf{m}\,\nu}^{B\d B\,\circ}(\rho y)-
t_{\mathbf{m}\,\nu}^{B\d B\,\circ}(y)=
\rho^5\,v^M_\nu(\rho y)-v^M_\nu(y)=
\frac{\hbar^2}{2}\,C_\mathrm{fish}\,\log\rho\,\,\,\d_\nu\delta(y)\ .
\ee

Taking \eqref{Z-2} into account we see that the terms \eqref{BdB} give
\begin{align}
Z_\rho^{(2)}&(L(g)^{\otimes 2})=C_\mathrm{fish}\,\ka^2\,\hbar^2\,\log\rho
\int dx_1 dx_2\,\,g(x_1)g(x_2)\notag\\
&\cdot\Bigl(\d_\nu\delta(x_1-x_2)\,B(x_1)\d^\nu B(x_2)
-g_{\la\nu}\,\delta(x_1-x_2)\,\d^\la B(x_1)\d^\nu B(x_2)+\ldots\Bigr)+\ldots\ ,
\end{align}
which yields
\be
z_\rho(L)=\hbar^{-1}\Bigl(L-\frac{iC_\mathrm{fish}\,\hbar\ka^2}{2}\,\log\rho\,\,(1+1)\,
(\d B)^2+\ldots\Bigr)
\ee
by using \eqref{Z-series} and \eqref{z-k}. We end up with
\be\label{b0}
b^{(1)}_{0\rho}=-2i\,C_\mathrm{fish}\,\log\rho=\frac{-1}{4\,\pi^2}\,\log\rho\ .
\ee

The conjecture can explicitly be verified: renormalizing $t^{B\d B\,\circ}$ or $t^{\d B\d B\,\circ}$ 
by means of method (B) the pertinent expressions
\eqref{dBdB-scal} and \eqref{BdB-scal}, respectively, vanish. Hence, also the values
$b^{(1)}_{0\rho}=\tfrac{-1}{8\,\pi^2}\,\log\rho$ and $b^{(1)}_{0\rho}=0$ can appear.

Note that
$$
\omega_0\Bigl(T_{2\,\mathbf{m}}(A\d\vf(x_1)\otimes A\d\vf(x_2))\Bigr)\, B(x_1) B(x_2)
$$
contributes only to $b^{(1)}_{1\rho}$ and not to $b^{(1)}_{0\rho}$, see \eqref{BB1} and \eqref{b1,b2}.

\subsection{Equality of certain coefficients}\label{ssec:e=e}

In this subsection we explain how some of the equations 
\eqref{e-geom-interpret1}-\eqref{e-geom-interpret12}
(which express the geometrical interpretability at all scales) can be fulfilled on 1-loop
level, by renormalizing such that certain Feynman diagrams, which go over into each other
by exchanging $B\leftrightarrow\vf$ for some lines, give the same contribution to the 
RG-flow (up to possibly different combinatorial factors).

\textbf{How to obtain $c^{(1)}_{0\rho}=b^{(1)}_{0\rho}$:} The terms contributing to $c^{(1)}_{0\rho}$
are obtained from \eqref{BdB}-\eqref{BdB-1} by exchanging $B\leftrightarrow \vf$ throughout.
The corresponding unrenormalized distributions $t^{\d\vf\d\vf\,\circ}$ and $t^{\vf\d\vf\,\circ}$ are given by 
\eqref{BdB-unren} with $\Delta^F_{m_H}$ replaced by $\Delta^F_m$. However, this modification does 
not show up in the pertinent massless parts $-g_{\la\nu}\,t_\mathrm{fish}^\circ$ \eqref{ut-u-sm} and 
$v_\nu^\circ$ \eqref{BdB-sm}, respectively. Since only the latter contribute to the RG-flow, 
we conclude that
\be\label{c0=b0}
c^{(1)}_{0\rho}=b^{(1)}_{0\rho}
\ee
can be obtained in the following way: 
\begin{itemize}
\item[$(\star)$]
\textit{Corresponding $t$-distributions
(or more precisely their massless parts) have to be renormalized all with method (A) or 
all with method (B)}. For the various $t$-distributions we may choose different renormalization 
mass scales $M$ when using method (A); or different linear combinations $M=\al_1 m+\al_2 m_H$ 
when using method (B). 
\end{itemize}

Taking Remark \ref{AB-PGI} into account, we see that this renormalization prescription is 
compatible with PGI of the initial $U(1)$-Higgs model.

Having obtained \eqref{c0=b0}, the equations \eqref{e-geom-interpret4}, \eqref{e-geom-interpret6}
and \eqref{e-geom-interpret8}-\eqref{e-geom-interpret9} simplify to
\be\label{e-geom-simplified}
l^{(1)}_{3\rho}=l^{(1)}_{4\rho}\ ,\quad l^{(1)}_{5\rho}=l^{(1)}_{6\rho}\ ,\quad
l^{(1)}_{7\rho}=l^{(1)}_{8\rho}=l^{(1)}_{9\rho}
\ee
on 1-loop level.

\textbf{Obtaining analogously $l^{(1)}_{1\rho}=l^{(1)}_{2\rho}$ \eqref{e-geom-interpret2}:}
There are contributions to $l^{(1)}_{1\rho}$ coming from $T_2(L_1\otimes L_2)$,
more precisely only the part $L^1_1:=BA\d\vf-\vf A\d B\ $ of $L_1$ contributes. These terms read
\begin{align}\label{l1:L1xL2}
2\Bigl(&\omega_0\Bigl(T_2\bigl(A_\la B(x_1)\otimes A_\nu B(x_2)\bigr)\Bigr)\,
\d^\la\vf(x_1)A^\nu(x_2)B(x_2)\notag\\
&-\,\omega_0\Bigl(T_2\bigl(A\d B(x_1)\otimes A_\nu B(x_2)\bigr)\Bigr)\,
\vf(x_1)A^\nu(x_2)B(x_2)\Bigr)+(x_1\leftrightarrow x_2)\ .
\end{align}
The corresponding contributions to $l^{(1)}_{2\rho}$ are obtained by exchanging $B\leftrightarrow\vf$
throughout. Proceeding similarly to the derivation of \eqref{c0=b0} (in particular the 
renormalization prescription $(\star)$ is used), we find that the contributions of these terms to 
$l^{(1)}_{1\rho}$ and $l^{(1)}_{2\rho}$ agree.

Note that similarly to \eqref{AdB1}-\eqref{sm:m-mH}, 
there is neither a contribution to $l^{(1)}_{1\rho}$ 
nor to $l^{(1)}_{2\rho}$ coming from the following $T_2(L^1_1\otimes L_2)$-term:
\be
-\frac{m_H^2}{m^2}\,\omega_0\Bigl(T_2\bigl((B\d_\la\vf-\vf\d_\la B)(x_1)\otimes B\vf(x_2)\bigr)\Bigr)\,
A^\la(x_1)B(x_2)\vf(x_2)+(x_1\leftrightarrow x_2)\ .
\ee

The contributions to $l^{(1)}_{1\rho}$, $l^{(1)}_{2\rho}$ coming from $T_3(L_1^{\otimes 3})$ use only the part
$L_1^1$ of $L_1$, the relevant terms of $T_3({L^1_1}^{\otimes 3})$ are triangle diagrams with 
$2$ or $3$ derivatives, they are of the form
\begin{align}\label{l1l2-triangle}
\Bigl(&\bigl(v^{\la\nu}_{11}(y_1,y_2)+r^{\la\nu}_{11}(y_1,y_2)\bigr)
\,A_\la(x_1)\d_\nu\vf(x_2)B(x_3)\notag\\
-&\bigl(v^{\la\nu}_{12}(y_1,y_2)+r^{\la\nu}_{12}(y_1,y_2)\bigr)
\,A_\la(x_1)\d_\nu B(x_2)\vf(x_3)\notag\\
+&\bigl(v^{\la}_{21}(y_1,y_2)+r^{\la}_{21}(y_1,y_2)\bigr)
\,A_\la(x_3)B(x_1)\vf(x_2)\notag\\
-&\bigl(v^{\la}_{22}(y_1,y_2)+r^{\la}_{22}(y_1,y_2)\bigr)
\,A_\la(x_3)\vf(x_1)B(x_2)\Bigr)+\Bigl(5\,\text{permutations of}\,x_1,x_2,x_3\Bigr)\ ,
\end{align}
where $y_j:=x_j-x_3$ and the first two lines are obtained from each other by
exchanging $B\leftrightarrow\vf$ throughout and the same for the third and fourth lines.
Moreover, we have inserted the sm-expansion. 
The remainders $r_{kl}$ ($k,l\in\{1,2\}$) do not 
contribute to the RG-flow, since they are renormalized by direct extension. The unrenormalized versions 
of the massless parts $v_{kl}$ agree pairwise: $v_k^\circ:=v_{k1}^\circ=v_{k2}^\circ
\in\Dcal^\prime(\RR^{8}\setminus\{0\})$; explicitly they read
\begin{align}\label{triangles}
v^{\la\nu\,\circ}_{1}(y_1,y_2)&=\hbar^3\Bigl(-\d^\nu D^F(y_1)\,D^F(y_2)\,\d^\la D^F(y_1-y_2)
+\d^\la \d^\nu D^F(y_1)\,D^F(y_2)\,D^F(y_1-y_2)\Bigr)\ ,\notag\\
v^{\la\,\circ}_{2}(y_1,y_2)&=-2\hbar^3\,
\d^\la \d_\nu D^F(y_1)\,\d^\nu D^F(y_2)\,D^F(y_1-y_2)\ . 
\end{align}
Obviously these $v^\circ$-distributions scale homogeneously in $(y_1,y_2)$. Renormalization breaks this symmetry
by terms of the form 
\begin{align}\label{breaking-scaling}
\rho^8\,v^{\la\nu}_{1l}(\rho y_1,\rho y_2)-v^{\la\nu}_{1l}(y_1,y_2)&=
\hbar^3\,\log\rho\,\,C_1\,g^{\la\nu}\,\delta(y_1,y_2)\ ,\notag\\
\rho^9\,v^{\la}_{2l}(\rho y_1,\rho y_2)-v^{\la}_{2l}(y_1,y_2)&=
\hbar^3\,\log\rho\,\,(C_{21}\d^\la_{y_1}+C_{22}\d^\la_{y_2})\delta(y_1,y_2)\ ,
\end{align}
where Lorentz covariance is taken into account.%
\footnote{In terms of the invariants $C_{j\triangle}$ 
computed in Appendix \ref{app:violation-scaling},
we have $C_1=-C_{1\triangle}+C_{2\triangle}=-2C_{1\triangle}$. The computation of the
invariants $C_{21}$ and $C_{22}$ is a more difficult task -- for our purposes, we do not 
need to know these numbers.\label{l1-computation}} 
According to the prescription $(\star)$ we have to choose the renormalization mass scales for $v_{k1}$
and $v_{k2}$ by the \textit{same} method.

Inserting these results into
\be\label{Z-3}
Z_\rho^{(3)}(L_1^\mathbf{m}(g)^{\otimes 3})=\si_\rho\circ T_{3\,\rho^{-1}\mathbf{m}}
\bigl(\si^{-1}_\rho(L_1^\mathbf{m}(g))^{\otimes 3}\bigr)-
T_{3\,\mathbf{m}}\bigl(L_1^\mathbf{m}(g)^{\otimes 3}\bigr)+\ldots\ ,
\ee
where $L^\mathbf{m}_1(g):=\int dx\,\,L^\mathbf{m}_1(x)\,g(x)$, we obtain the following contributions to 
$l^{(1)}_{1\rho}$ and $l^{(1)}_{2\rho}$, respectively: using method (A) throughout, we get
\be\label{l1}
l^{(1)}_{1\rho}=(-C_1-C_{21}+C_{22}-3i\,C_\mathrm{fish})\,\log\rho=l^{(1)}_{2\rho}\ ,
\ee
where the $C_\mathrm{fish}$-term is the contribution from \eqref{l1:L1xL2}.
When using (B) for
$v_1^\circ$ or $v_2^\circ$ (or for both), the constant $C_1$ or $(-C_{21}+C_{22})$, resp., (or both)
is/are replaced by zero, and analogously for the contribution from \eqref{l1:L1xL2}. In all these 
cases $l^{(1)}_{1\rho}=l^{(1)}_{2\rho}$ remains true.

\textbf{Obtaining analogously $l^{(1)}_{5\rho}=l^{(1)}_{6\rho}$ 
and $l^{(1)}_{7\rho}=l^{(1)}_{9\rho}$ \eqref{e-geom-simplified}:}
the terms contributing to $l^{(1)}_{5\rho}$ and $l^{(1)}_{7\rho}$ are listed in Appendix 
\ref{app:z(L)-coefficients}. The corresponding terms contributing to $l^{(1)}_{6\rho}$ 
and $l^{(1)}_{9\rho}$, respectively, are obtained by replacing $B\leftrightarrow \vf$ throughout. 
Proceeding as above, the renormalization prescription $(\star)$ implies
$l^{(1)}_{5\rho}=l^{(1)}_{6\rho}$ and $l^{(1)}_{7\rho}=l^{(1)}_{9\rho}$.

\textbf{Obtaining analogously $l^{(1)}_{3\rho}=l^{(1)}_{4\rho}$ 
and $l^{(1)}_{7\rho}=l^{(1)}_{8\rho}$ \eqref{e-geom-simplified}:} here, the combinatorics is 
somewhat involved -- there is not a $(1:1)$-correspondence of terms. In Appendix 
\ref{app:z(L)-coefficients} these two equations are verified by explicit computation of
the pertinent coefficients, under the assumption that all contributing terms are renormalized 
by method (A). From the calculations given there, we see that $l^{(1)}_{3\rho}=l^{(1)}_{4\rho}$ 
and $l^{(1)}_{7\rho}=l^{(1)}_{8\rho}$ hold true, also if the method (B) is used for corresponding terms.
For example, if we switch to method (B) in \eqref{l3-3} and \eqref{l4-3}, $C_{1\triangle}$ 
is replaced by zero in \eqref{l3} and \eqref{l4}, but $l^{(1)}_{3\rho}=l^{(1)}_{4\rho}$
remains true.

\subsection{Vanishing of the $A^4$-term due to maintenance of PC}\label{ssec:A4}

In this short subsection we explain, why the identity \eqref{e-geom-interpret11} 
holds true to 1-loop order. 

A byproduct of the calculations in Appendix \ref{app:PGI-tree} 
is the following (see Remark \ref{rem:A4}): 
working out stability of PC under the RG-flow,
\be\label{stability-PC}
 \lim_{\eps\downarrow 0}\,[Q,S(i\,z_\rho(L)(g_\eps))]_\star\approx 0\ ,
\ee
to order $\ka^4$, we obtain -- among other relations -- the equation
\be\label{lim[Q,A4]}
0\approx l^{(1)}_{11\rho}\,\lim_{\eps\downarrow 0}\int dx\,(g_\eps(x))^2\,[Q,(A^2)^2(x)]=
l^{(1)}_{11\rho}\,4i\,\lim_{\eps\downarrow 0}\int dx\,(g_\eps(x))^2\,A^2A\d u(x)\ .
\ee
Using results of Appendix A of \cite{DuetschSchroer2000} we may argue as follows:
since there does not exist a local field polynomial $W^\mu$ such that 
$A^2A\d u=\d_\mu W^\mu$, the equation \eqref{lim[Q,A4]} implies
\be\label{l11=0}
l^{(1)}_{11\rho}=0\ .
\ee

\subsection{Changing the running interaction by finite renormalizations}
\label{ssec:fin-ren}

To continue the 
fulfillment of the identities \eqref{e-geom-interpret1}-\eqref{e-geom-interpret12}
on 1-loop level, we take into account that the following finite 
renormalizations are admitted by the axioms of causal perturbation theory \cite{EG73,DF04}
and that they preserve PGI of the initial model:
to $T_2\bigl(L_1(x_1)\otimes L_1(x_2)\bigr)$ we may add
\begin{align}\label{fin-ren-1-1}
\hbar^2\,&\delta(x_1-x_2)\,\log\tfrac{m}{M}\,\Bigl(\al_1\,(\d\vf)^2(x_1)
+\al_2\,m_H^2\,\vf^2(x_1)+\al_3\,F^2(x_1)+\al_4\,(\d A+mB)^2\notag\\
& +\al_5\,\bigl(-m^2\,B^2(x_1)+(\d B)^2(x_1)\bigr)+\al_6\bigl(m^2\,A^2(x_1)-(\d A)^2(x_1)\bigr)
\notag\\
& +\al_7\,m^2\,\bigl(-2\,\ut u(x_1)+A^2(x_1)-B^2(x_1)\bigr)\Bigr)\ ,
\end{align}
where $\al_1,\ldots,\al_7\in\CC$ are arbitrary. 

The compatibility with PGI is obvious for the $\al_1$-, $\al_2$-, $\al_3$- 
and $\al_4$-term, because the commutator of $Q$ with the pertinent field polynomials is $\approx 0$.
For the other terms, the PGI-relation
\be\label{eq:PGI-fin-ren}
[Q,T_2\bigl(L_1(x_1)\otimes L_1(x_2)\bigr)]_\star\approx i\d_\nu^{x_1}
T_2\bigl(P^\nu_1(x_1)\otimes L_1(x_2)\bigr)+(x_1\leftrightarrow x_2)
\ee
is maintained, if we simultaneously renormalize $T_2\bigl(P^\nu_1(x_1)\otimes L_1(x_2)\bigr)$ by adding
\be
\hbar^2\,\delta(x_1-x_2)\,\log\tfrac{m}{M}\,\Bigl(\al_5\,2m\,u\d^\nu B(x_1)+
(\al_6+\al_7)\,2m^2\,A^\nu u(x_1)\Bigr)\ .
\ee

Proceeding analogously to the computation \eqref{Z-series}-\eqref{c2} of $c^{(1)}_{2\rho}$,
we find that the renormalizations \eqref{fin-ren-1-1} modify the 
1-loop coefficients $e^{(1)}_\rho$ appearing in $z_\rho(L)$ \eqref{z(L)} as follows:
\begin{align}
a_{0\rho}^{(1)}&\mapsto a_{0\rho}^{(1)}+2i\,\al_3\,\log\rho\ ,\label{fin-re-a0}\\
a_{1\rho}^{(1)}&\mapsto a_{1\rho}^{(1)}-i\,(\al_6+\al_7)\,\log\rho\ ,\label{fin-re-a1}\\
a_{2\rho}^{(1)}&\mapsto a_{2\rho}^{(1)}+i\,(\al_4-\al_6)\,\log\rho\ ,\label{fin-re-a2}\\
b_{0\rho}^{(1)}&\mapsto b_{0\rho}^{(1)}-i\,\al_5\,\log\rho\ ,\label{fin-re-b0}\\
b_{1\rho}^{(1)}&\mapsto b_{1\rho}^{(1)}+i\,(\al_4-\al_5-\al_7)\,\log\rho\ ,\label{fin-re-b1}\\
b_{2\rho}^{(1)}&\mapsto b_{2\rho}^{(1)}+i\,\al_4\,\log\rho\ ,\label{fin-re-b2}\\
c_{0\rho}^{(1)}&\mapsto c_{0\rho}^{(1)}-i\,\al_1\,\log\rho\ ,\label{fin-re-c0}\\
c_{1\rho}^{(1)}&\mapsto c_{1\rho}^{(1)}+i\,\al_2\,\log\rho\ ,\label{fin-re-c1}\\
c_{2\rho}^{(1)}&\mapsto c_{2\rho}^{(1)}-i\,\al_7\,\log\rho\ ,\label{fin-re-c2}
\end{align}
the other coefficients remain unchanged.

\begin{rem}\label{rem:fin-ren}
There are further, linearly independent (w.r.t. $\simeq$) possibilities for finite renormalization
which preserves PGI: 
\begin{itemize}
\item to $T_2\bigl(L_1(x_1)\otimes L_1(x_2)\bigr)$ we may add
\be
\hbar^2\,\delta(x_1-x_2)\,\log\tfrac{m}{M}\,\,
\beta_1\,\bigl(2\,\d\ut\d u(x_1)-(\d A)^2(x_1)+(\d B)^2(x_1)\bigr)\ ,
\ee
since $[Q,(2\,\d\ut\d u-(\d A)^2+(\d B)^2)]_\star\approx -2i\,\d^\mu(\d A\d_\mu u)$;
\item to $T_2\bigl(L_2(x_1)\otimes L_1(x_2)\bigr)$ we may add
\begin{align}\label{fin-ren-2-1}
\hbar^2\,&\delta(x_1-x_2)\,\log\tfrac{m}{M}\,\Bigl(\beta_2\,
\frac{m_H^2}{2m}\,\vf^3(x_1)\notag\\
&+\beta_3\Bigl[ m\,A^2\vf - m\,\ut u\vf + B(A\d\vf) 
 -\vf(A\d B) - \frac{m_H^2}{2m}\, \vf^3
  - \frac{m_H^2}{2m}\, B^2\vf\Bigr](x_1)\Bigr)\ ,
\end{align}
since $[\ldots]=L_1$ and $[Q,L_1]_\star\approx i\,\d_\nu P_1^\nu$ 
($P_1^\nu$ is given in \eqref{eq:LPoriginal});
\item to $T_2\bigl(L_2(x_1)\otimes L_2(x_2)\bigr)$ we may add
\be\label{fin-ren-2-2}
\hbar^2\,\delta(x_1-x_2)\,\log\tfrac{m}{M}\,\beta_4\,
\frac{m_H^2}{4m^2}\,\vf^4(x_1)\ .
\ee
\end{itemize}

However, the $\beta_1$- and $\beta_3$-renormalization 
add ``by hand'' novel kind of terms $\sim\d\ut\d u$
and $\sim m\,\ut u\vf$, respectively,
to $(z_\rho(L)-L)$ \eqref{z(L)} -- therefore, we do not take them into account. 
And, even if we would admit a $\d\ut\d u$- and a $(m\ut u\vf)$-term
in $(z_\rho(L)-L)$, the $\beta_1$- and $\beta_3$-renormalization cannot be used to fulfill 
the crucial identities \eqref{geom-int-crucial} or \eqref{geom-interpret-crucial-1},
because they do not change $a_{2\rho}^{(1)}-b_{0\rho}^{(1)}$ or
$l_{3\rho}^{(1)}-l_{0\rho}^{(1)}$, respectively. 

We may not use the  
$\beta_2$- and $\beta_4$-renormalization: they would destroy the relations 
$l_{3\rho}^{(1)}=l_{4\rho}^{(1)}$ and $l_{7\rho}^{(1)}=l_{8\rho}^{(1)}=l_{9\rho}^{(1)}$
since they would modify only $l_{3\rho}^{(1)}$ and $l_{7\rho}^{(1)}$, respectively.
\end{rem}

\subsection{Geometrical interpretation at all scales}\label{ssec:geom-all-scales}
\label{ssec:geom-interpret}

There are two necessary conditions for the geometrical interpretation at all scales, which 
are crucial, since they cannot be fulfilled by finite renormalizations.

\textbf{Verification of the first crucial necessary condition:}
The condition \eqref{geom-interpret-crucial} reads to 1-loop level
\be\label{geom-interpret-crucial-1}
l_{7\rho}^{(1)}-l_{3\rho}^{(1)}=l_{5\rho}^{(1)}-l_{0\rho}^{(1)}\ .
\ee
As discussed in Remark \ref{rem:fin-ren}, there is no possibility to fulfill this equation by  
finite renormalizations. Therefore, we investigate its validity by explicit calculation:
using the renormalization method (A) for all contributing terms, the results of Appendix
\ref{app:z(L)-coefficients} yield:
\begin{align}
\frac{l_{7\rho}^{(1)}-l_{3\rho}^{(1)}}{\log\rho}&=4\,C_{1\triangle}+
\frac{m^2}{m_H^2}\,8\,(iC_{2\square}-C_{2\triangle})\ ,\label{l7-l3}\\
\frac{l_{5\rho}^{(1)}-l_{0\rho}^{(1)}}{\log\rho}&=8\,C_{1\triangle}-4i\,C_{1\square}\ ;\label{l5-l0}
\end{align}
where cancellations of fish- with triangle-, fish- with square- and triangle- with 
square-diagrams are not used so far.
Using now relations among the invariants $C_{j\triangle}$ and $C_{j\square}$ 
(derived in Appendix \ref{app:violation-scaling}), we find that \eqref{geom-interpret-crucial-1}
holds indeed true: 
\be\label{geom-interpret-crucial-2}
l_{7\rho}^{(1)}-l_{3\rho}^{(1)}=4\,C_{1\triangle}=l_{5\rho}^{(1)}-l_{0\rho}^{(1)}\ .
\ee
The fact that we need cancellations of square- and triangle-contributions
shows that \eqref{geom-interpret-crucial-1} is of a deeper kind than the equalities 
derived in Sect.~\ref{ssec:e=e}.

The identity \eqref{geom-interpret-crucial-1} holds also if certain terms are renormalized 
by method (B), e.g.~all contributing triangle and square-diagrams with (B) and all 
contributing fish diagrams with (A), or vice versa. 

A further example, for which both sides of \eqref{geom-interpret-crucial-1}
vanish, is given below under the subtitle ``How to fulfill BRST-invariance of the running 
Lagrangian''.


\textbf{How to fulfill the second crucial necessary condition:} the condition
\eqref{e-geom-interpret12} reads to 1-loop order
\be
b_{2\rho}^{(1)}=\tfrac{1}2 \bigl(a_{2\rho}^{(1)}+b_{1\rho}^{(1)}-a_{1\rho}^{(1)}-b_{0\rho}^{(1)}\bigr)\ .
\label{geom-int-crucial}
\ee
Performing the finite renormalizations \eqref{fin-ren-1-1}, i.e.~inserting  
\eqref{fin-re-a0}-\eqref{fin-re-c2} into \eqref{geom-int-crucial}, we find that all
$\al_j$ drop out -- that is, the condition \eqref{geom-int-crucial} cannot be fulfilled by 
means of these finite renormalizations.

Inserting the explicit values \eqref{b0}, \eqref{a0,a1,a2} and \eqref{b1,b2} for the coefficients
$a_{j\rho}^{(1)},\,b_{j\rho}^{(1)}$, computed by using method (A), we obtain
\begin{align}\label{crucial-2}
\frac{1}{\log\rho}&\Bigl(\tfrac{1}2
(a_{2\rho}^{(1)}+b_{1\rho}^{(1)}-a_{1\rho}^{(1)}-b_{0\rho}^{(1)})-
b_{2\rho}^{(1)}\Bigr)\nn\\
&=iC_\mathrm{fish}\Bigl((2-\tfrac{1}4+1-3)+\frac{m_H^2}{m^2}(\tfrac{1}2-\tfrac{1}4)
+\frac{m_H^4}{m^4}(-\tfrac{1}2)\Bigr)\ .
\end{align}
Hence, using method (A) throughout, we have $\la_{12\rho}\not= 0$, i.e.~the geometrical interpretation 
is violated by terms $\sim A\d B$.

To fulfill the condition \eqref{geom-int-crucial}, we may proceed as follows: we use
\begin{itemize}
\item method (B) for the terms \eqref{BB}-\eqref{BB1} [i.e.~$b_{1\rho}^{(1)}$] and \eqref{AA1} 
[i.e.~$a_{0\rho}^{(1)}$ and part of $a_{1\rho}^{(1)}$];
\item and method (A) for \eqref{BdB-1} [i.e.~$b_{0\rho}^{(1)}$], \eqref{AA} [i.e.~part of $a_{1\rho}^{(1)}$]
and \eqref{AdB} [i.e.~$b_{2\rho}^{(1)}$].
\end{itemize}
With that the values \eqref{a0,a1,a2}-\eqref{b1,b2} are modified:
\be\label{ab-modified}
a_{0\rho}^{(1)}=0\ ,\quad a_{1\rho}^{(1)}=-4i\,C_\mathrm{fish}\,\log\rho\quad\text{and}\quad b_{1\rho}^{(1)}=0\ ,
\ee
and $a_{2\rho}^{(1)},\,b_{0\rho}^{(1)},\,b_{2\rho}^{(1)}$ remain unchanged.

\textbf{Fulfilling the remaining conditions by finite renormalizations:} 
to complete the fulfillment of the identities \eqref{e-geom-interpret1}-\eqref{e-geom-interpret12}
to 1-loop order, we show that we can reach by finite renormalizations that the numbers
$D_1,D_2,D_3$, defined by
\begin{align}
&D_1\,\log\rho:=l_{1\rho}^{(1)}-l_{0\rho}^{(1)}-\tfrac{1}2 \bigl(b_{0\rho}^{(1)}-a_{1\rho}^{(1)}\bigr)\ ,
\label{geom-int-1-loop1}\\
&D_2\,\log\rho:=l_{3\rho}^{(1)}-l_{0\rho}^{(1)}-\bigl(c_{1\rho}^{(1)}-a_{1\rho}^{(1)}\bigr)\ ,
\label{geom-int-1-loop2}\\
&D_3\,\log\rho:=l_{5\rho}^{(1)}-2\,l_{0\rho}^{(1)}+a_{1\rho}^{(1)}\ ,
\label{geom-int-1-loop3}
\end{align}
vanish. For the coefficients $e_{\rho}^{(1)}$ appearing in these definitions we use values which
fulfill the equations \eqref{e-geom-interpret2}, 
\eqref{c0=b0}, \eqref{e-geom-simplified}, \eqref{l11=0}, 
\eqref{geom-interpret-crucial-1} and \eqref{geom-int-crucial}. 

If $c_{1\rho}^{(1)},\,l_{0\rho}^{(1)},\,l_{3\rho}^{(1)}$ and $l_{5\rho}^{(1)}$ are 
renormalized by method (A) (see Appendix \ref{app:z(L)-coefficients}) and $a_{1\rho}^{(1)}$ as in
\eqref{ab-modified}, we have $D_3=0$ and $D_2=0$.%
\footnote{Since we have not computed $l_{1\rho}^{(1)}$ (see footnote \ref{l1-computation}), we 
cannot make a corresponding statement about the value of $D_1$.}
However, to be as general as possible, we admit 
arbitrary values of $D_1,D_2,D_3$ in the following.

Using \eqref{fin-re-a0}-\eqref{fin-re-c2}, we see that we have to solve the following system
of linear equations:
\begin{align}\label{geom-int-1-loop4}
D_1+\frac{i}2(\al_5-(\al_6+\al_7))&=0\nonumber\\
D_2-i(\al_2+(\al_6+\al_7))&=0\nonumber\\
D_3-i(\al_6+\al_7)&=0\ .
\end{align}
There is a unique solution for $(\al_2,\al_5,(\al_6+\al_7))$. To preserve $b_{0\rho}^{(1)}=c_{0\rho}^{(1)}$
we have to choose $\al_1=\al_5$ \eqref{fin-re-c0}. There remains a
3-dimensional freedom of renormalization: $\al_3,\,\al_4$ and $(\al_6-\al_7)$
are unrestricted. 

As a summary we explicitly give a particular solution for the coefficients $e_\rho^{(1)}$, which fulfills
the geometrical interpretation at all scales: using the method (B) only for the terms
specified before \eqref{ab-modified} (in order that we have the values \eqref{ab-modified})
and renormalizing all other terms with method (A), and then performing the 
$\al_5$-renormalization with $\al_5=2i\,D_1=2i\,l_1-4\,C_\mathrm{fish}$
\eqref{geom-int-1-loop4} and the pertinent $\al_1$-renormalization with $\al_1=\al_5$, we end up with:
\begin{align}\label{e-part-solu}
&a_{0\rho}^{(1)}=a_{2\rho}^{(1)}=l_{11\rho}^{(1)}=0\ ,\quad a_{1\rho}^{(1)}=
-4i\,C_\mathrm{fish}\,\log\rho\ ,\notag\\
&b_{0\rho}^{(1)}=c_{0\rho}^{(1)}=(2i\,C_\mathrm{fish}+2\,l_1)\,\log\rho\ ,\quad
b_{1\rho}^{(1)}=(4i\,C_\mathrm{fish}+2\,l_1)\,\log\rho\ ,\notag\\ 
&b_{2\rho}^{(1)}=3i\,C_\mathrm{fish}\,\log\rho\ ,\quad
c_{1\rho}^{(1)}=-i\,\Bigl(6\,\frac{m^2}{m_H^2}+5\,\frac{m_H^2}{m^2}\Bigr)\,C_\mathrm{fish}\,\log\rho\ ,
\quad c_{2\rho}^{(1)}=-i\,C_\mathrm{fish}\,\log\rho\ ,\notag\\
&l_{0\rho}^{(1)}=-3i\,C_\mathrm{fish}\,\log\rho\ ,\quad l_{1\rho}^{(1)}=l_{2\rho}^{(1)}=:l_1\,\log\rho\ ,\notag\\
&l_{3\rho}^{(1)}=l_{4\rho}^{(1)}=i\,\Bigl(1-6\,\frac{m^2}{m_H^2}-5\,\frac{m_H^2}{m^2}\Bigr)\,
C_\mathrm{fish}\,\log\rho\ ,\quad
l_{5\rho}^{(1)}=l_{6\rho}^{(1)}=-2i\,C_\mathrm{fish}\,\log\rho\ ,\notag\\
&l_{7\rho}^{(1)}=l_{8\rho}^{(1)}=l_{9\rho}^{(1)}=i\,\Bigl(2-6\,\frac{m^2}{m_H^2}-5\,\frac{m_H^2}{m^2}\Bigr)\,
C_\mathrm{fish}\,\log\rho\ ,
\end{align}
where $l_1$ is the number which one obtains on computing $l_{1\rho}^{(1)}=:l_1\,\log\rho$
by method (A) -- from \eqref{l1} and footnote \ref{l1-computation} we have
\be
l_1=-3i\,C_\mathrm{fish}+2\,C_{1\triangle}-C_{21}+C_{22}=-\frac{5i}2\,C_\mathrm{fish}-C_{21}+C_{22}\ .
\ee

\textbf{How to fulfill BRST-invariance of the running Lagrangian \eqref{s(z(L))=0}:}
we start with the values \eqref{e-part-solu}, except that we do not perform the finite 
renormalization with $\al_1=\al_5=-4\,C_\mathrm{fish}+2i\,l_1$, with that we have $b_{1\rho}^{(1)}=0$ and
$b_{0\rho}^{(1)}=c_{0\rho}^{(1)}=-2i\,C_\mathrm{fish}$. 

To fulfill the BRST-condition 
\be
b_{\rho}^{(1)}:=
b_{0\rho}^{(1)}=c_{0\rho}^{(1)}=a_{1\rho}^{(1)}=-b_{2\rho}^{(1)}=l_{0\rho}^{(1)}
=l_{1\rho}^{(1)}=l_{2\rho}^{(1)}=l_{5\rho}^{(1)}=l_{6\rho}^{(1)}
\ee
(see \eqref{parameter-BRS1}-\eqref{parameter-BRS2}),
there is the trivial possibility $b_{\rho}^{(1)}=0$, which is obtained by renormalizing all contributing terms
by method (B). However, there is also the solution 
$b_{\rho}^{(1)}=-3i\,C_\mathrm{fish}\,\log\rho$ which can be obtained 
from our starting values as follows: 
we perform finite renormalizations with $\al_1=\al_5=C_\mathrm{fish}$ and $\al_7=-C_\mathrm{fish}$; this yields 
$b_{0\rho}^{(1)}=c_{0\rho}^{(1)}=a_{1\rho}^{(1)}=-b_{2\rho}^{(1)}=l_{0\rho}^{(1)}=-3i\,C_\mathrm{fish}\,\log\rho$
and $c_{2\rho}^{(1)}=0$ \eqref{parameter-BRS3}, and does not change $c_{1\rho}^{(1)}$ and
$b_{1\rho}^{(1)}=0=a_{2\rho}^{(1)}$. In order that $l_{1\rho}^{(1)}=l_{2\rho}^{(1)}$ and
$l_{5\rho}^{(1)}=l_{6\rho}^{(1)}$ also get the value $-3i\,C_\mathrm{fish}\,\log\rho$ we switch the method from (A) to (B)
in the triangle terms of $l_{1\rho}^{(1)}=l_{2\rho}^{(1)}$ (i.e.~in \eqref{l1l2-triangle}) and in the term(s)
\eqref{l5-3} (or alternatively \eqref{l5-4} and \eqref{l5-6}) of $l_{5\rho}^{(1)}=l_{6\rho}^{(1)}$.

The condition $l_{\rho}^{(1)}:=c_{1\rho}^{(1)}=l_{3\rho}^{(1)}
=l_{4\rho}^{(1)}=l_{7\rho}^{(1)}=l_{8\rho}^{(1)}=l_{9\rho}^{(1)}$ \eqref{parameter-BRS2}
can non-trivially
be satisfied by replacing the terms $\mathcal{O}\bigl((\tfrac{m^2}{m_H^2})^0\bigr)$ by zero in the 
expressions \eqref{e-part-solu} for $l_{3\rho}^{(1)}
=l_{4\rho}^{(1)}$ and $l_{7\rho}^{(1)}=l_{8\rho}^{(1)}=l_{9\rho}^{(1)}$; that is, we switch the method from (A) 
to (B) in the terms \eqref{l3-3}, \eqref{l4-3}, \eqref{l7-3} and \eqref{l8-3}.

Taking into account that $a_{0\rho}^{(1)}$ is not restricted by BRST-invariance (i.e.~the finite
renormalization parameter $\al_3$ can freely be chosen), we get the following particular solution 
for the parameters \eqref{BRS-free-param}:
\be\label{BRS-solution}
\frac{a_{0\rho}^{(1)}}{\log\rho}\in\CC\,\,\,\text{arbitrary} \ ,\quad 
b_{\rho}^{(1)}=-3i\,C_\mathrm{fish}\,\log\rho\ ,\quad 
l_{\rho}^{(1)}=-i\,\Bigl(6\,\frac{m^2}{m_H^2}+5\,\frac{m_H^2}{m^2}\Bigr)\,C_\mathrm{fish}\,\log\rho\ .
\ee

\begin{rem}
We discuss whether there is a non-trivial renormalization of the gauge-fixing parameter to 1-loop 
order \eqref{gauge-ren}:
\begin{itemize}
\item if we fulfill the geometrical interpretation as described (i.e.~\eqref{ab-modified} and 
\eqref{geom-int-1-loop4} are satisfied) and choose $\al_3=0$ and $\al_4=\al_6$, we have 
$a_{0\rho}^{(1)}=0$ and $a_{2\rho}^{(1)}=0$ which yields
\be
\La_\rho=1+\mathcal{O}(\hbar^2\ka^4)\ .
\ee
\item In contrast, if we use the renormalization method (A) throughout and do not perform 
any finite renormalization, the values \eqref{a0,a1,a2} give
\be
\La_\rho=1-\frac{1}{24\pi^2}\,\log\rho\,\,\hbar\ka^2+\mathcal{O}(\hbar^2\ka^4)\ .
\ee
\end{itemize}
However, we recall that even BRST-invariance of $L_0+z_\rho(L)$ 
\eqref{s(z(L))=0} does not restrict $a_{0\rho}^{(1)}$
in any way; hence, we are free to modify $a_{0\rho}^{(1)}$ by a finite renormalization 
\eqref{fin-re-a0} and this changes $\La_\rho$ to 1-loop order.
\end{rem}

\section{PGI for tree diagrams for the running interaction}\label{sec:PGI-tree}

Besides the geometrical interpretability at all scales and BRST-invariance, 
there is a further
property which we will investigate for the running Lagrangian: PGI-tree.
Its restrictive power for a general renormalizable ansatz for the interaction
and the importance of that are pointed out in the Introduction. 
In \cite{Duetsch2005} it is generally proved that BRST-invariance of the Lagrangian
  (that is \eqref{s(z(L))=0} in our case) implies PGI-tree.  For interactions
which are only tri- and quadrilinear in the fields, it has turned out that PGI-tree restricts
the interaction as strong as BRST-invariance of the Lagrangian; however,
we will see that for $z_\rho(L)$, which contains also bilinear terms, PGI-tree is much
less restrictive.

\textbf{Definition of PGI-tree:}
to study PGI-tree, it suffices to consider the \textit{connected} tree diagrams. To
select the latter from the $S$-functionals appearing in the PGI-condition \eqref{PGI},
we first introduce the connected time-ordered products $(T^c_n)_{n\in\NN}$, by the (usual) 
recursive definition
\be\label{Tconnected}                 
  T^c_n(F_1\otimes...\otimes F_n):= T_n(F_1\otimes\ldots                                
  \otimes F_n)-\sum_{|P|\geq 2}\prod_{J\in P}T_{|J|}^c(F_{j_1}\otimes\ldots\otimes                            
  F_{j_{|J|}})\ ,        
  \ee                                                              
  where $\{j_1,\ldots,j_{|J|}\}=J$, $j_1<\ldots<j_{|J|}$,  the sum runs over
  all partitions $P$ of $\{1,...,n\}$ in at least two subsets and
  $\prod$ means the classical product. Analogously to \eqref{T-product},
let $S^c$ be the generating functional of the connected time-ordered products.
PGI \eqref{PGI} is equivalent to PGI for $S^c$, i.e.
\be\label{PGI-connected}
  [Q,S^c\bigl(i\, \mathcal{ L}(g)\bigr)]_\star\approx \frac{d}{d\eta}\vert_{\eta =0}\,
  S^c\bigl(i\, \mathcal{ L}(g)+\eta\, \mathcal{ P}^\nu(g;\d_\nu g)\bigr)\ ;
  \ee
this can be verified straightforwardly by using that 
  $[Q,\,\cdot\,]_{\star}$ is a graded derivation w.r.t.~classical product,
  see \cite[Lemma 1]{Duetsch2005}.

For a connected time ordered product $T^c_n\bigl(\mathcal{L}(g)^{\otimes n}\bigr)$, 
the tree diagrams are the terms of lowest order in $\hbar$,
if the interaction $\mathcal{L}(g)$ is homogeneous in $\hbar$, see e.g.~\cite{DF01b}.
If, as usual, $\mathcal{L}(g)\sim\hbar^{-1}$ and $\mathcal{P}(g;\d g)\sim\hbar^0$,
the tree diagrams of $S^c\bigl(i\, \mathcal{ L}(g)\bigr)$ [or $\frac{d}{d\eta}\vert_{\eta =0}\,
S^c\bigl(i\,z_\rho(L)(g)+\eta\,\mathcal{P}(g;\d g)\bigr)$] are precisely the terms 
$\sim\hbar^{-1}$ [or $\sim\hbar^0$, resp.], and all connected loop diagrams 
are of higher orders in $\hbar$. Taking into account that $[Q,F]_{\star}\sim\hbar$ 
if $F\sim\hbar^0$ (see again \cite[Lemma 1]{Duetsch2005}), we define: 
\textit{PGI-tree is the equation \eqref{PGI-connected} to lowest order in $\hbar$,
which is $\hbar^0$.}

But $z_\rho(L)$ is by itself a formal power series in $\hbar$. Therefore, we
  use a trick to select the tree diagrams from $S^c\bigl(i z_\rho(L)(g)\bigr)$ and 
$\frac{d}{d\eta}\vert_{\eta =0}\,S^c\bigl(i\,z_\rho(L)(g)+\eta\, \mathcal{P}^\nu(g;\d_\nu g)\bigr)$.
  Namely, {\it in all coefficients $e_\rho$ \eqref{powerseries}} 
(and nowhere else) {\it we replace $\hbar$ by another 
  parameter $\tau$}; however, in particular the factors $\hbar^{-1}$ for each vertex
  (see \eqref{z(L)}) and $\hbar$ for each propagator remain untouched. 
Note that this substitution concerns also 
the pertinent $Q$-vertex: $\hbar$ is replaced by $\tau$ in \eqref{L-j} \textit{and in \eqref{P-j}}.
  With that, we have $z_\rho(L)\sim\hbar^{-1}$ and $\mathcal{P}(g;\d g)\sim\hbar^0$, and we 
can apply the above given definition of PGI-tree to $S(i z_\rho(L)(g))$.
After the selection of the tree diagrams we reset $\tau:=\hbar$.

\begin{rem}\label{rem:BRS->PGI-tree}
 Writing the interaction $\mathcal{L}(g)=z_\rho(L)(g)$ and the 
pertinent $Q$-vertex $\mathcal{P}(g;f)$ by means of
the $\tau$-trick, the proof in \cite{Duetsch2005} that BRST-invariance of the Lagrangian
implies PGI-tree applies to $L_0+z_\rho(L)$ \eqref{s(z(L))=0}. In addition this proof yields
an explicit expression for the $Q$-vertex \cite[formula (3.23)]{Duetsch2005}), which gives
\be\label{Q-BRS}
P_0^{(1)\nu}  = 0  \ ,\quad 
P_1^{(1)\nu}  = b_\rho^{(1)}\,( m\,A^\nu u\vf-\d^\nu B\,u\vf +Bu\,\d^\nu\vf)  \ ,\quad
P_2^{(1)\nu}  = b_\rho^{(1)}\,(  A^\nu u \vf^2+A^\nu u B^2)   \ ,
\ee
if \eqref{s(z(L))=0} holds true. ($b_\rho^{(1)}$ is defined by \eqref{parameter-BRS1}.)
\end{rem}

\textbf{Restrictions on the 1-loop coefficients of $z_\rho(L)$ coming from PGI-tree:} here we assume that 
the coefficients $e_\rho$ of $z_\rho(L)$ are unknown.
 In Appendix \ref{app:PGI-tree} it is worked out that PGI-tree for $\mathcal{L}(g)=z_\rho(L)(g)$
can be fulfilled to order $\tau^1$ iff the following relations among the 1-loop coefficients 
$e^{(1)}_\rho$ hold true: 
  \begin{align}\label{parameter-PC}
  &a^{(1)}_{0\rho}\ ,\,\,b^{(1)}_{0\rho}\ ,\,\,b^{(1)}_{1\rho}\ ,
\,\,c^{(1)}_{0\rho}\ ,\,\, c^{(1)}_{2\rho}\ ,\,\,l^{(1)}_{7\rho}\ ,\,\,
l^{(1)}_{1\rho}\quad\text{are}\>\>\text{arbitrary}\ ,\nn\\
  &a^{(1)}_{2\rho}=0\ ,\quad l^{(1)}_{11\rho}=0\ ,\quad a^{(1)}_{1\rho}=b^{(1)}_{0\rho}+2c^{(1)}_{2\rho}-b^{(1)}_{1\rho}\ ,
  \quad -b^{(1)}_{2\rho}=l^{(1)}_{0\rho}=b^{(1)}_{0\rho}+c^{(1)}_{2\rho}-b^{(1)}_{1\rho}\ ,\nn\\
  &l^{(1)}_{2\rho}=l^{(1)}_{1\rho}\ ,\quad
  l^{(1)}_{3\rho}=l^{(1)}_{7\rho}+c^{(1)}_{2\rho}+2(b^{(1)}_{0\rho}-\tfrac{b^{(1)}_{1\rho}}{2}-l^{(1)}_{1\rho})\ ,\nn\\
  &l^{(1)}_{4\rho}=l^{(1)}_{7\rho}+c^{(1)}_{2\rho}+3b^{(1)}_{0\rho}-c^{(1)}_{0\rho}-b^{(1)}_{1\rho}-2l^{(1)}_{1\rho}
=l^{(1)}_{3\rho} +(b^{(1)}_{0\rho}-c^{(1)}_{0\rho})\ ,\nn\\
  &l^{(1)}_{5\rho}=2l^{(1)}_{1\rho}-b^{(1)}_{0\rho}\ ,\quad
  l^{(1)}_{6\rho}=2l^{(1)}_{1\rho}-c^{(1)}_{0\rho}=l^{(1)}_{5\rho}+(b^{(1)}_{0\rho}-c^{(1)}_{0\rho})\ ,\nn\\
  &l^{(1)}_{8\rho}=l^{(1)}_{7\rho}+(b^{(1)}_{0\rho}-c^{(1)}_{0\rho})\ ,\quad
  l^{(1)}_{9\rho}=l^{(1)}_{7\rho}+2(b^{(1)}_{0\rho}-c^{(1)}_{0\rho})\ ,\nn\\ 
  &c^{(1)}_{1\rho}=l^{(1)}_{7\rho} +2c^{(1)}_{2\rho}+4(b^{(1)}_{0\rho}-\tfrac{b^{(1)}_{1\rho}}{2}-l^{(1)}_{1\rho})
  +\tfrac{2m^2}{m_H^2}(l^{(1)}_{1\rho}-b^{(1)}_{0\rho}+\tfrac{b^{(1)}_{1\rho}}{2})\ .
  \end{align}
Let us compare these PGI-tree relations with the geometrical interpretability at all scales
on 1-loop level (i.e.~equations \eqref{e-geom-interpret1}-\eqref{e-geom-interpret12} to first order
in $\hbar\ka^2$): from the number of free parameters ($7$ versus $9$) we immediately see
that the geometrical interpretability cannot imply PGI-tree. Also the reversed statement does not
hold true: in order that \eqref{parameter-PC} implies the geometrical interpretability, precisely 
one additional relation is needed, namely
\be\label{add-rel}
   l^{(1)}_{1\rho}=b^{(1)}_{0\rho}-\tfrac{b^{(1)}_{1\rho}}{2}\ .
  \ee
However, note that \eqref{parameter-PC} implies the two crucial necessary conditions for the geometrical 
interpretability, \eqref{geom-int-crucial} and \eqref{geom-interpret-crucial-1}, \textit{without} this 
additional relation \eqref{add-rel}. Note also that the geometrical 
interpretability does not imply \eqref{add-rel}.

One verifies straightforwardly, that the particular solution \eqref{e-part-solu} for the 
1-loop coefficients $e_\rho^{(1)}$, generalized by an arbitrary finite renormalization of
$a_{0\rho}^{(1)}$ \eqref{fin-re-a0}, solves the system of linear
equations \eqref{parameter-PC}-\eqref{add-rel}, \textit{i.e.~there 
exists a way to renormalize such that PGI-tree and the geometrical interpretability
are satisfied.} In contrast to the latter, the system 
\eqref{parameter-PC}-\eqref{add-rel} fixes the values of the finite renormalization parameters 
$\al_3$ and $(\al_6-\al_7)$ uniquely (cf.~the discussion after \eqref{geom-int-1-loop4}), this reflects 
that \eqref{parameter-PC}-\eqref{add-rel} is more restrictive.

\textbf{Relation to minimal subtraction:}  dimensional 
regularization with minimal subtraction is a widespread scheme in conventional 
momentum space renormalization, which preserves BRST-invariance generically. 
Applied to the 1-loop diagrams of our initial model, this property implies 
that the resulting time-ordered products fulfill PGI.%
\footnote{We are not aware of a proof of this statement, but it is very plausible.
A corresponding statement for higher loop diagrams involves a partial
adiabatic limit, because 
such diagrams contain inner vertices, which are integrated out with $g(x)=1$ in  
conventional momentum space renormalization -- but PGI is formulated \textit{before}
the adiabatic limit $g\to 1$ is taken.} 
In the minimal subtraction scheme the mass scale(s) is/are
chosen in a way which belongs to the class ``use always method (A) and do not 
perform any finite renormalization''. Using the latter prescription,
neither PGI-tree nor the geometrical interpretability are maintained under the RG-flow,
because the second crucial necessary condition \eqref{geom-int-crucial} is violated.
Weakening this prescription by admitting the finite PGI-preserving renormalizations
\eqref{fin-re-a0}-\eqref{fin-re-c2}, the violation of \eqref{geom-int-crucial} cannot
be removed.%
\footnote{An alternative, simple argument that PGI-tree (and, hence, also PGI) can get lost under 
the RG-flow is the following: The $\al_1$-renormalization \eqref{fin-re-c0} maintains PGI of the initial 
model, but it can be used to violate the PGI-tree equations \eqref{parameter-PC}, since it modifies 
only $c_{0\rho}^{(1)}$ -- this argumentation works also for the $\al_2$-renormalization \eqref{fin-re-c1}.}

  \begin{rem}\label{bilin-interaction}
  For an ansatz for the interaction
  containing solely trilinear and quadrilinear terms it has been worked
  out for various models that PGI-tree determines the interaction essentially 
  uniquely\footnote{This holds also for our model. Namely, setting
  $a^{(1)}_{j\rho}=0,\,b^{(1)}_{j\rho}=0$ and $c^{(1)}_{j\rho}=0$
  (for all $j$), the restrictions from PGI-tree \eqref{parameter-PC}  and \eqref{sigma}
  yield $l^{(1)}_{k\rho}=0$ (for all $k$).} (see
  e.g.~\cite{StoraVienna1997,DS99,AsteDuetschScharf99,Scharf2001,GraciaBondia2010}).
  But here, for an ansatz containing also bilinear terms, we obtain crucially
  different results:\footnote{We are not aware of any other paper in which
  PGI has been studied for an interaction containing bilinear terms.}  
  \begin{itemize}
  \item BRST-invariance of the total Lagrangian does not determine the
    interaction uniquely (see \eqref{z(L)-BRS});
  \item PGI-tree is truly weaker than BRST-invariance
    of the Lagrangian. (Compared with \eqref{z(L)-BRS}, the relations
    \eqref{parameter-PC} leave $4$ additional parameters to be freely chosen.)

  This can be understood as follows: PGI presupposes that
  the free theory is BRST-invariant: $s_0L_0\simeq 0$. If we try to trace back the case
  of an interaction including bilinear terms to the case with solely
  tri- and quadrilinear terms, by renormalization of the wave functions
  and parameters \eqref{wf-ren}-\eqref{cc-parameter-ren}, 
  BRST-invariance of the free theory may get lost. Explicitly we 
obtain\footnote{Since $L_0^\rho=L_0+z_\rho(L)_\mathrm{bilinear}$, where 
$z_\rho(L)_\mathrm{bilinear}$ is the bilinear part of $z_\rho(L)$ \eqref{z(L)}
\textit{without} the $A\d B$-term, the easiest way to obtain the equivalence 
\eqref{s0L0} is to work out the condition $s_0\,z_\rho(L)_\mathrm{bilinear}\simeq 0$.}
\be\label{s0L0}
s_0L_0^\rho\simeq 0 \quad\Leftrightarrow\quad b_{0\rho}=0=a_{2\rho}\,\,\wedge\,\,
a_{1\rho}=b_{1\rho}=c_{2\rho}\ .
\ee
To 1-loop order we can simultaneously fulfil this condition and BRST-invariance of the 
Lagrangian \eqref{s(z(L))=0}: by using the renormalization method (B) for the relevant 
diagrams, we can reach that in the particular solution \eqref{BRS-solution} of \eqref{s(z(L))=0} 
the value for $b_{\rho}^{(1)}$ is replaced by $0$. But in general \eqref{s0L0} does not
hold true, see e.g.~the particular solution \eqref{e-part-solu}
of the geometrical interpretability. Moreover, there is the 
additional obstacle that, after the renormalization of the wave functions and parameters,
the interaction still contains the bilinear term $b_{2\rho}m\,A\d B$.

\item In \cite{DuetschSchroer2000} it is worked out for the model of three massive
  vector fields that, making a general renormalizable ansatz for the interaction,
 the condition of PC for tree diagrams (PC-tree)
restricts the interaction to the same extent as PGI-tree -- the essentially
unique solution is the $SU(2)$-Higgs-Kibble model. However, for our 
$S\bigl(iz_\rho(L)(g)\bigr)$, which contains also bilinear terms, PC-tree is 
significantly weaker than PGI-tree. This follows from our results: we have proved 
that PC (and, hence, also PC-tree) holds true, but in general PGI-tree is violated. 
 \end{itemize}
\end{rem}

\section{Summary and concluding remarks}

Defining the RG-flow by means of a scaling transformation \cite{HW03,DF04,BDF09} one can easily 
show that PC is maintained under the RG-flow. Hence, the $U(1)$-Higgs model is a consistent QFT-model 
at all scales. However, the somewhat stronger property of PGI gets lost in general,
and in particular if one uses a renormalization prescription corresponding to minimal subtraction.

Using the Epstein-Glaser axioms \cite{EG73,DF04}, completed by the requirement that the initial
model fulfills PGI, 
the RG-flow contains quite a large non-uniqueness, due to the following two facts:
\begin{itemize}
\item whether a certain Feynman diagram contributes to the RG-flow, 
depends on whether one chooses as renomalization mass scale a fixed mass (method (A)),
or a mass which is subject to our scaling transformation -- e.g.~the mass of 
one of the basic fields (method (B)).
\item By finite renormalizations \eqref{fin-ren-1-1}, which preserve PGI of the initial
model, one can modify the RG-flow.
\end{itemize}
To 1-loop level we have shown that, by using this non-uniqueness, one can achieve that
the geometrical interpretation is possible at all scales; one can even achieve that
the much stronger condition of BRST-invariance of the running Lagrangian is satisfied.
But this requires a quite (geometrical interpretation) or very (BRST-invariance) specific
prescription for the choice of the renormalization method ((A) or (B)) for the various 
Feynman diagrams, and for the finite renormalizations. If one uses always method 
(A) -- minimal subtraction is of this kind -- the geometrical interpretation is violated
by terms $\sim A\d B$; relaxing this prescription by admitting finite PGI-preserving 
renormalizations, these $A\d B$-terms cannot be removed.

Instead of a state independent renormalization scheme, as e.g.~minimal subtraction,
one may use state dependent renormalization conditions: e.g.~in the adiabatic limit 
the vacuum expectation values of certain time ordered products must agree with the
``experimentally'' known values for the masses of stable particles in the 
vacuum, and analogous conditions for parameters of certain vacuum correlation functions.
With such a scheme, quite a lot of diagrams are renormalized by method (A).
To 1-loop level, the geometrical interpretability at all scales amounts then mainly 
to the question, whether it is nevertheless possible to fulfill the second crucial necessary 
condition \eqref{geom-int-crucial}, which requires to renormalize certain diagrams
by method (B), see \eqref{crucial-2}-\eqref{ab-modified}. 
We postpone this question to future work, and we do so also for the dependence of our 
results on the initial value of the gauge-fixing parameter.

Returning to the fundamental question, already touched 
in the Introduction, whether masses are really generated by the Higgs mechanism, 
we may say that our results sow a germ of doubt.

Or -- one can keep the Higgs mechanism as 
a fundamental principle explaining the origin of mass at all scales (although it is 
not understood in a pure QFT framework), then our results forbid quite a lot of 
renormalization schemes, in particular minimal subtraction!

$\quad$

{\bf Acknowledgments.}  
During working at this paper the author was mainly at the Max Planck Institute 
for Mathematics in the Sciences, Leipzig; he thanks Eberhard Zeidler for the invitations to 
Leipzig and for enlightening discussions. In addition, the author profitted from invitations to give
a talk about the topic of this paper at the workshop 
``Algebraic Quantum Field Theory: Its status and its future'' at the Erwin Schr\"oder Institute in Vienna
(19.-23.05.2014) and at the conference ``Quantum Mathematical Physics'' in Regensburg (29.09.-02.10.2014).
The author thanks also the Vicerrector{\'i}a de Investigaci{\'o}n of the 
Universidad de Costa Rica for financial support. 
The question in the title of this paper was found during innumerable discussions with
J\"urgen Tolksdorf about his geometrical derivation of a value for the Higgs mass; the author
is very grateful to him, in particular for his detailed and patient explanations about his approach.
The author profitted also a lot from stimulating discussions with Klaus Fredenhagen, 
Jos{\'e} M. Gracia-Bond{\'i}a, Bert Schroer, Klaus Sibold and Joseph~C. V\'arilly.

\appendix

\section{Breaking of homogeneous scaling for some $1$-loop diagrams}
\label{app:violation-scaling}

In Sect.~\ref{ssec:explicit-comput} it is derived that the violation of homogeneous 
scaling of the massless fish diagram is
\be\label{scaling-fish}
(y\d_y+4)\,t_\mathrm{fish}^M(y)=
C_\mathrm{fish}\,\delta(y)\quad\text{with}\quad C_\mathrm{fish}=\frac{-i}{8\,\pi^2}\ ,
\quad y\d_y:=y_\la\d^\la_y\ ,
\ee
and that for the fish diagram $t^M_{m,m_H}$ \eqref{t-uu} (with different masses $m,m_H$) it holds
\be\label{scaling fish-mmH}
(y\d_y+4-m\d_m-m_H\d_{m_H})\,t^M_{m,m_H}(y)=C_\mathrm{fish}\,\delta(y)\ ,
\ee
where $M>0$ is a renormalization mass scale.

In this appendix we compute the breaking of homogeneous scaling 
for some massless triangle diagrams,
\begin{align}
t_{1\triangle}^{\mu\nu\circ}(y)&:=\d^\mu D^F(y_1)\,\d^\nu D^F(y_2)\,
D^F(y_1-y_2)\in\Dcal'(\RR^8\setminus\{0\})\ ,\label{triangle-1}\\
t_{2\triangle}^{\mu\nu\circ}(y)&:=D^F(y_1)\,D^F(y_2)\,
\d^\mu\d^\nu D^F(y_1-y_2)\in\Dcal'(\RR^8\setminus\{0\})\ ,\label{triangle-2}
\end{align}
some massless square diagrams,
\begin{align}
t_{1\square}^{\la\nu\circ}(y)&:=\d^\la\d^\nu D^F(y_1-y_2)\,\d_\mu D^F(y_2-y_3)\,
D^F(y_3)\,\d^\mu D^F(y_1)\in\Dcal'(\RR^{12}\setminus\{0\})\ ,\label{square-1}\\
t_{2\square}^{\la\nu\circ}(y)&:= D^F(y_1-y_2)\,\d^\nu\d_\mu D^F(y_2-y_3)\,
D^F(y_3)\,\d^\la\d^\mu D^F(y_1)\in\Dcal'(\RR^{12}\setminus\{0\})\ ,\label{square-2}\\
t_{3\square}^{\la\nu\circ}(y)&:=\d^\la D^F(y_1-y_2)\,\d^\nu\d_\mu D^F(y_2-y_3)\,
D^F(y_3)\,\d^\mu D^F(y_1)\in\Dcal'(\RR^{12}\setminus\{0\})\ ,\label{square-3}
\end{align}
and for some massive fish-like diagrams,
\begin{align}
t^{\mu\nu\,\circ}_{1\,m,m_H}(y):=\Delta^F_m(y)\,\d^\mu\d^\nu\Delta^F_{m_H}(y)
\in\Dcal'(\RR^4\setminus\{0\})\ ,\label{fish-1}\\
t^{\mu\nu\,\circ}_{2\,m,m_H}(y):=\d^\mu\Delta^F_m(y)\,\d^\nu\Delta^F_{m_H}(y)
\in\Dcal'(\RR^4\setminus\{0\})\ ,\label{fish-2}
\end{align}
by using the renormalization method (A) (see Sect.~\ref{ssec:explicit-comput}).
The point is that these computations can be traced back to the result \eqref{scaling-fish}.

\textbf{Massless triangle diagrams:}
first note that contraction of $t_{2\triangle}^{\mu\nu\circ}$ with $g_{\mu\nu}$ yields
\be
t_{2\triangle\mu}^{\mu\circ}=-i\,\delta(y_1-y_2)\,t^\circ_{\mathrm{fish}}(y_1)
\ee
by using $\square D^F(x) = -i\delta(x)$. Hence, for an arbitrary pair of almost 
homogeneous extensions to $\Dcal'(\RR^8)$, the difference is of the form
\be
t_{2\triangle\mu}^{\mu}(y)+i\,\delta(y_1-y_2)\,t_{\mathrm{fish}}(y_1)=C\,\delta(y)\ ,\quad C\in\CC\ ;
\ee
such a term scales homogeneously. We conclude that
\be
(y\d_y+8)\,t_{2\triangle\mu}^{\mu}(y)=-i\,C_{\mathrm{fish}}\,\delta(y)\ ,\quad\text{where}\quad
y\d_y:=y_1^\mu\d^{y_1}_\mu+y_2^\mu\d^{y_2}_\mu\ .
\ee
Due to Lorentz covariance, the expression $(y\d_y+8)\,t_{2\triangle}^{\mu\nu}(y)$
must be $\sim g^{\mu\nu}$; therefore, we obtain
\be\label{scaling-triangle-2}
\rho^8\,t_{2\triangle}^{\mu\nu}(\rho y)-t_{2\triangle}^{\mu\nu}(y)=
C_{2\triangle}\,g^{\mu\nu}\,\delta(y)\,\log\rho\quad\text{with}\quad
C_{2\triangle}=\frac{-i}{4}\,C_{\mathrm{fish}}\ .
\ee

To compute the violation of homogeneous scaling for $t_{1\triangle}^{\mu\nu}$, we introduce
\be
\tilde t_{\triangle}^{\mu}(y):=\d^\mu D^F(y_1)\,D^F(y_2)\,D^F(y_1-y_2)\ ,
\ee
which exists in $\Dcal'(\RR^8)$ by the direct extension (see footnote \ref{fn:extension}) 
and scales homogeneously: $(y\d_y+7)\,\tilde t_{\triangle}^{\mu}(y)=0$.
In $\Dcal'(\RR^8\setminus\{0\})$ we find
\be\label{triangle-relation}
(\d_{y_1}^\nu+\d_{y_2}^\nu)\tilde t_{\triangle}^{\mu\circ}(y)=
t_{2\triangle}^{\mu\nu\circ}(y_1-y_2,-y_2)+t_{1\triangle}^{\mu\nu\circ}(y)\ .
\ee
Therefore, arbitrary almost homogeneous extensions fulfill
\be
(\d_{y_1}^\nu+\d_{y_2}^\nu)\,\tilde t_{\triangle}^{\mu}(y)=
t_{2\triangle}^{\mu\nu}(y_1-y_2,-y_2)+t_{1\triangle}^{\mu\nu}(y)+\tilde C\,\delta(y)
\ee
for some $\tilde C\in\CC$. We conclude
\be 
0=(y\d_y+8)\,(\d_{y_1}^\nu+\d_{y_2}^\nu)\,\tilde t_{\triangle}^{\mu}(y)=
(y\d_y+8)\,t_{2\triangle}^{\mu\nu}(y_1-y_2,-y_2)+(y\d_y+8)\,t_{1\triangle}^{\mu\nu}(y)\ . 
\ee
Taking \eqref{scaling-triangle-2} into account we end up with
\be\label{scaling-triangle-1}
\rho^8\,t_{1\triangle}^{\mu\nu}(\rho y)-t_{1\triangle}^{\mu\nu}(y)=
C_{1\triangle}\,g^{\mu\nu}\,\delta(y)\,\log\rho\quad\text{with}\quad
C_{1\triangle}=-C_{2\triangle}=\frac{i}{4}\,C_{\mathrm{fish}}\ .
\ee

\textbf{Massless square diagrams:} proceeding analogously, we use that
\be
g_{\la\nu}\,t_{1\square}^{\la\nu\circ}(y)=
-i\delta(y_1-y_2)\,t_{1\triangle\mu}^{\mu\circ}(y_1,y_1-y_3)
\ee
and obtain
\be\label{scaling-square-1}
\rho^{12}\,t_{1\square}^{\la\nu}(\rho y)-t_{1\square}^{\la\nu}(y)=
C_{1\square}\,g^{\la\nu}\,\delta(y)\,\log\rho\quad\text{with}\quad
C_{1\square}=-i\,C_{1\triangle}=\frac{1}{4}\,C_{\mathrm{fish}}\ .
\ee

Taking into account that in $\Dcal'(\RR^{12}\setminus\{0\})$ it holds
\be
\d^\nu_{y_2}\Bigl(\d^\la D^F(y_1-y_2)\,\d_\mu D^F(y_2-y_3)\,
D^F(y_3)\,\d^\mu D^F(y_1)\Bigr)=-t_{1\square}^{\la\nu\circ}(y)+t_{3\square}^{\la\nu\circ}(y)\ ,
\ee
we conclude that
\be\label{scaling-square-3}
\rho^{12}\,t_{3\square}^{\la\nu}(\rho y)-t_{3\square}^{\la\nu}(y)=
C_{3\square}\,g^{\la\nu}\,\delta(y)\,\log\rho\quad\text{with}\quad
C_{3\square}=C_{1\square}\ .
\ee

Finally, by means of
\be
\d^\la_{y_1}\Bigl(D^F(y_1-y_2)\,\d^\nu\d_\mu D^F(y_2-y_3)\,
D^F(y_3)\,\d^\mu D^F(y_1)\Bigr)=t_{3\square}^{\la\nu\circ}(y)+t_{2\square}^{\la\nu\circ}(y)\ ,
\ee
we derive that
\be\label{scaling-square-2}
\rho^{12}\,t_{2\square}^{\la\nu}(\rho y)-t_{2\square}^{\la\nu}(y)=
C_{2\square}\,g^{\la\nu}\,\delta(y)\,\log\rho\quad\text{with}\quad
C_{2\square}=-C_{3\square}=-C_{1\square}\ .
\ee

Similarly to the massless fish diagram \eqref{diffren-scal}, the following holds also for the 
massless triangle diagrams 
\eqref{triangle-1}-\eqref{triangle-2} and for the 
massless square diagrams \eqref{square-1}-\eqref{square-3}:
the breaking of homogeneous scaling is \textit{equal} for all
almost homogeneous extensions. This must be so, because 
two almost homogeneous extensions differ by a term of the form $\sum_{|a|=\omega}C_a\,\d^a\delta$,
$C_a\in\CC$, which scales homogeneously.
(See footnote \ref{fn:extension} for the definition of $\omega$;
for the examples in hand we have $\omega=0$.)

\textbf{Massive fish-like diagrams:} as a preparation we first compute the violation of
homogeneous scaling of the renormalized version $t^{\mu\nu\,M}_2$ of the massless distribution 
\be\label{fish-2-0}
t^{\mu\nu\,\circ}_2(y):=\d^\mu D^F(y)\,\d^\nu D^F(y)
\in\Dcal'(\RR^4\setminus\{0\})\ .
\ee
This computation can be traced back to the result \eqref{scaling-fish} in
the following way: first we write $t^{\mu\nu\,\circ}_2$ as
\be
t^{\mu\nu\,\circ}_2(y)=\frac{y^\mu y^\nu}{48}\,\square_y\square_y 
t_\mathrm{fish}^\circ(y)\in\Dcal'(\RR^4\setminus\{0\})\ ,
\ee
which follows from the explicit formula $D^F(y)=\tfrac{-1}{4\pi^2\,(y^2-i0)}$
by straightforward calculation, taking into account that ``$y\not= 0$''.
Then, by differential renormalization we get
\be
t^{\mu\nu\,M}_2(y)=\frac{y^\mu y^\nu}{48}\,\square_y\square_y 
t_\mathrm{fish}^M(y)\in\Dcal'(\RR^4)\ ,
\ee
where $M>0$ is a fixed mass scale (method (A)).
From this relation we conclude that
\be\label{scaling-fish-2-0}
(y\d_y+6)\,t^{\mu\nu\,M}_2(y)=\frac{C_\mathrm{fish}}{48}\,y^\mu y^\nu
\,\square_y\square_y\delta(y)=
\frac{C_\mathrm{fish}}{12}\,(g^{\mu\nu}\,\square_y+2\,\d_y^\mu\d_y^\nu)\,\delta(y)\ ;
\ee
the second equality is obtained by straightforward calculation. 

We renormalize the massive fish-like diagrams \eqref{fish-1}, \eqref{fish-2} by using the
sm-expansion \cite{Duetsch2014}. Due to that we know that the violation of homogeneous 
scaling is of the form
\begin{align}
(y&\d_y+6-m\d_m-m_H \d_{m_H})\,t^{\mu\nu\,M}_{j\,m,m_H}(y)=
\Bigl(C_{j1}\,g^{\mu\nu}\,\square_y+C_{j2}\,\d_y^\mu\d_y^\nu\Bigr)\,\delta(y)\notag\\
&+g^{\mu\nu}\,\Bigl(m^2\,P_{j1}(\log\tfrac{m}{M},\log\tfrac{m_H}{M})
+m_H^2\,P_{j2}(\log\tfrac{m}{M},\log\tfrac{m_H}{M})\Bigr)\,\delta(y)\ ,\quad j=1,2,
\end{align}
where $P_{jl}(z_1,z_2)$ is a polynomial in $z_1$ and $z_2$. The term $\mathcal{O}(m^0)$
can be computed by setting $m:=0=:m_H$; hence, we know the values of the 
numbers $(C_{2l})_{l=1,2}$ from \eqref{scaling-fish-2-0}. We renormalize such that the relations
\begin{align}\label{relations-fish-m}
&\d_y^\mu\d_y^\nu t^\circ_{m,m_H}(y)=t^{\mu\nu\,\circ}_{1\,m,m_H}(y)+
t^{\mu\nu\,\circ}_{1\,m_H,m}(y)+t^{\mu\nu\,\circ}_{2\,m,m_H}(y)+
t^{\mu\nu\,\circ}_{2\,m_H,m}(y)\ ,\notag\\
&g_{\mu\nu}\,t^{\mu\nu\,\circ}_{1\,m,m_H}(y)=-m_H^2\,t^\circ_{m,m_H}(y)
\quad\text{and}\quad  t^{\mu\nu\,\circ}_{2\,m,m_H}(y)=
t^{\nu\mu\,\circ}_{2\,m_H,m}(y)
\end{align}
are maintained up to (local) terms which are in the kernel of the operator 
$(y\d_y+6-m\d_m-m_H \d_{m_H})$; for the first and the last relation this is a term of the form
$$
\bigl(C_1\,g^{\mu\nu}\,\square_y+C_2\,\d_y^\mu\d_y^\nu\bigr)\,\delta(y)
+g^{\mu\nu}\,\bigl(m^2\,C_3+m_H^2\,C_4\bigr)\,\delta(y)\ ,\quad C_k\in\CC\,\,\text{arbitrary.}
$$
Due to the sm-expansion, this renormalization prescription restricts only the local terms 
$\mathcal{O}(m^2,m_H^2)$: without this prescription the numbers $C_3$ and $C_4$
may be replaced by polynomials in $\log\tfrac{m}{M}$ and $\log\tfrac{m_H}{M}$.

By using the renormalized version of the relations \eqref{relations-fish-m} 
and \eqref{scaling fish-mmH} and 
\eqref{scaling-fish-2-0}, we determine the numbers $C_{1l}$ and
the polynomials $P_{jl}$. It results
\begin{align}
\rho^6\,&t^{\mu\nu\,M}_{1\,m/\rho,m_H/\rho}(\rho y)-t^{\mu\nu\,M}_{1\,m,m_H}(y)\notag\\
&=C_\mathrm{fish}\,\Bigl[\frac{1}{12}\,
\Bigl(-g^{\mu\nu}\,\square_y+4\,\d_y^\mu\d_y^\nu\Bigr)
-\frac{g^{\mu\nu}}4\,m_H^2\Bigr]\,\delta(y)\,\,\log\rho\ ,\label{scaling-fish-1-m}\\
\rho^6\,&t^{\mu\nu\,M}_{2\,m/\rho,m_H/\rho}(\rho y)-t^{\mu\nu\,M}_{2\,m,m_H}(y)\notag\\
&=C_\mathrm{fish}\,\Bigl[\frac{1}{12}\,
\Bigl(g^{\mu\nu}\,\square_y+2\,\d_y^\mu\d_y^\nu\Bigr)
+\frac{g^{\mu\nu}}8\,(m^2+m_H^2)\Bigr]\,\delta(y)\,\,\log\rho\ .\label{scaling-fish-2-m}
\end{align}

\section{Computation of some 1-loop coefficients of the running interaction} 
\label{app:z(L)-coefficients}

In this appendix we compute some 1-loop coefficients $e$ of $z_\rho(L)$, defined by
\eqref{z(L)}, \eqref{powerseries} and
\be\label{e-notation}
e\,\log\rho:=e^{(1)}_\rho\,,
\ee
by using the results of Appendix \ref{app:violation-scaling}.
We assume that for all contributing terms the renormalization mass scale $M$ is chosen according 
to method (A), see Sect.~\ref{ssec:explicit-comput}.
Working in Feynman gauge, we may use the following conventions:
\begin{align}\label{conventions}
&\om_0\Bigl(T_2\bigl(B(x)\otimes B(y)\bigr)\Bigr)=\hbar\,\Delta^F_m(x-y)\ ,\quad
\om_0\Bigl(T_2\bigl(\vf(x)\otimes \vf(y)\bigr)\Bigr)=\hbar\,\Delta^F_{m_H}(x-y)\ ,\notag\\
&\om_0\Bigl(T_2\bigl(u(x)\otimes \tilde u(y)\bigr)\Bigr)=\hbar\,\Delta^F_m(x-y)\ ,\quad
\om_0\Bigl(T_2\bigl(\tilde u(x)\otimes u(y)\bigr)\Bigr)=-\hbar\,\Delta^F_m(x-y)\ ,\notag\\
&\om_0\Bigl(T_2\bigl(A^\mu(x)\otimes A^\nu(y)\bigr)\Bigr)=-\hbar\, g^{\mu\nu}\,\Delta^F_m(x-y)\ ,\quad
(\square+m^2)\Delta^F_m(x)=-i\,\delta(x)\ .
\end{align}

\textbf{Coefficients of some bilinear fields:} to compute $a_0,\,a_1,\,a_2,\,b_1,\,b_2$, we have to take
into account the following terms of $T_2({L_1^2}^{\otimes 2})$ where
$L_1^2:=m\,A^2\vf+BA\d\vf-\vf A\d B-\tfrac{m_H^2}{2m}\,B^2\vf$: the most complicated is
\begin{align}\label{AA1}
&\om_0\Bigl(T_2\bigl((B\d^\mu\vf-\vf\d^\mu B)(x_1)\otimes 
(B\d^\nu\vf-\vf\d^\nu B)(x_2)\bigr)\Bigr)\,A_\mu(x_1)A_\nu(x_2)\notag\\
&=\Bigl(-t^{\mu\nu\,M}_{1\,m,m_H}(y)-t^{\mu\nu\,M}_{1\,m_H,m}(y)+t^{\mu\nu\,M}_{2\,m,m_H}(y)+
t^{\mu\nu\,M}_{2\,m_H,m}(y)\Bigr)\,A_\mu(x_1)A_\nu(x_2)\ ,
\end{align}
where $y:=x_1-x_2$. Using \eqref{scaling-fish-1-m}-\eqref{scaling-fish-2-m}, we find that 
\eqref{AA1} gives the following contribution to $Z^{(2)}_\rho(L(g)^{\otimes 2})$
\eqref{Z-2}:
\begin{align}\label{AA2}
&\ka^2\hbar^2\,C_\mathrm{fish}\int dx_1dx_2\,g(x_1)g(x_2)\,
\Bigl(\tfrac{1}3\,(g^{\mu\nu}\square_y-\d_y^\mu\d_y^\nu)+\tfrac{1}2\,g^{\mu\nu}(m^2+m_H^2)\Bigr)
\delta(y)\,A_\mu(x_1)A_\nu(x_2)\notag\\
&=\ka^2\hbar^2\,C_\mathrm{fish}\int dx\,(g(x))^2\,\Bigl(\tfrac{-1}6\,F^2(x)+
\tfrac{1}2\,(m^2+m_H^2)\,A^2(x)\Bigr)+\ldots\ ,
\end{align}
where the dots stand for terms with derivatives of $g$, which do not contribute to the adiabatic limit.
The further contributing terms are
\begin{align}
&m^2\,4\,\om_0\Bigl(T_2\bigl(A^\mu\vf(x_1)\otimes 
A^\nu\vf(x_2)\bigr)\Bigr)\,A_\mu(x_1)A_\nu(x_2)\ ,\label{AA}\\
&\frac{m_H^4}{4\,m^2}\,4\,\om_0\Bigl(T_2\bigl(B\vf(x_1)\otimes 
B\vf(x_2)\bigr)\Bigr)\,B(x_1)B(x_2)\ ,\label{BB}\\
&\om_0\Bigl(T_2\bigl(A\d\vf(x_1)\otimes A\d\vf(x_2)\bigr)\Bigr)\,B(x_1)B(x_2)
=g_{\mu\nu}\,t^{\mu\nu\,M}_{1\,m,m_H}(y)\,B(x_1)B(x_2)\ ,\label{BB1}\\
&-m\,2\,\om_0\Bigl(T_2\bigl(A^\mu\vf(x_1)\otimes 
A^\nu\vf(x_2)\bigr)\Bigr)\,A_\mu(x_1)\d_\nu B(x_2)+(x_1\leftrightarrow x_2)\notag\\
&+m\,2\,\om_0\Bigl(T_2\bigl(A^\mu\vf(x_1)\otimes 
A\d\vf(x_2)\bigr)\Bigr)\,A_\mu(x_1)B(x_2)+(x_1\leftrightarrow x_2)\ ,\label{AdB}\\
&-\frac{m_H^2}{2\,m}\,2\,\om_0\Bigl(T_2\bigl((B\d^\mu\vf-\vf\d^\mu B)(x_1)\otimes 
B\vf(x_2)\bigr)\Bigr)\,A_\mu(x_1)B(x_2)+(x_1\leftrightarrow x_2)\ .\label{AdB1}
\end{align}
The last term does not contribute to the RG-flow, because in the sm-expansion 
of the pertinent unrenormalized expression the leading terms 
(which are the corresponding massless distributions) cancel, 
\be\label{sm:m-mH}
\Delta^F_m(y)\d^\mu\Delta^F_{m_H}(y)-\d^\mu\Delta^F_m(y)\Delta^F_{m_H}(y)=
0+\mathcal{O}(m^2,m_H^2)\ ,
\ee
and the terms $\mathcal{O}(m^2,m_H^2)$ have singular order $\om\leq -1$.

Now from \eqref{AA2} and \eqref{AA} we obtain
\be\label{a0,a1,a2}
a_0=\frac{i}3\,C_\mathrm{fish}\ ,\quad a_1=i\Bigl(\frac{1}2(1+\frac{m_H^2}{m^2})-4\Bigr)\,
C_\mathrm{fish}\ ,\quad a_2=0\ ,
\ee
and from \eqref{BB}, \eqref{BB1} and \eqref{AdB} we get
\be\label{b1,b2}
b_1=i\Bigl(\frac{m_H^2}{m^2}-\frac{m_H^4}{m^4}\Bigr)\,C_\mathrm{fish}\ ,\quad
b_2=3i\,C_\mathrm{fish}\ ;
\ee
the computation of $b_2$ is analogous to the computation of $b_0$ given in 
Sect.~\ref{ssec:explicit-comput}. 

There are $5$ terms contributing to $c_1$: one term is obtained from \eqref{BB1} by
$\vf\leftrightarrow B$ and four terms are 
$\sim\om_0\Bigl(T_2\bigl(\phi(x_1)\otimes\phi(x_2)\bigr)\Bigr)
\,\vf(x_1)\vf(x_2)$ where $\phi=\vf^2,\,B^2,\,A^2$ and $\phi=\ut u$. We find
\be\label{c1}
c_1=-i\Bigl(6\,\frac{m^2}{m_H^2}+5\,\frac{m_H^2}{m^2}\Bigr)\,C_\mathrm{fish}\ .
\ee

\textbf{Coefficients of some trilinear fields:}
the contributions to $l_0,\,l_3$ and $l_4$ come from fish diagrams (without derivatives) belonging to
$T_2(L_1^0\otimes L_2)$ and from triangle diagrams (with two derivatives) belonging to 
$T_3(L_1^{1\,\otimes 2}\otimes L_1^0)$,
where $L_1^0:=m\,A^2\vf-\tfrac{m_H^2}{2m}\,(\vf^3+B^2\vf)$ and $L_1^1:=BA\d\vf-\vf A\d B$.
To compute $l_0$ we have to take into account the terms
\begin{align}
&4m\,\om_0\Bigl(T_2\bigl(A_\la\vf(x_1)\otimes A_\nu\vf(x_2)\bigr)\Bigr)\,A^\la(x_1)
A^\nu\vf(x_2)+1\,\text{permutation}\ ,\label{l0-1}\\
&\frac{-m_H^2}{4m}\,\om_0\Bigl(3\,T_2\bigl(\vf^2(x_1)\otimes\vf^2(x_2)\bigr)
+T_2\bigl(B^2(x_1)\otimes B^2(x_2)\bigr)\Bigr)\,\vf(x_1)A^2(x_2)+1\,\text{permutation}\ ,\label{l0-2}
\end{align}
\begin{align}
&\frac{-m_H^2}{2m}\,\om_0\Bigl(T_3\bigl((B\d^\mu\vf-\vf\d^\mu B)(x_1)\otimes
(B\d^\nu\vf-\vf\d^\nu B)(x_2)\otimes(3\vf^2+B^2)(x_3)\bigr)\Bigr)\notag\\
&\quad\quad\quad\quad\quad\cdot A_\mu(x_1)A_\nu(x_2)\vf(x_3)
+2\,\text{permutations}\ ,\label{l0-3}\\
&-2m\,\om_0\Bigl(T_3\bigl(A^\mu\vf(x_1)\otimes A\d B(x_2)\otimes
(B\d^\la\vf-\vf\d^\la B)(x_3)\bigr)\Bigr)\notag\\
&\quad\quad\quad\quad\quad\cdot A_\mu(x_1)\vf(x_2)A_\la(x_3)
+5\,\text{permutations}\ ,\label{l0-4}
\end{align}
where permutations of the vertices are meant. These terms yield  
\be\label{l0} 
l_0=-4i\,C_\mathrm{fish}+
\frac{m_H^2}{m^2}\,(-2i\,C_\mathrm{fish})+
\frac{m_H^2}{m^2}\,2(3+1)\,C_{1\triangle}+4\,C_{1\triangle}=-3i\,C_\mathrm{fish}\ ,
\ee
where in the first step only $C_{2\triangle}=-C_{1\triangle}$ is used, and
the $k$-th summand comes from the $k$-th term in \eqref{l0-1}-\eqref{l0-4} ($k=1,2,3,4$).
In \eqref{l3},\eqref{l4}, \eqref{l5}, \eqref{l7} 
and \eqref{l8} the summands are ordered correspondingly.

Turning to $l_3$, the terms
\begin{align}
&\frac{m}2\,\om_0\Bigl(T_2\bigl(A^2(x_1)\otimes A^2(x_2)\bigr)\Bigr)\,\vf(x_1)
\vf^2(x_2)+1\,\text{permutation}\ ,\label{l3-1}\\
&\frac{m_H^4}{8m^3}\,\om_0\Bigl(9\,T_2\bigl(\vf^2(x_1)\otimes\vf^2(x_2)\bigr)
+T_2\bigl(B^2(x_1)\otimes B^2(x_2)\bigr)\Bigr)\,\vf(x_1)\vf^2(x_2)+1\,\text{permutation}\ ,\label{l3-2}
\end{align}
\begin{align}
&\frac{-m_H^2}{2m}\,\om_0\Bigl(T_3\bigl(A\d B(x_1)\otimes
A\d B(x_2)\otimes B^2(x_3)\bigr)\Bigr)\, \vf(x_1)\vf(x_2)\vf(x_3)
+2\,\text{permutations}\ ,\label{l3-3}\\
&m\,\om_0\Bigl(T_3\bigl(A\d B(x_1)\otimes A\d B(x_2)\otimes A^2(x_3)\bigr)\Bigr)\,
\vf(x_1)\vf(x_2)\vf(x_3)+2\,\text{permutations}\ ,\label{l3-4}
\end{align}
give  
\begin{align}\label{l3} 
l_3=&\frac{m^2}{m_H^2}\,(-8i\,C_\mathrm{fish})+
\frac{m_H^2}{m^2}\,(-5i\,C_\mathrm{fish})
+4\,C_{1\triangle}+\frac{m^2}{m_H^2}\,(-8\,C_{2\triangle})\notag\\
=&iC_\mathrm{fish}\,\Bigl(1-6\,\frac{m^2}{m_H^2}-5\,\frac{m_H^2}{m^2}\Bigr)\ .
\end {align}

The contributions to $l_4$ come from the terms
\begin{align}
&\frac{m}2\,\om_0\Bigl(T_2\bigl(A^2(x_1)\otimes A^2(x_2)\bigr)\Bigr)\,\vf(x_1)
B^2(x_2)+1\,\text{permutation}\ ,\label{l4-1}\\
&\frac{m_H^4}{8m^3}\,\om_0\Bigl(3\,T_2\bigl(\vf^2(x_1)\otimes\vf^2(x_2)\bigr)
+3\,T_2\bigl(B^2(x_1)\otimes B^2(x_2)\bigr)\Bigr)\,\vf(x_1)B^2(x_2)+1\,\text{permutation}\notag\\
&+\frac{m_H^4}{8m^3}\,8\,\om_0\Bigl(T_2\bigl(\vf B(x_1)\otimes\vf B(x_2)\bigr)\Bigr)\,B(x_1)\vf B(x_2)
+1\,\text{permutation}\ ,\label{l4-2}\\
&\frac{-m_H^2}{2m}\,3\,\om_0\Bigl(T_3\bigl(A\d\vf(x_1)\otimes
A\d\vf(x_2)\otimes \vf^2(x_3)\bigr)\Bigr)\, B(x_1)B(x_2)\vf(x_3)
+2\,\text{permutations}\notag\\
&+\frac{m_H^2}{2m}\,2\,\om_0\Bigl(T_3\bigl(A\d B(x_1)\otimes
A\d\vf(x_2)\otimes \vf B(x_3)\bigr)\Bigr)\, \vf(x_1)B(x_2)B(x_3)
+5\,\text{permutations}\ ,\label{l4-3}\\
&m\,\om_0\Bigl(T_3\bigl(A\d\vf(x_1)\otimes A\d\vf(x_2)\otimes A^2(x_3)\bigr)\Bigr)\,
B(x_1)B(x_2)\vf(x_3)+2\,\text{permutations}\ ,\label{l4-4}
\end{align}
which yield  
\be\label{l4} 
l_4=\frac{m^2}{m_H^2}\,(-8i\,C_\mathrm{fish})+
\frac{m_H^2}{m^2}\,(-(3+2)i\,C_\mathrm{fish})
+(12-8)\,C_{1\triangle}
+\frac{m^2}{m_H^2}\,(-8\,C_{2\triangle})=l_3\ .
\ee

\textbf{Coefficients of some quadrilinear fields:}
the contributions to $l_5,\,l_7$ and $l_8$ come from fish diagrams (without derivatives) belonging to
$T_2(L_2\otimes L_2)$, from triangle diagrams (with two derivatives) belonging to 
$T_3(L_1^{1\,\otimes 2}\otimes L_2)$  and from square diagrams (with four derivatives) belonging 
to $T_4(L_1^{1\,\otimes 4})$. The following terms contribute to $l_5$: 
\begin{align}
&\frac{-m_H^2}{8\,m^2}\Bigl[3\,\omega_0\Bigl(T_2\bigl(\vf^2(x_1)\otimes \vf^2(x_2)\bigr)\Bigr)\,\notag\\
&\quad\quad\quad\quad\quad +\omega_0\Bigl(T_2\bigl(B^2(x_1)\otimes B^2(x_2)\bigr)\Bigr)\Bigr]\,
A^2(x_1)\vf^2(x_2)+1\,\text{permutation}\ ,\label{l5-1}\\
&4\,\omega_0\Bigl(T_2\bigl(A^\mu\vf(x_1)\otimes A^\nu\vf(x_2)\bigr)\Bigr)\,
A_\mu\vf(x_1)A_\nu\vf(x_2)\ ,\label{l5-2}\\
&\frac{1}{2}\,\om_0\Bigl(T_3\bigl(A\d B(x_1)\otimes
A\d B(x_2)\otimes B^2(x_3)\bigr)\Bigr)\, \vf(x_1)\vf(x_2)A^2(x_3)
+2\,\text{permutations}\ ,\label{l5-3}\\
&-2\,\om_0\Bigl(T_3\bigl(A\d B(x_1)\otimes
(B\d^\mu\vf-\vf\d^\mu B)(x_2)\otimes A^\nu\vf(x_3)\bigr)\Bigr)\notag\\
&\quad\quad\quad\quad\quad\cdot \vf(x_1)A_\mu(x_2)A_\nu\vf(x_3)
+5\,\text{permutations}\ ,\label{l5-4}\\
&\frac{-m_H^2}{4m^2}\,\om_0\Bigl(T_3\bigl((B\d^\mu\vf-\vf\d^\mu B)(x_1)\otimes
(B\d^\nu\vf-\vf\d^\nu B)(x_2)\otimes (3\vf^2+B^2)(x_3)\bigr)\Bigr)\notag\\
&\quad\quad\quad\quad\quad\cdot A_\mu(x_1)A_\nu(x_2)\vf^2(x_3)
+2\,\text{permutations}\ ,\label{l5-5}\\
&\om_0\Bigl(T^c_4\bigl((B\d^\nu\vf-\vf\d^\nu B)(x_1)\otimes (B\d^\mu\vf-\vf\d^\mu B)(x_2)\otimes 
A\d B(x_3)\otimes A\d B(x_4)\bigr)\Bigr)\notag\\
&\quad\quad\quad\quad\quad\cdot A_\nu(x_1)A_\mu(x_2)\vf(x_3)\vf(x_4)+5\,\text{permutations}\ ,\label{l5-6}
\end{align}
where the upper index 'c' means connected. We obtain
\be\label{l5} 
l_5=\frac{m_H^2}{m^2}\,(-2i\,C_{\mathrm{fish}})
-4i\,C_\mathrm{fish}+4\,C_{1\triangle}
+8\,C_{1\triangle}+\frac{2(3+1)\,m_H^2}{m^2}\,C_{1\triangle}
-4i\,C_{1\square}=-2i\,C_\mathrm{fish}\ ,
\ee
and in the first step only $C_{2\triangle}=-C_{1\triangle}$ and $C_{1\square}=-C_{2\square}=C_{3\square}$ are used.

The contributions to $l_7$ come from 
\begin{align}
&\Bigl[\tfrac{1}4\,\omega_0\Bigl(T_2\bigl(A^2(x_1)\otimes A^2(x_2)\bigr)\Bigr)+
36\,\bigl(\tfrac{m_H^2}{8m^2}\bigr)^2\,
\omega_0\Bigl(T_2\bigl(\vf^2(x_1)\otimes \vf^2(x_2)\bigr)\Bigr)\notag\\
&\quad\quad\quad+\bigl(\tfrac{m_H^2}{4m^2}\bigr)^2\,\omega_0\Bigl(T_2\bigl(B^2(x_1)
\otimes B^2(x_2)\bigr)\Bigr)\Bigr]\,\vf^2(x_1)\vf^2(x_2)\ ,\label{l7-1}\\
&\frac{1}2 \,\om_0\Bigl(T_3\bigl(A\d B(x_1)\otimes A\d B(x_2)\otimes A^2(x_3)\bigr)\Bigr)\,
\vf(x_1)\vf(x_2)\vf^2(x_3)+2\,\text{permutations}\ ,\label{l7-2}\\
&\frac{-m_H^2}{4m^2}\,\om_0\Bigl(T_3\bigl(A\d B(x_1)\otimes
A\d B(x_2)\otimes B^2(x_3)\bigr)\Bigr)\, \vf(x_1)\vf(x_2)\vf^2(x_3)
+2\,\text{permutations}\ ,\label{l7-3}\\
&\om_0\Bigl(T^c_4\bigl(A\d B(x_1)\otimes A\d B(x_2)\otimes A\d B(x_3)\otimes A\d B(x_4)\bigr)\Bigr)\,
\vf(x_1)\vf(x_2)\vf(x_3)\vf(x_4)\ ;\label{l7-4}
\end{align}
it results  
\begin{align}\label{l7} 
l_7=&-i\,C_\mathrm{fish}\,\Bigl(8\frac{m^2}{m_H^2}+\frac{(9+1)}2\,\frac{m_H^2}{m^2}\Bigr)
+\frac{m^2}{m_H^2}\,(-16\,C_{2\triangle})+8\,C_{1\triangle}
+\frac{m^2}{m_H^2}\,8i\,C_{2\square}\notag\\
=&iC_\mathrm{fish}\,\Bigl(2-6\,\frac{m^2}{m_H^2}-5\,\frac{m_H^2}{m^2}\Bigr)\ .
\end{align}

Finally, to compute $l_8$ we have to take account of
\begin{align}
&\vf^2(x_1)B^2(x_2)\,
\Bigl[\tfrac{1}4\,\omega_0\Bigl(T_2\bigl(A^2(x_1)\otimes A^2(x_2)\bigr)\Bigr)\notag\\
&\quad +\frac{m_H^4}{8\cdot 4\,m^4}\,
6\,\omega_0\Bigl(T_2\bigl(\vf^2(x_1)\otimes \vf^2(x_2)\bigr)
+T_2\bigl(B^2(x_1)\otimes B^2(x_2)\bigr)\Bigr)\Bigr]+1\,\text{permutation}\notag\\
&+\vf B(x_1)\vf B(x_2)\,16\,\bigl(\frac{m_H^2}{4m^2}\bigr)^2\,
\omega_0\Bigl(T_2\bigl(\vf B(x_1)\otimes \vf B(x_2)\bigr)\Bigr)\ ,\label{l8-1}
\end{align}
\begin{align}
&\frac{1}2\,\Bigl[\om_0\Bigl(T_3\bigl(A\d B(x_1)\otimes A\d B(x_2)\otimes A^2(x_3)\bigr)\Bigr)\,
\vf(x_1)\vf(x_2)B^2(x_3)\notag\\
&+\,\om_0\Bigl(T_3\bigl(A\d\vf(x_1)\otimes A\d\vf(x_2)\otimes A^2(x_3)\bigr)\Bigr)\,
B(x_1)B(x_2)\vf^2(x_3)\Bigr]+2\,\text{permutations}\ ,\label{l8-2}\\
&\frac{-m_H^2}{8m^2}\,6\,\Bigl[\om_0\Bigl(T_3\bigl(A\d B(x_1)\otimes
A\d B(x_2)\otimes B^2(x_3)\bigr)\Bigr)\, \vf(x_1)\vf(x_2)B^2(x_3)\notag\\
&+\om_0\Bigl(T_3\bigl(A\d\vf(x_1)\otimes
A\d\vf(x_2)\otimes \vf^2(x_3)\bigr)\Bigr)\, B(x_1)B(x_2)\vf^2(x_3)\Bigr]
+2\,\text{permutations}\notag\\
&+\frac{m_H^2}{4m^2}\,4\,\om_0\Bigl(T_3\bigl(A\d B(x_1)\otimes
A\d\vf(x_2)\otimes B\vf(x_3)\bigr)\Bigr)\, \vf(x_1)B(x_2)\vf B(x_3)
+5\,\text{permutations}\ ,\label{l8-3}\\
&\omega_0\Bigl(T^{c}_4\bigl( A\d B(x_1)\otimes  A\d B(x_2)
\otimes A\d \vf(x_3)\otimes A\d \vf(x_4)\bigr)\Bigr)\notag\\
&\quad\quad\quad\quad\quad\cdot\vf(x_1)\vf(x_2)B(x_3)B(x_4)+5\,\text{permutations}\ ,\label{l8-4}
\end{align}
and we get
\be\label{l8} 
l_8=-i\,C_\mathrm{fish}\,\Bigl(8\frac{m^2}{m_H^2}+(3+2)\,\frac{m_H^2}{m^2}\Bigr)
-\frac{m^2}{m_H^2}\,(8+8)\,C_{2\triangle}+(12+12-16)\,C_{1\triangle}
+\frac{m^2}{m_H^2}\,8i\,C_{2\square}=l_7\ .
\ee

Note that \eqref{l3-4}, \eqref{l4-4}, \eqref{l7-2} and \eqref{l8-2} can be viewed 
also as fish diagram contributions, since their
unrenormalized versions are $\sim -g_{\mu\nu}\,\d^\mu\d^\nu D^F(x_1-x_2)D^F(x_1-x_3)D^F(x_2-x_3)
=i\,\delta(x_1-x_2)\,t^\circ_\mathrm{fish}(x_1-x_3)$; however in 
Sect.~\ref{ssec:geom-interpret} we treat them as triangle diagram contributions.

\section{Working out PGI-tree for the running interaction}
\label{app:PGI-tree}

In this appendix we work out PGI-tree for the interaction $\mathcal{L}(g)=z_\rho(L)(g)$, as defined after 
\eqref{PGI-connected}. We use that $\mathcal{ L}(g)$ is of the form \eqref{Lcal(g)} with the explicit 
expressions \eqref{L-j} and \eqref{L-k-j}, with unknown coefficients $e_\rho^{(j)}$ in
the $L_k^{(j)}$ ($k=0,1,2$) for $j\geq 1$. About the $Q$-vertex $\mathcal{P}^{\nu}(g;f)$
\eqref{P(g)} we only know that it is of the form \eqref{P-j}, the field polynomials $P_k^{(j)}$
are completely unknown.

 It is well-known (see
e.g.~\cite{DS99,Scharf2001,GraciaBondia2010}) that in the inductive Epstein-Glaser construction of 
the time ordered products, PGI can be violated only by local terms. Hence, we need to study only the local 
contributions. However, in principle the splitting of a
distribution into a local and a non-local part is non-unique; hence,
some caution is called for. Let $x_1,...,x_n$ be the vertices of the considered connected tree
diagram. Everywhere in our calculations we replace $(\d)\square
\Delta_m^F$ by $(-m^2(\d)\Delta_m^F -i (\d)\delta)$. Then, outside the total diagonal
$x_1=x_2=...=x_n$ only terms with at least one propagator $\Delta_m^F$, $\d_\mu \Delta_m^F$, 
$\d_\nu \d_\mu \Delta_m^F$ and $\d_\nu \d_\mu \d_\la\Delta_m^F$ (with no contraction of Lorentz indices) 
contribute. Since these terms cancel outside the total diagonal, they
cancel also on the total diagonal. The remaining terms are the local terms, they are
linear combinations of $\d^a\delta(x_1-x_n,...,x_{n-1}-x_n)$.
We write $T_\mathrm{tree}$ for the contribution of the connected
tree diagrams and $T(...)\vert_\mathrm{loc}$ means the selection of the local terms.
The latter is a rather delicate issue. Considering
\be
\d_\nu^x\,T_\mathrm{tree}\bigl(P^\nu(x)\otimes L(y)\bigr)\vert_\mathrm{loc}\ ,
\ee
there appear the following possibilities how the divergence $\d^x_\nu$ generates local terms
(cf.~\cite{DS99,GraciaBondia2010}):
\begin{enumerate}
\item[Type 1] If $P^\nu = \d^\nu \phi \,F +\cdots$ and 
$L = \phi \,E +\cdots$, then the contraction of
$\d^\nu \phi(x)$ with $\phi(y)$ gives a propagator 
$\hbar\,\d^\nu \Delta^F_m(x - y)$, and on computing its divergence we
find the local contribution $-i\hbar\,\delta(x-y)\,F(x)E(x)$.
\item[Type 2] If $P^\nu $ is as before and $L = \d^\mu\phi \,E +\cdots$, then 
analogously to type 1 we
obtain the local contribution $i\hbar\,\d^\mu\delta(x-y)\,F(x)E(y)$.
\item[Type 3] If $P^\nu = A^\nu \,F +\cdots$ and 
$L = (\d_\mu A^\mu) \,E +\cdots$, then the contraction of
$A^\nu(x)$ with $\d_\mu A^\mu(y)$ gives a propagator 
$\hbar\,g^{\nu\mu}\d_\mu \Delta^F_m(x - y)$, and we
get the local contribution $-i\hbar\,\delta(x-y)\,F(x)E(x)\ $.
\end{enumerate}

\begin{rem}
\item[(1)] Usually interactions for spin-1 fields do not contain a $\d_\mu A^\mu$-field;
hence, the type 3 mechanism is non-standard, however it has been used already in the 
application of PGI to spin-2 gauge theories \cite{Scharf2001}.
\item[(2)] In the literature about PGI mostly a different normalization of the 
time ordered products is used (denoted by $T^N$ in the following). Considering 
$S\bigl(i\, \mathcal{ L}(g)\bigr)$, where $\mathcal{ L}(g)$ is of the form \eqref{Lcal(g)}, 
the arguments of $T^N$ are only the vertices
$\mathcal{L}_{(1)}(x_j)$ which are of first order in $g$. A higher order vertex
$\int dx\,(g(x))^n\, \mathcal{L}_{(n)}(x)$ ($n\geq 2$) is taken into account as a local contribution
\be
n!\,(-i)^{n-1}\,\delta(x_1-x_n,...,x_{n-1}-x_n)\,\mathcal{ L}_{(n)}(x_n)\quad\text{to}\quad
T_{n,\mathrm{tree}}^N\bigl(\otimes_{j=1}^n \mathcal{L}_{(1)}(x_j)\bigr)\ .
\ee
Analogously a higher order $Q$-vertex $\int dx\,(g(x))^{(n-1)}\,
\mathcal{ P}_{(n)}^\nu (x)\,f(x)$ ($n\geq 2$) appears as a local contribution
\be
(n-1)!\,(-i)^{n-1}\,\delta(x_1-x_n,...,x_{n-1}-x_n)\,\mathcal{P}_{(n)}(x_n)\quad\text{to}\quad
T_{n,\mathrm{tree}}^N\bigl(\mathcal{P}_{(1)}(x_1)\otimes(\otimes_{j=2}^n \mathcal{L}_{(1)}(x_j))\bigr)
\ee
integrated out with $f(x_1)\,\prod_{j=2}^ng(x_j)$. The relation between the time ordered products 
$T^N$ and $T$ can generally be described in terms of the Main Theorem, see 
\cite[formula (2.29)]{Duetsch2005}.
\end{rem}

Now we are going to work out  PGI-tree. Selecting the local terms which are of order $\hbar^0$ 
and of a certain order in $\tau$ and $\kappa$, we obtain the following equations:
\begin{align}
\hbar^0\tau^0\ka^1:\quad &\tfrac{i}{\hbar}\,[Q,L_1(g)]\approx  -(\d P_1)(g)\ ,\label{PGI1}\\
\hbar^0\tau^0\ka^2:\quad &\tfrac{i}{\hbar}\,[Q,L_2(g^2)]\approx -\tfrac{i}{2}(\d P_2)(g^2)\nn\\
&\quad - \tfrac{i}{\hbar}\,\int dxdy\,g(x)g(y)\,\d_x
T_\mathrm{tree}\bigl(P_1(x)\otimes L_1(y)\bigr)\vert_\mathrm{loc}\ ,\label{PGI2}
\end{align}
\begin{align}
\hbar^0\tau^0\ka^3:\quad & 0\approx -\tfrac{i}{\hbar}\,\int dxdy\,g(x)(g(y))^2\,\Bigl(\d_x\,
T_\mathrm{tree}\bigl(P_1(x)\otimes L_2(y)\bigr)\vert_\mathrm{loc}\nn\\
&\quad\quad\quad +\tfrac{1}{2}\d_y\,
T_\mathrm{tree}\bigl(L_1(x)\otimes P_2(y)\bigr)\vert_\mathrm{loc}\Bigr)\ ,\label{PGI3}
\end{align}
\begin{align}
\hbar^0\tau^1\ka^2:\quad &\tfrac{i}{\hbar}\,[Q,L_0^{(1)}(g^2)]\approx  
-\tfrac{1}{2}\,(\d P_0^{(1)})(g^2)\ ,\label{PGI4}\\
\hbar^0\tau^1\ka^3:\quad &\tfrac{i}{\hbar}\,[Q,L_1^{(1)}(g^3)]\approx -\tfrac{1}{3}(\d P_1^{(1)})(g^3)\nn\\
&\quad - \tfrac{i}{\hbar}\,\int dxdy\,(g(x))^2g(y)\,\Bigl(\d_y
T_\mathrm{tree}\bigl(L_0^{(1)}(x)\otimes P_1(y)\bigr)\vert_\mathrm{loc}\nn\\
&\quad\quad\quad +\tfrac{1}{2}\d_x\,
T_\mathrm{tree}\bigl(P_0^{(1)}(x)\otimes L_1(y)\bigr)\vert_\mathrm{loc}\Bigr)\ ,\label{PGI5}
\end{align}
\begin{align}
\hbar^0\tau^1\ka^4:\quad &\tfrac{i}{\hbar}\,[Q,L_2^{(1)}(g^4)]\approx -\tfrac{1}{4}(\d P_2^{(1)})(g^4)\nn\\
&\quad - \tfrac{i}{\hbar}\,\int dxdy\,(g(x))^3g(y)\,\Bigl(\tfrac{1}{3}\,\d_x
T_\mathrm{tree}\bigl(P_1^{(1)}(x)\otimes L_1(y)\bigr)\vert_\mathrm{loc}\nn\\
&\quad\quad\quad +\d_y T_\mathrm{tree}\bigl(L_1^{(1)}(x)\otimes P_1(y)\bigr)\vert_\mathrm{loc}\Bigr)\nn\\
&\quad - \tfrac{i}{2\,\hbar}\,\int dxdy\,(g(x))^2(g(y))^2\,\Bigl(\d_x\,
T_\mathrm{tree}\bigl(P_0^{(1)}(x)\otimes L_2(y)\bigr)\vert_\mathrm{loc}\nn\\
&\quad\quad\quad +\d_y T_\mathrm{tree}\bigl(L_0^{(1)}(x)\otimes P_2(y)\bigr)\vert_\mathrm{loc}\Bigr)\nn\\
&\quad + \tfrac{1}{\hbar^2}\,\int dydx_1dx_2\,g(y)(g(x_1))^2g(x_2)\,\d_y\,
T_\mathrm{tree}\bigl(P_1(y)\otimes L_0^{(1)}(x_1)\otimes L_1(x_2)\bigr)\vert_\mathrm{loc}\ ,\label{PGI6}
\end{align}
\begin{align}
\hbar^0\tau^1\ka^5:\quad & 0\approx - \tfrac{i}{\hbar}\,\int dxdy\,(g(x))^3(g(y))^2\,\Bigl(\tfrac{1}{3}\,\d_x
T_\mathrm{tree}\bigl(P_1^{(1)}(x)\otimes L_2(y)\bigr)\vert_\mathrm{loc}\nn\\
&\quad\quad\quad +\tfrac{1}{2}\,\d_y T_\mathrm{tree}\bigl(L_1^{(1)}(x)\otimes P_2(y)\bigr)\vert_\mathrm{loc}\Bigr)\nn\\
&\quad - \tfrac{i}{\hbar}\,\int dxdy\,g(x)(g(y))^4\,\Bigl(\d_x\,
T_\mathrm{tree}\bigl(P_1(x)\otimes L_2^{(1)}(y)\bigr)\vert_\mathrm{loc}\nn\\
&\quad\quad\quad +\tfrac{1}{4}\,\d_y 
T_\mathrm{tree}\bigl(L_1(x)\otimes P_2^{(1)}(y)\bigr)\vert_\mathrm{loc}\Bigr)\nn\\
&\quad + \tfrac{1}{\hbar^2}\,\int dydx_1dx_2\,g(y)(g(x_1))^2(g(x_2))^2\,\d_y\,
T_\mathrm{tree}\bigl(P_1(y)\otimes L_0^{(1)}(x_1)\otimes L_2(x_2)\bigr)\vert_\mathrm{loc}\ .\label{PGI7}
\end{align}
This list contains {\it all} local terms of \eqref{PGI-connected} 
which are of order $\hbar^0\tau^0\ka^l$ or 
$\hbar^0\tau^1\ka^l$ for $l$ arbitrary. On computing the terms appearing in this list, we replace
$\square\phi$ by $-m_\phi^2\,\phi$.
\begin{itemize}

\item {\it PGI-equations \eqref{PGI1}-\eqref{PGI3}}. The $\tau^0$-equations 
express PGI-tree for the $(\rho=1)$-theory, 
they have a unique solution for $P_1$ and $P_2$ given in \eqref{eq:LPoriginal} (cf.~\cite{ADS97,Pepe2008}).
\item {\it PGI-equation~\eqref{PGI4}. (Tree diagrams with 2 external lines.)} Throughout this appendix 
we use the notation $e\,\log\rho:=e_\rho^{(1)}$ \eqref{e-notation}. 
With that \eqref{PGI4} is equivalent to
\be
a_1-a_2+b_2-c_2  = 0\quad\wedge\quad
b_2+b_0-b_1+c_2  = 0 \label{PGI4-1}
\ee
and a non-unique expression for $P_0^{(1)}$:
\be
\tfrac{1}{2}\,P_0^{(1)\nu}=(c_2+a_2)\,m^2\,A^\nu u+(b_2+b_0)
(\sigma\,m\,u\d^\nu B+(1-\sigma)\,m\,B\d^\nu u)\ ,\label{PGI4-3}
\ee
where $\si\in\CC$ is an arbitrary number.

\item {\it PGI-equation \eqref{PGI5}. (Tree diagrams with 3 external lines.)} A type 3 term appears only in 
$\d_y T_\mathrm{tree}\bigl(L_0^{(1)}(x)\otimes P_1(y)\bigr)\vert_\mathrm{loc}$. The equation \eqref{PGI5} 
is equivalent to the following relations: $P_1^{(1)}$ is of the form
\be
\tfrac{1}{3}\,P_1^{(1)\nu}=\alpha\,\vf B\d^\nu u+\beta\,\vf u\d^\nu B+
\gamma\,uB\d^\nu\vf+\lambda\,m\,A^\nu u\vf\ ,
\quad \alpha,\beta,\gamma,\lambda\in\CC\ ;\label{PGI5-0}
\ee
and
\begin{align}
m\,A\d u\vf: & \quad 0= -2l_0+l_2+\la-\tfrac{\si}{3}\,(b_2+b_0)-\tfrac{2}{3}\,b_2\ ,\label{PGI5-1}\\ 
m\,Au\d \vf: & \quad 0= -l_1+\la+\tfrac{5\si}{3}\,(b_2+b_0)-\tfrac{2}{3}\,b_2 \ ,\label{PGI5-2}\\
m\,\d Au\vf: & \quad 0= \la+\tfrac{2\si}{3}\,(b_2+b_0)-a_2+\tfrac{1}{3}\,b_2\ ,\label{PGI5-3}\\
B\d u\d\vf:  & \quad 0= -l_1+\alpha+\gamma+\tfrac{2}{3}\,c_0\ ,\label{PGI5-4}\\
\d B\d u\vf: & \quad 0= l_2+\alpha+\beta-\tfrac{2}{3}\,b_0\ ,\label{PGI5-5}\\
\d Bu\d\vf:  & \quad 0= \beta+\gamma+\tfrac{2}{3}\,c_0-\tfrac{2}{3}\,b_0\ ,\label{PGI5-6}\\
uB\vf:& \quad 0= m_H^2(l_4-\gamma-\si\,(b_2+b_0)-c_1+\tfrac{1}{3}\,c_0)\nn\\
 & \quad\quad +m^2 (-\alpha-\beta-(1-\si)(b_2+b_0)+b_1-\tfrac{1}{3}\,b_0)\ .\label{PGI5-7}
\end{align}
The equations \eqref{PGI5-1}-\eqref{PGI5-7} are obtained by
setting the coefficient of the indicated field monomial to zero.


\item {\it PGI-equation \eqref{PGI6}. (Tree diagrams with 4 external lines.)} A type 3 contribution appears only
in $\d_y T_\mathrm{tree}\bigl(L_0^{(1)}(x)\otimes P_4(y)\bigr)\vert_\mathrm{loc}$. 
There is only one type 1 contribution coming from a contraction of $\d u$ with $\ut$, namely in 
$\d_x T_\mathrm{tree}\bigl(P_1^{(1)}(x)\otimes L_1(y)\bigr)\vert_\mathrm{loc}$.
The last term in \eqref{PGI6} is the most difficult one; we explain the computation:
the local contributions come from terms of the form
\be\label{Bsp}
\tfrac{1}{\hbar^2}\,\int dydx_1dx_2\,g(y)(g(x_1))^2g(x_2)\,\d_\nu^y\,
T_\mathrm{tree}\bigl((G\d^\nu\phi)(y)\otimes \tfrac{1}{2}(\d\phi)^2(x_1)
\otimes (F_\tau\d^\tau\phi)(x_2)\bigr)\vert_\mathrm{loc}\ ,
\ee
where $\phi=B$ or $\phi=\vf$.
The contraction of $\d^\nu\phi(y)$ with $\d^\mu\phi(x_1)$ is of type 2; the contraction of $\d_\mu\phi(x_1)$ 
with $\d^\tau\phi(x_2)$ gives a propagator $i\hbar\,\d_\mu\d^\tau\Delta^F(x_1-x_2)$. With that \eqref{Bsp}
is equal to
\begin{align}
= & -i\int dydx_1dx_2\,g(y)(g(x_1))^2g(x_2)\,G(y)\,\d^\mu\delta(y-x_1)\,\d_\mu\d^\tau\Delta^F(x_1-x_2)\,
F_\tau(x_2)\vert_\mathrm{loc}\nn\\
= &i \int dydx_2\,\tfrac{\d_y^\mu(g(y))^3}{3}\,g(x_2)\,G(y)\,\d_\mu\d^\tau\Delta^F(y-x_2)\,
F_\tau(x_2)\vert_\mathrm{loc}\nn\\
= & \tfrac{-1}{3}\,\int dydx_2\,(g(y))^3\,g(x_2)\,G(y)\,\d^\tau\delta(y-x_2)\,F_\tau(x_2)\nn\\
= & \tfrac{1}{3}\,\int dy\,\Bigl(\tfrac{3}{4}\,\d_\tau (g(y))^4\,G(y)F_\tau(y)+
(g(y))^4\,\d^\tau G(y)F_\tau(y)\Bigr)\nn\\
= & \int dy\,(g(y))^4\,\Bigl(\tfrac{1}{12}\,\d^\tau G(y)\,F_\tau(y)-\tfrac{1}{4}\,G(y)\d^\tau F_\tau(y)\Bigr)\ ,
\end{align}
where non-local terms are omitted.
If the $x_2$-vertex is of the simpler form $(F\phi)(x_2)$, then $\d^\tau\Delta^F(x_1-x_2)\,F_\tau(x_2)$
is replaced by $-\Delta^F(x_1-x_2)\,F(x_2)$ and it results $\tfrac{1}3\,\int dy\,(g(y))^4\,G(y)F(y)\ $.

Proceeding as for \eqref{PGI5}, the PGI-equation \eqref{PGI6} is equivalent to the following: 
$P_2^{(1)}$ is of the form
\be
\tfrac{1}{4}\,P_2^{(1)\nu}=\Upsilon\,u\vf^2 A^\nu+\Xi\,uB^2 A^\nu\ ,\quad
\Upsilon,\Xi\in\CC\ ,\label{PGI6-0}
\ee
and
\begin{align}
A^2A\d u: &\quad 0 =  l_{11}\ ,\label{PGI6-1}\\
B\vf^2 u: &\quad 0 =-m\alpha+
\tfrac{m_H^2}{2m}(l_8-3l_3+2l_4-\si(b_2+b_0)-3\gamma-2\beta+c_0-\tfrac{2}{3}b_0)
\ ,\label{PGI6-3}\\
m\,A^2Bu: &\quad 0 =  -l_6+l_0+\si(b_2+b_0)+\gamma-\tfrac{1}{3}c_0\ ,\label{PGI6-4}\\
\tfrac{m_H^2}{2m}\,B^3 u: &\quad 0 =  l_9-l_4-\si(b_2+b_0)-\gamma+\tfrac{1}{3}c_0\ ,\label{PGI6-5}\\
A\vf^2\d u: &\quad 0 =  \Upsilon-l_5+\tfrac{3}{4}l_2-\tfrac{1}{4}\beta-\tfrac{1}{12}b_0\ ,\label{PGI6-6}\\
\d A\,\vf^2 u: &\quad 0 =  \Upsilon-\tfrac{1}{4}l_2-\tfrac{1}{2}a_2
+\tfrac{3}{4}\beta+\tfrac{1}{4}b_0\ ,\label{PGI6-7}\\
2Au\vf\d\vf: &\quad 0 =  \Upsilon+\tfrac{1}{4}l_2-\tfrac{1}{2}l_1+\tfrac{3}{4}\beta+\tfrac{1}{4}b_0\ ,\label{PGI6-8}\\
AB^2\d u: &\quad 0 =  \Xi-l_6+\tfrac{3}{4}l_1+\tfrac{1}{4}\gamma-\tfrac{1}{12}c_0\ ,\label{PGI6-9}\\
\d A\,B^2 u: &\quad 0 =  \Xi-\tfrac{1}{4}l_1-\tfrac{1}{2}a_2-\tfrac{3}{4}\gamma+\tfrac{1}{4}c_0\ ,\label{PGI6-10}\\
2AuB\d B: &\quad 0 =  \Xi+\tfrac{1}{4}l_1-\tfrac{1}{2}l_2-\tfrac{3}{4}\gamma+\tfrac{1}{4}c_0\ .\label{PGI6-11}
\end{align}


\item {\it PGI-equation~\eqref{PGI7}. (Tree diagrams with 5 external lines.)}
Note that
\be
\d_y T_\mathrm{tree}\bigl(L_1^{(1)}(x)\otimes P_2(y)\bigr)\vert_\mathrm{loc}=0=
\d_y T_\mathrm{tree}\bigl(L_1(x)\otimes P_2^{(1)}(y)\bigr)\vert_\mathrm{loc}\ ,
\ee
since $A^\nu(y)$ (appearing in $P_2(y)$ and $P_2^{(1)}(y)$) has no partner 
field $\d_\mu A^\mu(x)$ which is needed for a type 3 contribution. Proceeding as above we get
\begin{align}
uBA^2\vf: & \quad 0  = l_5-l_6+\beta+\gamma+\tfrac{1}{3}(b_0-c_0)\ ,\label{PGI7-1}\\
\tfrac{m_H^2}{2m^2}u\vf^3B: & \quad 0 = l_8-l_7-\beta-\gamma+\tfrac{1}{3}(c_0-b_0)\ ,\label{PGI7-2}\\
\tfrac{m_H^2}{2m^2}u\vf B^3: & \quad 0 = l_9-l_8-\beta-\gamma+\tfrac{1}{3}(c_0-b_0)\ .\label{PGI7-3}
\end{align}
\end{itemize}

The system of equations \eqref{PGI4-1}, \eqref{PGI5-1}-\eqref{PGI5-7},
\eqref{PGI6-1}-\eqref{PGI6-11} and \eqref{PGI7-1}-\eqref{PGI7-3} 
contains a lot of redundancies; the most general
solution is given in \eqref{parameter-PC}. To complete this result we add 
\begin{align}
&\alpha=0\ ,\quad
\beta =-l_1+\tfrac{2b_0}{3}\ ,\quad 
\gamma =l_1-\tfrac{2c_0}{3}\ ,\quad
\la = b_0+c_2-b_1-\tfrac{2l_1}{3}\ ,\nn\\
&\Upsilon =l_1-\tfrac{3b_0}{4}\ ,\quad
\Xi = l_1-\tfrac{3c_0}{4}
\end{align}
and the relation determining $\si$,
\be\label{sigma}
\si(b_1-c_2)=l_1-b_0+b_1-c_2\ .
\ee

The most general solution of the BRST-condition \eqref{s(z(L))=0} (given in 
\eqref{parameter-BRS1}-\eqref{parameter-BRS3} and 
for the pertinent $Q$-vertex see \eqref{Q-BRS}) is a {\it true} subset
of the PGI-tree solution computed here, due to Remark \ref{rem:BRS->PGI-tree}. 
This subset property is a good check of the calculations in this appendix.

The result \eqref{parameter-PC} gives the restrictions from PGI-tree on the
1-loop coefficients $e_\rho^{(1)}$. The corresponding restrictions on the 
higher loop coefficients  $e_\rho^{(2)},\ e_\rho^{(3)},\ldots $ can be obtained by continuing the 
calculations in this appendix: one has to select the local terms of \eqref{PGI-connected} 
which are of order $\hbar^0\tau^r\ka^l$ for $l$ arbitrary and $r=2,3,\ldots\ $.

\begin{rem}\label{rem:A4} 
We now are able to see, why the claim \eqref{lim[Q,A4]} holds true.
First note that PC for $S\bigl(iz_\rho(L)(g)\bigr)$ \eqref{stability-PC} implies PC for the 
\textit{connected} time-ordered products:
\be\label{PC-connected}
\lim_{\eps\downarrow 0}\,\, [Q,S^c\bigl(iz_\rho(L)(g_\eps)\bigr)]_{\star}\approx 0\ ,
  \ee
this follows analogously to \eqref{PGI-connected}. Now,
using the $\tau$-trick in this equation, the terms 
$\sim\tau^0$ vanish separately, because they are the $U(1)$-Higgs model, which fulfills PGI 
and, hence, also \eqref{PC-connected}. 
Therefore, taking $\tau =\hbar$ into account, there cannot be a cancellation of 
terms $\sim\hbar^0\tau^1\ka^4$ with terms $\sim\hbar^1\tau^0\ka^4$; hence, the terms 
$\sim\hbar^0\tau^1\ka^4$ (which are tree-terms) must fulfill \eqref{PC-connected} separately.
Moreover, as explained above, the non-local connected tree terms fulfill PGI separately and, hence, 
they fulfill also \eqref{PC-connected} separately. Now, as we see from \eqref{PGI6-1}, there is only 
one local connected (tree) term $\sim\hbar^0\tau^1\ka^4\,A^2A\d u$
contributing to the l.h.s.~of  \eqref{PC-connected}, namely the r.h.s. of
\eqref{lim[Q,A4]}; therefore, the latter must be $\approx 0$ individually.
\end{rem}

\bibliographystyle{amsalpha}
\bibliography{Literatur1}

\end{document}